%% file: Bond011510.tex
\documentclass[manuscript]{aastex}
\usepackage{graphicx}

\def\kms   {\hbox{km s$^{-1}$}}

\shorttitle{Milky Way Tomography III: Stellar Kinematics}
\shortauthors{Bond et al.}

\begin{document}

\title{The Milky Way Tomography with SDSS: III. Stellar Kinematics}

\input{abstract}

\input{introduction}

\input{methodology}

\input{analysisPM}

\input{analysis6D}

\input{model}

\input{conclusions}

\input{appendix}

\bibliographystyle{apj}

\bibliography{apj-jour,Bond011510}

\input{tables}

\input{figuresDR7}

\input{figuresApp}

\end{document}

%% file: abstract.tex
\author{
Nicholas A. Bond\altaffilmark{\ref{Rutgers}},
\v{Z}eljko Ivezi\'{c}\altaffilmark{\ref{Washington}},
Branimir Sesar\altaffilmark{\ref{Washington}},
Mario Juri\'{c}\altaffilmark{\ref{IAS}},
Jeffrey A. Munn\altaffilmark{\ref{USNOFlagstaff}}, 
Adam Kowalski\altaffilmark{\ref{Washington}},
Sarah Loebman\altaffilmark{\ref{Washington}},
Rok Ro\v{s}kar\altaffilmark{\ref{Washington}},
Timothy C. Beers\altaffilmark{\ref{JINA}},
Julianne Dalcanton\altaffilmark{\ref{Washington}},
Constance M. Rockosi\altaffilmark{\ref{UCSC}}, 
Brian Yanny\altaffilmark{\ref{FNAL}},
Heidi J. Newberg\altaffilmark{\ref{Rensselaer}}, 
Carlos Allende Prieto\altaffilmark{\ref{Texas}, \ref{Mullard}},
Ron Wilhelm\altaffilmark{\ref{TexasTech}},
Young Sun Lee\altaffilmark{\ref{JINA}},
Thirupathi Sivarani\altaffilmark{\ref{JINA},\ref{India}},
Steven R. Majewski\altaffilmark{\ref{UVa}},
John E. Norris\altaffilmark{\ref{RSAA}},
Coryn A.L. Bailer-Jones\altaffilmark{\ref{MPAstro}},
Paola Re Fiorentin\altaffilmark{\ref{MPAstro},\ref{Ljubljana}},
David Schlegel\altaffilmark{\ref{LBNL}},
Alan Uomoto\altaffilmark{\ref{JHU}},
Robert H. Lupton\altaffilmark{\ref{Princeton}},
Gillian R. Knapp\altaffilmark{\ref{Princeton}}, 
James E. Gunn\altaffilmark{\ref{Princeton}},
Kevin R. Covey\altaffilmark{\ref{Harvard}},
J. Allyn Smith\altaffilmark{\ref{AustinPeay}},
Gajus Miknaitis\altaffilmark{\ref{FNAL}},
Mamoru Doi,\altaffilmark{\ref{UT}},
Masayuki Tanaka\altaffilmark{\ref{UT2}},
Masataka Fukugita\altaffilmark{\ref{UT3}},
Steve Kent\altaffilmark{\ref{FNAL}},
Douglas Finkbeiner\altaffilmark{\ref{Harvard}},
Tom R. Quinn\altaffilmark{\ref{Washington}},
Suzanne Hawley\altaffilmark{\ref{Washington}},
Scott Anderson\altaffilmark{\ref{Washington}},
Furea Kiuchi\altaffilmark{\ref{Washington}},
Alex Chen\altaffilmark{\ref{Washington}},
James Bushong\altaffilmark{\ref{Washington}},
Harkirat Sohi\altaffilmark{\ref{Washington}},
Daryl Haggard\altaffilmark{\ref{Washington}},
Amy Kimball\altaffilmark{\ref{Washington}},
Rosalie McGurk\altaffilmark{\ref{Washington}},
John Barentine\altaffilmark{\ref{APO}}, 
Howard Brewington\altaffilmark{\ref{APO}}, 
Mike Harvanek\altaffilmark{\ref{APO}}, 
Scott Kleinman\altaffilmark{\ref{APO}}, 
Jurek Krzesinski\altaffilmark{\ref{APO}}, 
Dan Long\altaffilmark{\ref{APO}}, 
Atsuko Nitta\altaffilmark{\ref{APO}}, 
Stephanie Snedden\altaffilmark{\ref{APO}}, 
Brian Lee\altaffilmark{\ref{LBNL}},
Jeffrey R. Pier\altaffilmark{\ref{USNOFlagstaff}}, 
Hugh Harris\altaffilmark{\ref{USNOFlagstaff}}, 
Jonathan Brinkmann\altaffilmark{\ref{APO}}, 
Donald P. Schneider\altaffilmark{\ref{PennState}}
}

\altaffiltext{1}{Physics and Astronomy Department, Rutgers University
    Piscataway, NJ 08854-8019, U.S.A.
\label{Rutgers}}
\altaffiltext{2}{Department of Astronomy, University of Washington, Box 351580, Seattle, WA 98195
\label{Washington}}
\altaffiltext{3}{Institute for Advanced Study, 1 Einstein Drive, Princeton, NJ 08540
\label{IAS}}
\altaffiltext{4}{University of California--Santa Cruz, 1156 High St., Santa Cruz, CA 95060
\label{UCSC}}
\altaffiltext{5}{Fermi National Accelerator Laboratory, P.O. Box 500, Batavia, IL 60510
\label{FNAL}}
\altaffiltext{6}{Department of Physics, Applied Physics, and Astronomy,
Rensselaer Polytechnic Institute, 110 8th St., Troy, NY 12180
\label{Rensselaer}}
\altaffiltext{7}{Dept. of Physics \& Astronomy and JINA: Joint Institute for Nuclear Astrophysics, Michigan State University, 
East Lansing, MI  48824, USA 
\label{JINA}}
\altaffiltext{8}{Department of Astronomy, University of Virginia,
       P.O. Box 400325, Charlottesville, VA 22904-4325
\label{UVa}}
\altaffiltext{9}{McDonald Observatory and Department of Astronomy, 
University of Texas, Austin, TX 78712
\label{Texas}}
\altaffiltext{10}{Department of Physics, Texas Tech University, Box 41051, Lubbock, TX 79409
\label{TexasTech}}
\altaffiltext{11}{Research School of Astronomy \& Astrophysics, The Australian National 
University, Cotter Road, Weston, ACT 2611, Australia 
\label{RSAA}}
\altaffiltext{12}{Max Planck Institut f\"{u}r Astronomie, K\"{o}nigstuhl 17,
69117 Heidelberg, Germany
\label{MPAstro}}
\altaffiltext{13}{Department of Physics, University of Ljubljana, Jadranska 19, 
1000 Ljubljana, Slovenia
\label{Ljubljana}}
\altaffiltext{14}{Lawrence Berkeley National Laboratory, One Cyclotron Road, 
MS 50R5032, Berkeley, CA, 94720 
\label{LBNL}}
\altaffiltext{15}{Department of Physics and Astronomy, 
The John Hopkins University, 3701 San Martin Drive, Baltimore, MD 21218
\label{JHU}}
\altaffiltext{16}{Princeton University Observatory, Princeton, NJ 08544
\label{Princeton}}
\altaffiltext{17}{Harvard-Smithsonian Center for Astrophysics, 60 Garden Street,
                          Cambridge, MA 02138
\label{Harvard}}
\altaffiltext{18}{Dept. of Physics \& Astronomy, Austin Peay State University, 
Clarksville, TN 37044
\label{AustinPeay}}
\altaffiltext{19}{Institute of Astronomy, University of Tokyo, 2-21-1 Osawa,
Mitaka, Tokyo 181-0015, Japan
\label{UT}}
\altaffiltext{20}{Dept. of Astronomy, Graduate School of Science, University of Tokyo,
Hongo 7-3-1, Bunkyo-ku, Tokyo, 113-0033, Japan
\label{UT2}}
\altaffiltext{21}{
Institute for Cosmic Ray Research, University of Tokyo, Kashiwa, Chiba, Japan
\label{UT3}}
\altaffiltext{22}{U.S. Naval Observatory, Flagstaff Station, P.O. Box 1149, Flagstaff, AZ 86002
\label{USNOFlagstaff}}
\altaffiltext{23}{Apache Point Observatory, 2001 Apache Point Road, P.O. Box 59, 
Sunspot, NM 88349-0059
\label{APO}}
\altaffiltext{24}{Department of Astronomy
and Astrophysics, The Pennsylvania State University, University Park, PA 16802
\label{PennState}}
\altaffiltext{25}{Mullard Space Science Laboratory, University College London,
Holmbury St. Mary, Dorking, Surrey, RH5 6NT, UK
\label{Mullard}}
\altaffiltext{26}{Indian Institute of Astrophysics, Bangalore, 560034, India
\label{India}}

\begin{abstract}

We study Milky Way kinematics using a sample of 18.8 million main-sequence
stars with $r<20$ and proper-motion measurements derived from SDSS and POSS
astrometry, including $\sim$170,000 stars with radial-velocity measurements
from the SDSS spectroscopic survey. Distances to stars are
determined using a photometric parallax relation, covering a 
distance range from $\sim$100 pc to 10 kpc over a quarter of the sky at high
Galactic latitudes ($|b|>20^\circ$).  We find that in the region defined by 1 kpc $<Z<$ 5 kpc and 3 kpc $<R<$
13 kpc, the rotational velocity for disk stars smoothly decreases, and all three
components of the velocity dispersion increase, with distance from the Galactic
plane. In contrast, the velocity ellipsoid for halo stars is aligned with
a spherical coordinate system and appears to be spatially invariant within the probed volume.
The velocity distribution of nearby ($Z<1$ kpc) K/M stars is complex, and
cannot be described by a standard Schwarzschild ellipsoid. For stars in a
distance-limited subsample of stars ($<$100 pc), we detect
a multimodal velocity distribution consistent with that seen by HIPPARCOS. 
This strong non-Gaussianity significantly affects the
measurements of the velocity ellipsoid tilt and vertex deviation when using the
Schwarzschild approximation. We develop and test a simple descriptive
model for the overall kinematic behavior that captures these features over most
of the probed volume, and can be used to search for substructure in
kinematic and metallicity space. We use this model to predict further
improvements in kinematic mapping of the Galaxy expected from Gaia and LSST. 
\end{abstract}
\keywords{
methods: data analysis ---
stars: statistics ---
Galaxy: disk, halo, kinematics and dynamics, stellar content, structure
}

%% file: introduction.tex
\section{                        INTRODUCTION                             }

The Milky Way is a complex and dynamic structure that is constantly being shaped
by the infall of matter from the Local Group and mergers with neighboring
galaxies. From our vantage point inside the disk of the Milky Way, we have a unique
opportunity to study an $\sim L^*$ spiral galaxy in great detail. By measuring
and analyzing the properties of large numbers of individual stars, we can map
the Milky Way in a nine-dimensional space spanned by the three spatial
coordinates, three velocity components, and three stellar parameters --
luminosity, effective temperature, and metallicity.

In this paper, the third in a series of related studies, we use data obtained by
the Sloan Digital Sky Survey \citep{SDSS} to study in detail the distribution of
tens of millions of stars in this multi-dimensional space. In \citet[][hereafter
J08]{Juric08}, we examined the spatial distribution of stars in the Galaxy, and
in \citet[][hereafter I08]{Ivezic08} we extended our analysis to include the
metallicity distribution. In this paper, working with a kinematic data set
unprecedented in size, we investigate the distribution of stellar velocities. Our data
include measurements from the SDSS astrometric, photometric, and spectroscopic
surveys: the SDSS Data Release 7 \citep{DR7} radial-velocity sample includes $\sim 170,000$
main-sequence stars, while the proper-motion sample includes $18.8$ million stars,
with about $6.8$ million F/G stars for which photometric metallicity estimates
are also available. These stars sample a distance range from $\sim 100$~pc to
$\sim 10$~kpc, probing much farther from Earth than the HIPPARCOS sample, which covers only the nearest $\sim 100$~pc \citep[e.g.,][]{DB98,Nordstrom04}.
With the SDSS data set, we are offered for the first time an opportunity to
examine {\it in situ} the thin/thick disk and disk/halo boundaries over a large
solid angle, using millions of stars. 

In all three of the papers in this series, we have employed a set of photometric
parallax relations, enabled by accurate SDSS multi-color measurements, to
estimate the distances to main-sequence stars. With these distances, accurate to
$\sim 10-15$\%, the stellar distribution in the multi-dimensional phase space
can be mapped and analyzed without any additional assumptions. The primary aim
of this paper is thus to develop quantitative understanding of the large-scale
kinematic behavior of the disk and halo stars. From the point of view of an observer,
the goal is to measure and describe the radial-velocity and proper-motion
distributions as functions of the position in, for example, the $r$ vs. $g-r$
color-magnitude diagram, and as functions of the position of the analyzed sample
on the sky. From the point of view of a theorist, we seek to directly quantify the
behavior of the probability distribution function, $p(v_R, v_\phi, v_Z | R,
\phi, Z, T, L, [{\rm Fe}/{\rm H}])$, where $(v_R, v_\phi, v_Z)$ are the three velocity components
in a cylindrical coordinate system, $(R, \phi, Z)$ describe the position of a
star in the Galaxy, and $T$, $L$, and $[{\rm Fe/H}]$ are its temperature, luminosity, and metallicity, respectively (``$|$'' means ``given'').

This a different approach than that taken by the widely-used ``Besan\c{c}on''
Galaxy model \citep[][and references therein]{Robin86,Robin03}, which attempts to
generate model stellar distributions from ``first principles'' (such as an
adopted initial mass function) and requires dynamical self-consistency. Instead,
we simply seek to describe the directly observed distributions of kinematic and
chemical quantities without imposing any additional constraints. If these
distributions can be described in terms of simple functions, then one can try to
understand and model these simple abstractions, rather than the full voluminous
data set.  

As discussed in detail by J08 and I08, the disk and halo components have spatial
and metallicity distributions that are well-fit by simple analytic models
within the volume probed by SDSS (and outside regions with strong substructure,
such as the Sgr dwarf tidal stream and the Monoceros stream). In this paper, we develop
analogous models that describe the velocity distributions of disk and halo
stars. 

Questions we ask include: 
\begin{itemize}
\item What are the limitations of the Schwarzschild ellipsoidal approximation 
\citep[a three-dimensional Gaussian distribution,][]{Schwarzschild79} for describing the velocity distributions? 
\item Given the increased distance range compared to older data sets, can we 
detect spatial variation of the best-fit Schwarzschild ellipsoid parameters, 
including its orientation? 
\item Does the halo rotate on average? 
\item Is the kinematic difference between disk and halo stars as
remarkable as the difference in their metallicity distributions? 
\item Do large spatial substructures, which are also traced in metallicity 
space, have distinctive kinematic behavior?
\end{itemize}

Of course, answers to a number of these questions are known to some extent
\citep[for excellent reviews, see][for context and references, see also the
first two papers in this series]{GWK89,Majewski93,Helmi08}. For example, it has
been known at least since the seminal paper of \citet{ELS62} that
high-metallicity disk stars move on nearly circular orbits, while many
low-metallicity halo stars move on eccentric, randomly oriented orbits. However,
given the order of magnitude increase in the number of stars compared to previous work, larger distance limits, and accurate and diverse measurements
obtained with the same facility, the previous results (see I08 for a summary of
kinematic results) can be significantly improved and expanded.

The main sections of this paper include a description of the data and methodology
(\S 2), followed by analysis of the various stellar subsamples. In \S 3, we
begin by analyzing the proper-motion sample and determining the dependence of
the azimuthal (rotational) and radial-velocity distributions on position for
halo and disk subsamples selected along $l=0^\circ$ and $l=180^\circ$. The
spectroscopic sample is used in \S 4 to obtain constraints on the behavior of the
vertical-velocity component, and to measure the velocity-ellipsoid tilts. The
resulting model is then compared to the full proper-motion sample and 
radial-velocity samples in \S 5. Finally, in \S 6, we summarize and discuss our
results, including a comparison with prior results and other work based on SDSS
data.

%% file: methodology.tex
.
\section{     DATA AND METHODOLOGY}
\label{sec:methodology}

The characteristics of the SDSS imaging and spectroscopic data relevant to this
work \citep{F96, Gunn98, Hogg02, Smith02, EDR, Pier03, Ivezic04, Tucker06, Gunn06, 
DR7, Yanny09} are described in detail in the first two papers in the series (J08, I08).
Here, we only briefly summarize the photometric-parallax and photometric-metallicity
methods, and then describe the proper-motion data and their error analysis. The
subsample definitions are described at the end of this section.

\subsection{ The Photometric Parallax Method \label{subsec:photpar}}

The majority of stars in the SDSS imaging catalogs ($\sim 90$\%) are on the main sequence (J08 and references therein) and, using the broadband colors measured by SDSS, it is possible to estimate their absolute magnitude.  Briefly, the $r$-band absolute magnitude, $M_r$, of a star can be estimated from its position on the stellar locus of the $M_r$ vs. $g-i$ color-magnitude diagram.  The position of this stellar locus is in turn sensitive to metallicity, so we must apply an additional correction to the absolute magnitude.  A maximum-likelihood implementation of this method was introduced and discussed in detail in J08. The method was further refined by I08, who calibrated its dependence on metallicity using globular clusters.

We estimate absolute magnitudes using equation A7 in I08, which corrects for age effects, 
and equation A2 in the same paper to account for the impact of metallicity. The resulting 
distance range covered by the photometric parallax relation depends upon color and metallicity, 
but spans $\sim 100$~pc to $\sim 10$~kpc.  Based on an analysis of stars in globular clusters, 
I08 estimate that the probable systematic errors in absolute magnitudes determined using these 
relations are about 0.1 mag, corresponding to $5$\% systematic distance errors (in addition to 
the $10-15$\% random distance errors).  In addition, \citet[][hereafter SIJ08]{Sesar08} used 
a large sample of candidate wide-binary stars to show that the expected error distribution is 
mildly non-Gaussian, with a root-mean-square (rms) scatter in absolute magnitude of $\sim 0.3$~mag. 
They also quantified biases in the derived absolute magnitudes due to unresolved binary stars.

\subsection{ The Photometric Metallicity Method}

Stellar metallicity can significantly affect the position of a star in the
color-magnitude diagram (there is a shift of $\sim 1$~mag between the median
halo metallicity of $[{\rm Fe/H}] \sim -1.5$ and the median disk metallicity of $[{\rm Fe/H}]
\sim -0.2$). SDSS spectroscopy is only available for a small fraction of the
stars in our sample, so we adopt a photometric metallicity method based on SDSS
$u-g$ and $g-r$ colors. This relation was originally calibrated by I08 using
SDSS spectroscopic metallicities. However, the calibration of SDSS spectroscopic
metallicity changed at the high-metallicity end after SDSS Data Release 6 \citep{DR6}.
Therefore, we recalibrate their expressions as described in the Appendix. The
new calibration, given in equation (A1), is applicable to F/G stars with $0.2 <
g-r < 0.6$ and has photometric-metallicity errors that approximately follow a
Gaussian distribution with a width of $0.26$~dex. In addition, the $\sim
0.1$~dex systematic uncertainties in SDSS spectroscopic metallicity
\citep{Beers06,Allende06,Lee08a, Allende08} are inherited by the photometric
metallicity estimator. We emphasize that photometric metallicity estimates are
only robust in the range $-2<[{\rm Fe/H}]<0$ (see Appendix for details). 

For stars with $g-r > 0.6$, we assume a constant metallicity of $[{\rm Fe/H}]=-0.7$,
motivated by results for the disk metallicity distribution presented in I08 and
the fact that SDSS data are too shallow to include a large fraction of red halo
stars. A slightly better approach would be to use the disk metallicity distribution
from I08 to solve for best-fit distance iteratively, but the resulting changes
in the photometric distances are negligible compared to other systematic errors.

\subsection{ The SDSS-POSS Proper Motion Catalog}

We take proper-motion measurements from the \citet{Munn04} catalog (distributed as a part of the public SDSS data releases), which is based on a comparison of astrometric measurements between SDSS and a collection of Schmidt photographic surveys. Despite the sizable random and systematic astrometric errors in the Schmidt surveys, the combination of a long baseline ($\sim 50$~years for the POSS-I survey), and a recalibration of the photographic data using the positions of SDSS galaxies (see Munn et al. for details), results in median random proper-motion errors (per component) of only $\sim 3$~mas~yr$^{-1}$ for $r<18$ and $\sim 5$~mas~yr$^{-1}$ for $r<20$. As shown below, systematic errors are typically an order of magnitude smaller. At a distance of $1$~kpc, a random error of $3$~mas~yr$^{-1}$ corresponds to a velocity error of $\sim 15$~\kms, which is comparable to the radial velocity accuracy delivered by the SDSS stellar spectroscopic survey ($\sim5.3$~\kms\ at $g=18$ and $20$~\kms\ at $g=20.3$; Schlaufman et al. 2009). At a distance of $7$~kpc, a random error of $3$~mas~yr$^{-1}$ corresponds to a velocity error of $100$~\kms, which still represents a usable measurement for large samples, given that systematic errors are much smaller ($\sim 20$~\kms\ at a distance of $7$~kpc). The small and well-understood proper-motion errors, together with the large distance limit and sample size (proper-motion measurements are available for about $38$ million stars with $r<20$ from SDSS Data Release 7) make this catalog an unprecedented resource for studying the kinematics of Milky Way stars.

{\it We warn the reader that proper-motion measurements made publicly available prior
  to SDSS Data Release 7 are known to have significant systematic errors}. Here
  we use a revised set of proper-motion measurements \citep{Munn08}, which are
  publicly available only since Data Release 7. As described in the next
  section, we can assess the error properties of this revised proper motion
  catalog using objects with known zero proper motion -- that is, distant quasars.

\subsubsection{Determination of Proper Motion Errors Using Quasars }

All known quasars are sufficiently distant that their proper motions are
vanishingly small compared to the expected random and systematic errors in the
Munn et al. catalog. The large number of spectroscopically-confirmed SDSS
quasars \citep{Schneider07} which were not used in the recalibration of POSS
astrometry can therefore be used to derive robust independent estimates of these
errors. In SDSS Data Release 7, there are $69,916$ quasars with $14.5<r<20$,
redshifts in the range $0.5<z<2.5$, and available proper motions (see Appendix
for the SQL query used to select and download the relevant data from the SDSS
CAS). Within this sample of quasars, the proper motions have a standard
deviation of $\sim 3.1$~mas~yr$^{-1}$ for each component (determined from
the inter-quartile range), with medians differing from zero by less than
$0.2$~mas~yr$^{-1}$. The dependence of the random error on $r$-band magnitude is
well-described by 
\begin{equation} 
    \sigma_\mu = 2.7 + 2.0 \times 10^{0.4\,(r-20)} \, {\rm mas~yr^{-1}}
\label{QSOpmErr}
\end{equation}  
fitting only to quasars in the range $15 <r < 20$. When the measurements of each
proper-motion component are normalized by $\sigma_\mu$, the resulting
distribution is approximately Gaussian, with only $\sim 1.8$\% of the quasar
sample deviating by more than $3\sigma$ from zero proper motion. In addition to
their dependence on magnitude, the random proper-motion errors also depend on
position on the sky, but the variation is relatively small ($\sim 20$\%, see
right panels in Figure~\ref{qsoPMerrors}). Finally, we find that the correlation
between the errors in the two components is negligible compared to the total
random and systematic errors.

The median proper motion for the full quasar sample is $\sim 0.2$~mas~yr$^{-1}$,
but the systematic errors can be larger by a factor of $2-3$ in small sky
patches, as illustrated in Figure~\ref{qsoPMerrors}. We find that the
distribution of systematic proper-motion errors in $\sim 100$~deg$^2$ 
patches of sky has a width of $\sim 0.67$~mas~yr$^{-1}$ in each component, about twice
as large as that expected from purely statistical noise (per bin, using
equation~(\ref{QSOpmErr})). As the figure shows, a few regions of the sky have coherent
systematic errors at a level $\sim 1$~mas~yr$^{-1}$ (e.g., the median $\mu_l$ towards
$l\sim 270^\circ$, or $\mu_b$ towards the inner Galaxy). Therefore, the
kinematics measured using proper motions in these regions should be treated with
caution.

The largest systematic errors, $\sim 1$~mas~yr$^{-1}$ for $\mu_l$, are seen
toward $l \sim 270^\circ$ in the top left panel in Figure~\ref{qsoPMerrors}, which
corresponds to $\delta \la 10^\circ$. In this region, the systematic deviation
of quasar proper motions from zero is approximately parallel to lines of
constant right ascension, suggesting that the data may be suffering from
systematic effects due to atmospheric refraction and spectral differences
between quasars and galaxies used in the recalibration of POSS astrometry. This
effect would be strongest for observations obtained at high airmass, as are
typical for fields at low declination (the POSS data were obtained at a latitude
of $+33^\circ$). We find that the median quasar proper motion in the $\delta$
direction is well-described by
\begin{equation}
\langle \mu_{\delta} \rangle = -0.72 + 0.019\, \delta \,\,\, {\rm mas~yr^{-1}}
\end{equation}
for $-5^\circ < \delta < 30^\circ$, where $\delta$ is in degrees.  At
$\delta > 30^\circ$, we find $\langle \mu_{\delta}
\rangle\la0.2$~mas~yr$^{-1}$.

The observed direction and magnitude of this systematic offset (corresponding to
an astrometric displacement of up to $\sim 30$~mas) are consistent with detailed
studies of atmospheric dispersion effects on observations of quasars
\citep{Kaczmarczik09}. Therefore, it is possible that the systematic errors in {\it
stellar} proper motions (whose spectral energy distributions differ less from
galaxy spectral energy distributions than is the case for quasars) are smaller
than implied by Figure~\ref{qsoPMerrors}. Nevertheless, we will conservatively
adopt the quasar proper-motion distributions as independent estimates of
systematic and random proper motion errors for stars analyzed in this work. In
particular, we adopt $0.6$~mas~yr$^{-1}$ as an estimate for typical systematic
proper-motion error.

The quasar sample has a much narrower color range than that seen in
main-sequence stars ($96$\% of the quasar sample satisfies $-0.2 < g-r < 0.6$),
and provides a better estimate of systematic proper-motion errors for the blue
stars than for the red stars. Within the above well-sampled color range, we find
a median proper-motion gradient with respect to the $g-r$ color of $\la
0.1$~mas~yr$^{-1}$~mag$^{-1}$ (per component). When the fit is extended to
$g-r<1.6$ (using a much smaller number of quasars), the gradient is still
smaller than $0.5$~mas~yr$^{-1}$~mag$^{-1}$. Hence, the proper-motion
systematics have a color dependence that is smaller than, or at most comparable
to, their dependence on sky position.

A systematic error in proper motion of $0.6$~mas~yr$^{-1}$ corresponds to a
systematic velocity error of $3$~\kms\ at $1$~kpc, and $\sim 20$~\kms\ at
$7$~kpc. In addition, the $\sim 5$\% systematic distance errors discussed in
\S~\ref{subsec:photpar} are responsible for a $\sim 5$\% systematic velocity
uncertainty. Hence, for a disk-like heliocentric tangential velocity of
$20$~\kms, proper-motion systematics dominate at distances beyond $\sim 1$~kpc.
Similarly, for a halo-like heliocentric tangential velocity of $200$~\kms,
proper motion systematics will dominate at distances greater than $7$~kpc. At
smaller distances, the dominant systematic in our tangential-velocity estimates
comes from systematic distance errors. For most of the Galaxy volume analyzed in
this work, the systematic distance errors dominate over systematic proper-motion
errors.

\subsection{Comparison of Proper Motions with Independent Measurements}

As further tests of the proper-motion errors, we have analyzed two independent
sets of measurements. As shown below, they confirm the results based on our
analysis of the quasar sample.

We have compared the SDSS-POSS proper motions to proper-motion measurements by
\citet{Majewski92} for a sample of 326 stars observed towards the North Galactic
Pole. The measurements in the Majewski sample have random errors that are three times smaller, and
comparable, but most likely, different systematic errors. The median 
proper-motion differences between the two data sets are below $1$~mas~yr$^{-1}$, with
an rms scatter $3-4$~mas~yr$^{-1}$ (per coordinate). Hence, this comparison is
consistent with our error estimates discussed in the preceding section, and with
the estimates of \citet{Majewski92}.

We have also compared the SDSS-POSS proper motions with proper motions from the
SDSS stripe 82 region. In \citet{Bramich08}, proper motions are computed using
only SDSS data, and thus they are expected to have different, and probably
smaller, systematic errors than the SDSS-POSS proper motions (random errors for
the stripe 82 proper motions are larger by about a factor of two). For
$\sim$500,000 stars with both SDSS-POSS and Bramich et al. proper-motion
measurements, we find the median differences and the rms scatter to agree with
expectation. A single worrisome result is that the median difference between
the two data sets is a function of magnitude: we find a gradient of
$0.8$~mas~yr$^{-1}$ between $r=15$ and $r=20$. It is more likely that this
gradient is due to systematic errors in centroiding sources on photographic
plates, rather than a problem with SDSS data. This gradient corresponds to a
systematic velocity error as a function of distance, $\Delta v \sim$ 4 
($D$/kpc) \kms. For example, a halo star at 5 kpc, with a relative velocity of 200
\kms, would have a systematic velocity uncertainty of 10\%. This systematic
error is comparable to other sources of systematic errors discussed above, and
has to be taken into account when interpreting our results below.

\subsection{ The Main Stellar Samples }

When using proper motions, random errors in the inferred velocities have a
strong dependence on magnitude, and therefore distance, while systematic errors
are a function of position on the sky, as discussed above. Random errors in
radial-velocity measurements also depend on magnitude, as fits to spectral
features become more difficult at lower signal-to-noise ratios. As such, when
radial-velocity and proper-motion measurements are analyzed simultaneously, the
systematic and random errors combine in a complex way -- care is needed when
interpreting the results of such an analysis.

In order to minimize these difficulties, we separately analyze the proper-motion
sample and the much smaller sample of stars with radial velocities. Furthermore,
motivated by the metallicity distribution functions presented in I08, we
separately treat the low-metallicity ``halo'' stars and the high-metallicity ``disk''
stars. For these two samples, we require $g-r<0.6$, the regime in which the
photometric metallicity estimator is believed to be accurate. Finally, we 
discuss a sample of ``red'' stars with $g-r>0.6$ (roughly, $g-i>0.8$), which are
dominated by nearby ($<2$~kpc) disk stars. 

These samples are selected from SDSS Data Release 7 using the 
following common criteria:
\begin{enumerate}
\item Unique unresolved sources that show subarcsecond parallax: binary processing flags
      DEBLENDED\_AS\_MOVING, SATURATED, BLENDED, BRIGHT, and NODEBLEND
      must be false, and parameter nCHILD=0 
\item The interstellar extinction in the $r$ band, $A_r <0.3$~mag
\item Dust-corrected magnitudes in the range $14.5 < r < 20$~mag
\item High galactic latitudes: $|b|>20^\circ$
\item Proper motion available,
\end{enumerate}
yielding $20.1$ million stars.  The dust corrections, $A_r$, were computed
using the \citet{SFD98} dust maps, with conversion coefficients derived assuming an $R_V=3.1$
dust model.  The intersection of the following color criteria then
selects stars from the main stellar locus:
\begin{itemize}
\item Blue stars ($6.9$ million):
  \begin{enumerate}
    \item $0.2 < (g-r) < 0.6$
    \item $0.7 < (u-g) < 2.0$ and $-0.25 < (g-r) - 0.5(u-g) < 0.05$
    \item $-0.2 < 0.35(g-r) - (r-i) < 0.10$
  \end{enumerate}
\item Red stars ($11.9$ million):
  \begin{enumerate}
    \item $0.6 < (g-r) < 1.6$
    \item $-0.15 < -0.270 r + 0.800 i - 0.534 z + 0.054 < 0.15$,
  \end{enumerate}
\end{itemize}
where the last condition is based on a ``principal color'' orthogonal
to the stellar locus in the $i-z$ vs. $r-i$ color-color diagram, as
defined in \citet{Ivezic04}.  This condition allows for a
$0.15$~mag offset from the stellar locus. 

During the analysis, ``blue'' stars are often further divided by photometric
metallicity (see below for details) into candidate ``halo'' stars ($[{\rm Fe/H}]<
-1.1$) and candidate ``disk'' stars ($[{\rm Fe/H}]>-0.9$). Subsamples with
intermediate metallicities include non-negligible fractions of both halo and
disk stars. Although the reduced proper-motion diagram is frequently used for
the separation and analysis of these two populations, we find it inadequate for
our purposes; the vertical gradient in
rotational velocity blurs the kinematic distinction between disk and halo (for a
discussion, see SIJ08), and thus this method is applicable only to stars with
significant proper motion (leading to severe selection effects). Although metallicities are not available for red stars, results
from I08 imply that they are dominated by the disk population (red stars can
only be seen out to $\sim 2$~kpc).

For each of the subsamples defined above, we further separate those objects with 
SDSS spectroscopic data (see Appendix for a sample SQL query) into independent 
subsamples. In total, these spectroscopic subsamples include $172,000$ stars (out 
of $352,000$ stars with spectra), after an additional requirement to select only 
main-sequence stars; that is, stars with log$(g)>3.5$\footnote{Note that the majority 
of stars with $g-r>1.2$ do not have reliable measurements of log$(g)$ -- we assume 
that all stars with $g-r>1.2$ are main-sequence stars.}.  Of the stars with 
spectroscopic data, $111,000$ are blue ($0.2<g-r<0.6$) and $61,000$ are red 
($0.6<g-r<1.6$). When separating low- and high-metallicity stars with spectra, we 
use the spectroscopic metallicity (see \citet{Allende06} for details). Due to 
increased difficulties with measuring $[{\rm Fe/H}]$ for red stars ($g-r>0.6$) from 
SDSS spectra, we adopt $[{\rm Fe/H}]=-0.7$ for all such stars; this value is the 
median spectroscopic $[{\rm Fe/H}]$ for stars with $0.6<g-r<1.3$ ($\sigma$=0.4 dex). 
For over 90\% of $\sim$30,000 stars with $g-r>1.3$, $[{\rm Fe/H}]$ is not 
successfully measured.

\subsection{ Coordinate Systems and Transformations }

Following J08 and I08, we use a right-handed, Cartesian 
Galactocentric coordinate system defined by the following 
set of coordinate transformations:
\begin{eqnarray}
    X =  R_\odot - D\,\cos(l)\,\cos(b) \nonumber \\ 
    Y = -D\,\sin(l)\,\cos(b) \\ 
    Z =  D\,\sin(b), \nonumber
\end{eqnarray}
where $R_\odot = 8$~kpc is the adopted distance to the Galactic center, $D$ is
distance of the star from the Sun, and $(l,b)$ are the Galactic coordinates. Note that the $Z=0$ plane
passes through the Sun, not the Galactic center (see J08), the $X$ axis is
oriented toward $l=180^\circ$, and the $Y$ axis is oriented toward $l=270^\circ$
(the disk rotates toward $l\sim90^\circ$). The main reason for adopting a
Galactocentric coordinate system, rather than a traditional heliocentric system,
is that new data sets extend far beyond the solar neighborhood. 

We also employ a cylindrical coordinate system defined
by
\begin{eqnarray}
        R=\sqrt{X^2+Y^2}, \quad
     \phi=\tan^{-1}\left(\frac{Y}{X}\right). 
\end{eqnarray}

The tangential velocity, $v$, is obtained from the proper motion, $\mu$, 
and the distance $D$ by
\begin{equation}
 v  = 4.74 \,\, {\mu \over {\rm mas~yr^{-1}}} \,\, {D \over {\rm kpc} } \,\, {\rm \kms}.
\end{equation}

Given the line-of-sight radial velocity, $v_{{\rm rad}}$, and the two components
of tangential velocity aligned with the Galactic coordinate system, $v_l$
and $v_b$, the observed heliocentric Cartesian velocity components are
given by:
\begin{eqnarray} 
\label{measV}
 v_X^{{\rm obs}} = -v_{{\rm rad}} \cos(l) \cos(b) + v_b \cos(l) \sin(b) + v_l \sin(l) \nonumber \\
 v_Y^{{\rm obs}} = -v_{{\rm rad}} \sin(l) \cos(b) + v_b \sin(l) \sin(b) - v_l \cos(l)  \\
 v_Z^{{\rm obs}} = -v_{{\rm rad}} \sin(b) + v_b \cos(b).
 \,\,\,\,\,\,\,\,\,\,\,\,\,\,\,\,\,\,\,\,\,\,
 \,\,\,\,\,\,\,\,\,\,\,\,\,\,\,\,\,\,\,\,\,\,           \nonumber
\end{eqnarray} 
These components are related to the traditional $UVW$ nomenclature
by, $v_X=-U$, $v_Y=-V$, and $v_Z=W$, e.g., \citet{BM98}.   

In order to obtain the Galactocentric cylindrical velocity components,
we must first correct for the solar motion.  
Taking into account HI measurements of the Galactic rotation
curve \citep{GKT79} and Hipparcos measurements of Cepheid proper motions \citep{FW97}, we adopt $v_{{\rm LSR}}=220$~\kms\ for the
motion of the local standard of rest and $R_\odot=8$~kpc \citep[for an analysis of other recent
measurements, see][]{BHR09}.
The adopted value of $R_\odot$ is motivated by geometrical measurements 
of the motions of stars around Sgr A$^\ast$, which yield 
$R_\odot=7.94\pm0.42$ kpc \citep{E03}. For the solar peculiar motion,
we adopt the HIPPARCOS result, $v_X^{\odot,{\rm pec}}=-10.0\pm0.4$~\kms,
$v_Y^{\odot,{\rm pec}}=-5.3\pm0.6$ \kms, and
$v_Z^{\odot,{\rm pec}}=7.2\pm0.4$~\kms\ (\citet{DB98}, also see \citet{Hogg05}). Using these
values, along with equation~(\ref{measV}), we obtain the Galactocentric
velocity components:
\begin{equation}
  v_i = v_i^{{\rm obs}} + v_i^\odot, \,\,\,\,\, i=X,Y,Z,
\end{equation}
where $v_X^\odot=-10$~\kms, $v_Y^\odot=-225$~\kms, and 
$v_Z^\odot=7$~\kms\ (note that $v_Y^\odot = -v_{{\rm LSR}} + v_Y^{\odot,{\rm pec}}$).
Below, we discuss attempts to directly determine the solar 
peculiar motion (\S~\ref{sec:solmotion}) and $v_{{\rm LSR}}$ 
(\S~\ref{sec:vLSR}) from our data.

Finally, the cylindrical components, $v_R$ and $v_\phi$, can be 
computed using a simple coordinate rotation,
\begin{eqnarray} 
 v_R = v_X {X \over R} + v_Y {Y \over R}  \nonumber \\
 v_\phi = -v_X {Y \over R} + v_Y {X \over R}. 
\end{eqnarray} 
Note that, in our adopted system, the disk has a prograde rotation $v_\phi=-220$
\kms; retrograde rotation is indicated by $v_\phi>0$. Stars with $v_R>0$ move away
from the Galactic center, and stars with $v_Z>0$ move toward the
North Galactic Pole. 

\subsection{  Analysis Philosophy }

Such a massive data set, extending to a large distance limit and probing a large
fraction of the Galaxy volume, can be used to map the kinematics of stars in great
detail. It can also be used to obtain best-fit parameters of an appropriate
kinematic model. However, it is not obvious what model (functional form) to
chose without at least some preliminary analysis. Hence, in the next two
sections, we first discuss various projections of the multi-dimensional space of
the available observable quantities and obtain a number of constraints on the spatial
variation of stellar kinematics. We then synthesize all of the constraints into a model
described in \S~\ref{sec:models}. Before proceeding with our analysis, we
provide a brief summary of the first two papers in this series, whose results
inform our subsequent analysis. 
 
\subsection{ A Summary of Papers I and II}

Using photometric data for 50 million stars from SDSS Data 
Release 4 \citep{DR4}, sampled over a distance range from 100 pc to 15 kpc,
J08 showed that the stellar number density distribution, 
$\rho(R,Z,\phi)$ can be well-described (apart from local 
overdensities; the J08 best-fit was refined using residual 
minimization algorithms) as a sum of two cylindrically 
symmetric components:
\begin{equation} 
      \rho(R,Z,\phi) = \rho_D(R,Z) + \rho_H(R,Z).
\end{equation}

The disk component can be modeled as a sum of two exponential disks
\begin{eqnarray} 
\rho_D(R,Z)=\rho_D(R_\odot) 
     \left({\rm e}^{-\frac{|Z+Z_\odot|}{H_1} - \frac{(R-R_\odot)}{L_1}} 
   + \epsilon_D {\rm e}^{-\frac{|Z+Z_\odot|}{H_2} - \frac{(R-R_\odot)}{L_2}} \right),
\end{eqnarray} 
while the halo component requires an oblate power-law model
\begin{equation} 
 \rho_H(R,Z)= \rho_D(R_\odot)\,\epsilon_H\, \left({R_\odot^2 \over R^2 
               + (Z/q_H)^2}\right)^{n_H/2}.
\end{equation} 

The best-fit parameters are discussed in detail by J08. We have adopted
the following values for the parameters relevant to this work (second column 
in Table 10 from J08): $Z_\odot$=25 pc, $H_1=245$ pc, $H_2=743$ pc, 
$\epsilon_D=0.13$, $\epsilon_H=0.0051$, $q_H=0.64$, and $n_H=2.77$.
The normalization $\rho_D(R_\odot)$ (which is essentially the local luminosity 
function for main-sequence stars) is listed in J08 as a function of
color.

Using a photometric metallicity estimator for F/G main-sequence stars,
I08 obtained an unbiased, three-dimensional metallicity 
distribution of $\sim$2.5 million F/G stars at heliocentric distances 
of up to $\sim$8 kpc. They found that the metallicity distribution 
functions (MDF) of the halo and disk stars are clearly distinct.
The median metallicity of the disk exhibits a vertical (with 
respect to the Galactic plane, $Z$) gradient, and no gradient in the 
radial direction (for $Z > 0.5$ kpc and $6 < R < 10$~kpc). 

Similarly to the stellar number density distribution, $\rho(R,Z)$, the
overall behavior of the MDF $p([{\rm Fe/H}]|R,Z)$ for disk stars can be 
well-described as a sum of two components
\begin{equation} 
  p(x=[{\rm Fe/H}]|R,Z,\phi) = [1-f_H(R,Z)]\,p_D(x|Z) + f_H(R,Z)\,p_H(x),
\end{equation}
where the halo star-count ratio is simply, 
\begin{equation}
f_H(R,Z)=\frac{\rho_H(R,Z)}{\rho_D(R,Z)+\rho_H(R,Z)}. 
\end{equation}

The halo metallicity distribution, $p_H([{\rm Fe/H}])$, is spatially
invariant within the probed volume, and well-described by a Gaussian 
distribution centered on $[{\rm Fe/H}]=-1.46$, with an intrinsic 
(corrected for measurement errors) width $\sigma_H=0.30$ dex. 
For $|Z|\la10$ kpc, an upper limit on the halo radial 
metallicity gradient is 0.005 dex/kpc. 

The disk metallicity distribution varies with $Z$ such that its shape 
remains fixed, while its median, $\mu_D$, varies as 
\begin{equation}
 \label{muD}
   \mu_D(Z) = \mu_\infty + \Delta_\mu \,{\rm e}^{-\frac{|Z|}{H_\mu}},
\label{muZ}
\end{equation}
with the best-fit parameter values $H_\mu=0.5$ kpc, $\mu_\infty=-0.82$, and 
$\Delta_\mu=0.55$. The {\it shape} of the disk metallicity distribution 
can be modeled as
\begin{eqnarray} 
\label{pDmetal}
  p_D(x=[{\rm Fe/H}]|Z)= 0.63\,G[x|\mu=a(Z), \sigma=0.2] + \nonumber \\
    0.37\,G[x|\mu=a(Z)+0.14, \sigma=0.2], 
\end{eqnarray} 
where the position $a$ and the median $\mu_D$ are related via 
$a(Z)=\mu_D(Z)-0.067$ (unless measurement errors are very large).

The main result of this third paper in the series is the extension 
of these results for number density and metallicity distributions 
to include kinematic quantities.

%% file: analysisPM.tex
\section{       ANALYSIS OF THE PROPER MOTION SAMPLE   }
\label{sec:aPM}

We begin by analyzing the proper-motion measurements of stars observed toward
the North Galactic Pole. In this region, the Galactocentric azimuthal velocity,
$v_\phi$, and radial velocity, $v_R$, can be determined directly from the
proper-motion measurements (that is, without knowledge of the
spectroscopically-determined radial velocity, $v_{{\rm rad}}$). In this way, we can
study the kinematic behavior of stars as a function of metallicity and distance
from the Galactic plane, $Z$. We then extend our analysis to the entire
meridional $Y=0$ plane, and study the variation of stellar kinematics with $R$
and $Z$. In the following section, we only consider the northern Galactic
hemisphere, where most of the proper-motion data are available.

\subsection{Kinematics Towards the North Galactic Pole}
\label{subsec:NGP}

We select three stellar subsamples in the region $b>80^\circ$, including $14,000$
blue disk stars at $Z<7$~kpc, $23,000$ blue halo stars at $Z<7$~kpc, and a
sample of $105,000$ red stars at $Z<1$~kpc. In Figure~\ref{fig:vPhivRblue}, we
plot the distribution of $v_\phi$ vs. $v_R$ for $\sim 6,000$ blue disk and halo
stars at $Z=4-5$~kpc. In this and all subsequent two-dimensional projections of
the velocity distribution we plot smoothed, color-coded maps, where the velocity
distributions are estimated using the Bayesian density estimator of
\citet[][see their Appendix for the derivation and a discussion]{Ivezic05}.
At a given position, the density is evaluated as 
\begin{equation}
     \rho = {C \over \sum_{i=1}^N d_i^2},
\end{equation}
where $d_i$ is the distance in the velocity-velocity
plane, and we sum over the $N=10$ nearest neighbors. The normalization constant,
$C$, is easily evaluated by requiring that the density summed over all pixels is
equal to the total number of data points divided by the total area. The grid
size is arbitrary, but the map resolution is determined by the density of points
-- we choose pixel size equal to one half of the mean velocity error. As shown
by \citet{Ivezic05}, this method is superior to simple Gaussian smoothing. For
comparison, we also plot linearly-spaced density contours.

The six panels of Figure~\ref{fig:vPhivRblue} demonstrate the variation of
kinematics with metallicity, with the full range of metallicities
($-3<[{\rm Fe/H}]<0$) plotted in the upper left panel and subsamples with increasing
metallicity running from left-to-right, top-to-bottom. The mean azimuthal
velocity varies strongly with metallicity, from a non-rotating low-metallicity
subsample with large velocity dispersion (top center panel) to a rotating
high-metallicity sample with much smaller dispersion (bottom right panel). This
strong metallicity-kinematic correlation is qualitatively the same as discussed
in the seminal paper by \citet{ELS62}, except that here it is reproduced {\it in
situ} with a $\sim$100 times larger, nearly-complete sample, thus extending it
beyond the solar neighborhood. There are some indications of substructure in the
velocity distribution, but much of it remains unresolved due to the large
velocity measurement errors.

The substructure becomes more apparent in Figure~\ref{fig:vPhivRred}, where we
plot the same velocity-space projection for $60,000$ stars within $Z<2.5$~kpc.
In this figure, the panels show subsamples of increasing distance from the
Galactic plane, beginning with $Z=0.1-0.2$~kpc in the upper left panel (note the
changing axes between the top and bottom rows). The substructure seen in the
closest bin probed by red stars is very similar to the substructure seen in the
local HIPPARCOS sample \citep{Dehnen98}. These results were
based on a maximum-likelihood analysis over the entire sky,  while our result arises from a
direct mapping of the velocity distribution of stars selected from a small
region ($\sim 300$ deg$^2$). Using a subsample of $\sim17,000$ HIPPARCOS stars
with full three-dimensional velocity information, \citet{Nordstrom04,Famaey05,Holmberg07,Holmberg09} have
detected the same kinematic morphology. The similarity between these
HIPPARCOS-based velocity distributions and ours, including the multi-modal
behavior reminiscent of moving groups \citep{Egg96}, is quite encouraging, given the
vastly different data sources. The similarity of observed substructure with
moving groups is even more striking for stars from a closer distance bin
($Z=50-100$~pc), matched to distances probed by the HIPPARCOS sample (see
Figure~\ref{fig:Eggen}). As suggested by \citet{DeSim04}, these moving groups
may arise from irregularities in the Galactic gravitational potential.

The remainder of our analysis will focus on blue stars, which sample a much
larger distance range. For a detailed study of the velocity distribution of
nearby red stars, including a discussion of non-Gaussianity, vertex deviations,
and difficulties with traditional thin/thick disk separation, we refer the
reader to Kowalski et al. (in prep.).

The dependence of the rotational velocity on height above the Galactic plane is
shown in Figure~\ref{fig:VvsZphi}. The two subsamples display remarkably
different kinematic behavior \citep[first seen locally by][]{ELS62} with halo
stars exhibiting a small constant rotational motion ($\sim -20$~km~s$^{-1}$),
and disk stars exhibiting a large rotational-velocity component ($\sim
-200$~km~s$^{-1}$ at $Z \sim 1$~kpc) that decreases with height above the
Galactic plane.

We have performed the same analysis using proper motions based only on POSS data, with SDSS positions not included in the proper-motion fit (not publicly available\footnote{Available from J. Munn on request}). While random proper-motion errors become larger when SDSS data are not used, the median rotational velocity for halo stars decreases to only $5$~km~s$^{-1}$, suggesting that the apparent rotational motion in the halo subsample is influenced by systematic errors. These tests also suggest that the leading contribution to systematic proper-motion errors could be a difference between the SDSS (digital data) and POSS (digitized photographic data) centroid-determination algorithms. In addition, \citet{Smith09} did not detect halo rotation using a smaller sample, but with more robust proper motion measurements based on only SDSS data, while \citet{Allende06} found no evidence for halo rotation using SDSS DR3 radial velocities. We conclude that the net halo rotation in the direction of the North Galactic Pole is $\left|v_{{\rm rot}}\right|\lesssim 10$~km~s$^{-1}$. In addition, the measured halo velocity dispersion increases with $Z$, but when random measurement errors are taken into account, the data are consistent with a constant dispersion of $\sigma_{\phi}^H=85 \pm 5$~km~s$^{-1}$ (derived using the test described in \S~\ref{sec:models}).

The decrease of rotational velocity with $Z$ for disk stars (often referred to
as asymmetric drift, velocity lag, or velocity shear; see \S 3.4 of I08 for
more details and references to related work) is in agreement with a preliminary
analysis presented in I08. We find that the observed behavior in the $Z=1-4$~kpc
range can be described by
\begin{equation} 
\langle v_\phi \rangle = -205 + 19.2 \,\left|\frac{Z}{\rm kpc}\right|^{1.25} \,\,\, {\rm km~s^{-1}}.
\label{eq:Dlag}
\end{equation} 
The measured rotational velocity dispersion of disk stars increases
with $Z$ faster than can be attributed to measurement errors. Using a
functional form $\sigma=a+b|Z|^c$, we obtain an {\it intrinsic} velocity
dispersion fit of
\begin{equation}
    \sigma_{\phi}^D = 30 + 3.0 \,\left|\frac{Z}{\rm kpc}\right|^{2.0} \,\,\, {\rm km~s^{-1}}.
\label{eq:Ddisp}
\end{equation} 

This function and the best-fit rotational velocity for halo stars are shown as dotted lines in the bottom right panel of Figure~\ref{fig:VvsZphi} (see Table 1 for a summary of all best-fit parameters). I08 fit a linear model to $v_{\phi}$ vs.  $Z$, but the difference between this result and their equation~(15) never exceeds $5$~km~s$^{-1}$ for $Z<3$~kpc. The errors on the power-law exponents of equations~(\ref{eq:Dlag}) and (\ref{eq:Ddisp}) are $\sim 0.1$ and $\sim0.2$, respectively.

However, a description of the velocity distribution based solely on
the first and second moments (equations~(\ref{eq:Dlag}) and (\ref{eq:Ddisp}))
does not fully capture the detailed behavior of our data.  As already
discussed by I08, the rotational-velocity distribution for disk stars
is strongly non-Gaussian (see their figure~13). It can be formally
described by a sum of two Gaussians, with a fixed normalization ratio
and a fixed offset of their mean values for $\left|Z\right|<5$~kpc,
\begin{eqnarray} 
\label{pDvPhi}
 p_D(x=v_\phi|Z)=0.75\,G[x|v_n(Z),\sigma_1]+  
                 0.25\,G[x|v_n(Z)-34 \, {\rm km~s^{-1}},\sigma_2], 
\end{eqnarray} 
where 
\begin{equation} 
\label{wZ}
      v_n(Z) = -194+19.2\,\left|\frac{Z}{\rm kpc}\right|^{1.25} \,\,\, {\rm km~s^{-1}.} 
\label{drift}
\end{equation} 

The intrinsic velocity dispersions, $\sigma_1$ and $\sigma_2$, are
modeled as $a+b|Z|^c$, with best-fit parameters listed in
Table~\ref{Tab:kinD} (see $\sigma_{\phi}^1$ and $\sigma_{\phi}^2$).
Closer to the plane, in the $0.1<Z<2$~kpc range probed by red stars,
the median rotational velocity and velocity dispersion are consistent
with the extrapolation of fits derived here using much more luminous blue
stars.

Figure~\ref{fig:VphiHist} shows the $v_\phi$ distribution for four
bins in $Z$ (analogous to figure~13 from I08), overlaying
two-component Gaussian fits with the measurement errors and $v_n(Z)$
as free parameters. The mean velocity and velocity dispersion exhibit
$\sim 10$~km~s$^{-1}$ variations relative to their expected values;
while such deviations could be evidence of kinematic substructure,
they are also consistent with the plausible systematic errors. We
conclude that equations~(\ref{pDvPhi}) and (\ref{wZ}) provide a good
description of the disk kinematics for stars observed toward the North
Galactic Pole, within the limitations set by the random and systematic
errors in our data set. 

The fits to the observed velocity distributions for halo and disk stars are shown in Figure~\ref{fig:VphiHist} and demonstrate that the vertical gradients in median rotational velocity and velocity dispersion for disk stars seen in Figure~\ref{fig:VvsZphi} are not due to contamination by halo stars.  To quantitatively assess the impact of ``population mixing'' as a result of the adopted metallicity-based classification ($[{\rm Fe/H}]<-1.1$ for ``halo'' stars and $[{\rm Fe/H}]>-0.9$ for ``disk'' stars) on our measurements of these gradients, we have performed a series of Monte Carlo simulations. Assuming that the fits shown in Figure~\ref{fig:VphiHist} accurately depict the intrinsic velocity distributions, and adopting analogous fits for their metallicity distributions (I08, their Figure 7), we have estimated the expected bias in median velocity and velocity dispersion for each population as a function of their relative normalization. As shown in Figure 6 from I08, the fraction of halo stars increases with distance from the plane, from about 0.1 at $Z=1$~kpc to about 0.9 at $Z=5$~kpc. We find that the median rotation velocity and velocity dispersion biases are $<10$~\kms\ for disk stars at $Z<3$ kpc, as well as for halo stars at $Z>3.5$~kpc. Furthermore, the biases are $<20$~\kms\ for $Z<4$~kpc for disk stars and to $Z>3$~kpc for halo stars. At $Z=3$~kpc, the contamination of both disk and halo subsamples by the other population is typically $\sim10-15$\%. This small contamination and the use of the median (as opposed to the mean) and dispersion computed from the interquartile range results is reasonably small biases. With the adopted metallicity cutoffs, the sample contamination reaches 50\% at $Z=1-1.5$~kpc for halo stars and at $Z=4.5$~kpc for disk stars.

The dependence of the Galactocentric radial velocity on $Z$ is shown
for halo and disk subsamples in Figure~\ref{fig:VvsZR}. The median
values (bottom left panel) are consistent with zero, within the
plausible systematic errors ($10-20$~km~s$^{-1}$), at all $Z$. The
intrinsic dispersion for halo stars is consistent with a constant
value of $\sigma_{R}^H=135 \pm 5$~km~s$^{-1}$.  For disk stars, the
best-fit functional form $\sigma=a+b|Z|^c$ is
\begin{equation}
    \sigma_{R}^D = 40 + 5 \,\left|\frac{Z}{\rm kpc}\right|^{1.5} \,\,\, {\rm km~s^{-1}}.
\label{eq:DdispR}
\end{equation} 
The $\sigma_{R}^D/\sigma_{\phi}^D$ ratio has a constant value
of $\sim1.35$ for $Z<1.5$~kpc, and decreases steadily at larger
$Z$ to $\sim 1$ at $Z\sim4$~kpc. 

\subsection{Kinematics in the Meridional $Y\sim0$ Plane }

The analysis of the rotational-velocity component can be extended to the
meridional plane defined by $Y=0$, for which the longitudinal proper motion
depends only on the rotational-velocity component and the latitudinal proper motion,
$v_b$, is a linear combination of radial and vertical components,
\begin{equation}
      v_b= \sin(b) v_R + \cos(b) v_Z.
\end{equation}
Figure~\ref{fig:RZpanelsGPblue} plots $v_\phi$ and $v_b$ as functions of $R$ and
$Z$ for halo and disk stars within $10^\circ$ of the meridional plane. The
median $v_b$ is close to zero throughout most of the plotted region, as would be
expected if the median $v_R$ and $v_Z$ are zero (the behavior of $v_Z$ is
discussed in the next section). One exception is a narrow feature with $v_b \sim
-100$~km~s$^{-1}$\ for $R<4$~kpc. While a cold stellar stream could produce such
a signature, its narrow geometry points directly at the observer. This behavior
is also consistent with a localized systematic proper-motion error. Indeed, the
bottom left panel in Figure~\ref{qsoPMerrors} shows that the systematic
latitudinal proper-motion error at $l\sim 0^\circ$, $b\sim45^\circ$ is about
$1$~mas~yr$^{-1}$, corresponding to a velocity error of $\sim100$~km~s$^{-1}$ at
a distance of $7$~kpc.

As seen in the upper left panel of Figure~~\ref{fig:RZpanelsGPblue}, the median
$v_\phi$ for halo stars is close to zero for $R<12$~kpc. In the region with
$R>12$~kpc and $Z<6$~kpc, the median indicates a surprising prograde rotation in
excess of $100$~km~s$^{-1}$. This behavior is also seen in disk stars, and is
likely due to the Monoceros stream, which has a metallicity intermediate between
disk and halo stars and rotates faster than disk stars (see \S 3.5.1 and
\S 3.5.2 in I08). There is also an indication of localized retrograde rotation for
halo stars with $Z \sim 9$~kpc and $R\sim15$~kpc (corresponding to
$l\sim180^\circ$, $b\sim50^\circ$, and a distance of $\sim 11$~kpc). Stars with
$Z=8-10$ kpc and $R=15-17$ kpc have median $v_\phi$ larger by $40$~\kms\
(a $\sim1\sigma$ effect) and median $[{\rm Fe/H}]$ larger by $0.1$~dex ($\sim5\sigma$
effect) than stars with $Z=8-10$~kpc and $R=7-13$~kpc. A systematic error in
$\mu_l$ of $\sim0.8$~mas~yr$^{-1}$ is required to explain this kinematic feature
as a data problem (although this would not explain the metallicity offset).
However, the top right panel in Figure~\ref{qsoPMerrors} shows that the
systematic $\mu_l$ errors in this sky region are below $0.5$~mas~yr$^{-1}$, so
this feature may well be real. We note that in roughly the same sky region and at
roughly the same distance, \citet{GD06} have detected a narrow stellar stream.  

In order to visualize the extent of ``contamination'' by the Monoceros stream, we
replace the rotational velocity for each disk star by a simulated value drawn
from the distribution described by equation~(\ref{pDvPhi}). We subtract this model
from the data, and the residuals are shown in the right-hand panel of
Figure~\ref{fig:RZpanelsGPdiskModel}. The position of the largest deviation is
in excellent agreement with the position of Monoceros stream quantified in I08
($R = 15-16$~kpc and $\left|Z\right| \sim 3 - 5$~kpc). Further evidence for the
presence of the Monoceros stream is shown in Figure~\ref{fig:vPhiFeHpanelsGPblue},
in which we analyze $v_\phi$ vs. $[{\rm Fe/H}]$, as a function of $R$, for blue stars at
$Z=4-6$~kpc. As is evident in the bottom right panel, there is a significant excess
of stars at $R>17$~kpc with $-1.5 < [{\rm Fe/H}] < -0.5$ that rotate in a prograde
direction with $\sim200$~km~s$^{-1}$.

%% file: analysis6D.tex
\section{ ANALYSIS OF THE SPECTROSCOPIC SAMPLE }
\label{sec:a6D}

Despite its smaller size, the SDSS DR-7 spectroscopic sample of $\sim100,000$ main-sequence stars is invaluable, because it enables a direct\footnote{Statistical deprojection methods, such as that recently applied to a subsample of M stars discussed by \citet{Fuchs09}, can be used to indirectly infer the three-dimensional kinematics from proper motion data.} study of the three-dimensional velocity distribution. The sample extends to a distance of $\simeq 10$~kpc, at which it can deliver velocity errors as small $\sim10$~\kms\ (corresponding tangential velocity errors are $\sim150$~\kms\ at a distance of $10$~kpc). For each object in the SDSS spectroscopic survey, its spectral type, radial velocity, and radial-velocity error are determined by matching the measured spectrum to a set of stellar templates, which were calibrated using the ELODIE stellar library \citep{ELODIE}. Random errors on the radial-velocity measurements are a strong function of spectral type and signal-to-noise ratio, but are usually $< 5$~\kms\ for stars brighter than $g\sim18$, rising sharply to $\sim15$~\kms\ for stars with $r=20$. We model the behavior of the radial-velocity errors as \begin{equation}
  \sigma_{{\rm rad}} = 3 + 12 \times 10^{0.4\,(r-20)} \,\,\,\, {\rm \kms}.
\end{equation}  

We begin our analysis with blue disk and halo stars, and then briefly
discuss the kinematics of nearby red M stars.

\subsection{ Velocity Distributions }
\label{subsec:Blue6D}

We select $111,000$ stars with $0.2<g-r<0.6$ ($74,000$ have $b>20^\circ$) and,
using their spectroscopic metallicity, separate them into $47,000$ candidate halo
stars with $[{\rm Fe/H}]<-1.1$, and $\sim53,000$ disk stars with $[{\rm Fe/H}]> -0.9$.
Assuming the spectroscopic metallicities accurately separate disk
from halo stars, the use of photometric metallicity for the same
selection would result in a contamination rate of $6$\% for the
halo subsample, and $12$\% for the disk subsample.

The dependence of the median vertical velocity, $v_z$, and its dispersion on
height above the disk, is shown in Figure~\ref{fig:VvsZZ} for the halo and disk
subsamples. The median values of $v_Z$ are consistent with zero to better than
$10$~\kms\ at $Z<5$~kpc, where statistical fluctuations are small.  

As with $\sigma_{\phi}$ and $\sigma_R$, the vertical velocity dispersion can be
modeled using a constant dispersion for halo stars ($\sigma_{Z}^H=85$~\kms),
while for disk stars, the best-fit functional form is
\begin{equation}
    \sigma_{Z}^D = 25 + 4 \,\left|\frac{Z}{\rm kpc}\right|^{1.5} \,\,\, {\rm \kms}.
\label{DdispZ}
\end{equation} 

The other two velocity components behave in a manner consistent with
equations~(\ref{eq:Dlag}), (\ref{eq:Ddisp}), and
(\ref{eq:DdispR}), just as they did in the proper-motion sample. This is
encouraging, because the spectroscopic sample is collected over the
entire northern hemisphere, unlike the proper-motion subsample studied
in \S~\ref{subsec:NGP}, which is limited to $b>80^\circ$.

The availability of all three velocity components in the spectroscopic sample
makes it possible to study the orientation of the halo velocity ellipsoid.
Figure~\ref{fig:VVHVT} shows two-dimensional projections of the velocity
distribution for subsamples of candidate halo stars with $0.2<g-r<0.4$. The top
row corresponds to stars above the Galactic plane at $3<Z/$kpc~$<4$, while the
bottom row is for stars the same distance below the plane. The velocity
ellipsoid is clearly tilted in the top- and bottom-left panels, with a tilt angle
consistent with ${\rm tan}^{-1}(v_Z/v_R)=R/z$. While the tilt-angle errors are too large to
obtain an improvement over existing measurements of $R_\odot$, it is remarkable
that the northern and southern subsamples agree so well\footnote{A plausible, if
somewhat optimistic, tilt-angle uncertainty of $1^\circ$ corresponds to
an $R_\odot$ error of $0.5$~kpc; extending the sample to $|Z|=8$~kpc could deliver
errors of $0.3$~kpc per bin of a similar size.}. In addition, when the
$Z=3-5$~kpc sample is divided into three subsamples with $7<R/$kpc~$<11$, the tilt
angle varies by the expected $\sim8^\circ$ in the correct direction (see
Figure~\ref{fig:VVHVT2}). For all bins in the $R-Z$ plane, the best-fit tilt
angle is statistically consistent (within $5^\circ$) with ${\rm tan}^{-1}(v_Z/v_R)=R/z$. The
other two projections of the velocity distribution for halo stars do not exhibit
significant tilts to within $\sim3^\circ$ .

If we transform the velocities to a spherical coordinate system,
\begin{eqnarray} 
\label{eq:vSphere}
 v_r = v_R {R \over R_{\rm gc}} + v_Z {Z \over R_{\rm gc}}  \nonumber \\
 v_\theta = v_R {Z \over R_{\rm gc}} - v_Z {R \over R_{\rm gc}},  
\end{eqnarray} 
where $r=R_{{\rm gc}}=(R^2+Z^2)^{1/2}$ is the spherical Galactocentric radius,
we find no statistically significant tilt in any of the
two-dimensional velocity-space projections for halo stars (with 
tilt-angle errors ranging from $\sim 1^\circ$ to $\sim 5^\circ$).

As shown in Figure~\ref{fig:VVDVT}, we see no evidence for a velocity ellipsoid
tilt in $v_Z$ vs. $v_R$ for the disk stars. The plotted subsamples are again
selected to have colors $0.2<g-r<0.4$, but are selected closer to the Galactic
plane, $|Z|=1.5-2.5$~kpc, in order to improve statistics and reduce
contamination from halo stars. The velocity-ellipsoid tilt is consistent with
zero within $\sim1\sigma$, and alignment of the velocity ellipsoid with the
spherical coordinate system of equation~(\ref{eq:vSphere}) is ruled out at a
$\sim2\sigma$ or greater confidence level for each of five analyzed $R-Z$ bins
($R=6-11$ kpc, with $\Delta R=1$~kpc). We conclude that there is no evidence for
a velocity-ellipsoid tilt in the disk subsample, but caution that, due to the
small $Z$ range, the data cannot easily distinguish between cylindrical and spherical
alignment. A model prediction for velocity ellipsoid tilt is discussed below. 

The vertex deviation is analogous to the velocity-ellipsoid tilt discussed above,
but is defined in the $v_\phi$ vs. $v_R$ plane instead of the $v_Z$ vs. $v_R$
plane. The same plots for red ($g-r>0.6$, median $1.2$) disk stars are shown in
the center top and bottom panels in Figure~\ref{fig:VVDredVT}. These stars can
be traced closer to the plane, $|Z|=0.6-0.8$~kpc; in both hemispheres, the
data are consistent with a vertex deviation of $\sim20^\circ$, with an
uncertainty of $\sim10^\circ$. This result is consistent with the vertex
deviation obtained by \citet{Fuchs09}.

Another interpretation for the $v_\phi$ vs. $v_R$ distribution of disk
stars invokes a two-component velocity distribution, which can result
in a similar deviation even if each component is perfectly symmetric
in the cylindrical coordinate system.  Kowalski et al. (in prep.) find
that the $v_\phi$ vs. $v_R$ distribution for red stars toward the
North Galactic Pole, with $0.1<Z/{\rm kpc}<1.5$, can be fit by a sum
of two Gaussian distributions that are offset from each other by
$\sim10$~\kms\ in each direction. This offset results in a non-zero
vertex deviation if the sample is not large enough, or if measurements
are not sufficiently accurate, to resolve the two Gaussian components. This
double-Gaussian structure would be at odds with the classical
description based on the Schwarzschild approximation -- we refer the
interested reader to the Kowalski et al. study for more
details. Unfortunately, the spectroscopic samples are not large enough to
distinguish a two-component model from the standard interpretation.

\subsubsection{A Model Prediction for Velocity Ellipsoid Tilt} 
\label{sec:tilt} 

The tilt of the velocity ellipsoid tilt in the $v_Z$-$v_R$ plane is calculated using methods developed in \citet{KdZ91}, who show that the tilt of the ellipsoid at any point in the Milky Way depends not only on the gravitational potential of the Galaxy but also on the isolating integrals of a particular orbit and thus (weakly) on the distribution function of stellar velocities. We use the model of the Milky Way gravitational field from \citet{CI87}, scaled to a solar radius of $8$~kpc and a circular velocity of $220$~\kms.  At a particular point in the Galaxy, we integrate eight orbits, each one launched from that point, with velocities in each of the $R$, $Z$, and $\phi$ coordinates that are $\pm 1\sigma$ from the systemic velocity. We use the observed velocity distributions to obtain the systemic velocity and the $1\sigma$ values. Furthermore, using the ``least-squares fitting'' method, we determine the parameters of the prolate spheroidal coordinate system that best matches each orbit -- from this, we can compute its contribution to the velocity ellipsoid.  For some orbits (especially those of halo stars), the orbits are not local to the solar neighborhood, and the local fitting method cannot fit the entire orbit.  For these cases, we confine the orbits to radii greater than $4$~kpc.  Since the only purpose of the fit is to model the kinematics in a small volume of space, the fact that the fit is no longer global is of no consequence.  The scatter in tilt angle among the individual orbits at each position depends only weakly on position ($\sim 10$\%).

Table~\ref{Tab:tilt} gives the predicted tilt angles for each position plotted 
in Figures~\ref{fig:VVHVT}-\ref{fig:VVDredVT}. For the halo stars plotted in 
Figures~\ref{fig:VVHVT}-\ref{fig:VVHVT2}, where the ellipsoid
shows the most distinctive tilt, the predictions match the observed tilts
quite well, but are slightly steeper at $Z>3$ kpc than the 
$Z/R$ relation predicted for spherical symmetry. For the disk stars in 
Figures~\ref{fig:VVDVT}-\ref{fig:VVDredVT}, the predicted tilts lie between 
the cases of cylindrical and spherical symmetry.   The observed ellipsoid is 
sufficiently round, however, that no definitive comparison can be made.

\subsection{Direct Determination of the Solar Peculiar Motion}
\label{sec:solmotion}

If there is no net streaming motion in the $Z$ direction in the solar
neighborhood, the median heliocentric, $v_Z^{\rm obs}$, for nearby stars should be
equal to $v_Z^\odot$ \citep[$7$~\kms, based on an analysis of HIPPARCOS results
by][]{DB98}. We do not expect a large velocity gradient within $\sim 1$~kpc from
the Sun, so we select from the spectroscopic sample $\sim13,000$ M dwarfs with
$2.3<g-i<2.8$. In the northern hemisphere, we have $5,700$ stars with a median
heliocentric velocity, $\langle v_Z^{\rm obs}\rangle=-1.8$~\kms, while for stars
in the southern hemisphere we obtain $\langle v_Z^{\rm obs}\rangle=-11.0$~\kms.
This difference is likely due to a systematic radial-velocity error. If we
simultaneously vary an assumed radial-velocity error, $\Delta_{\rm rad}$, and
the solar peculiar motion, $v_Z^\odot$, while requiring that the median
$v_Z^{\rm obs}$ should be the same for both hemispheres, we obtain $\Delta_{\rm
rad}=5.0 \pm 0.4$~\kms\ and $v_Z^\odot=6.5 \pm 0.4$~\kms. This value for
$v_Z^\odot$ is in excellent agreement with the HIPPARCOS value of
$7.2\pm0.4$~\kms\ \citep{DB98}. This systematic offset in SDSS radial velocities
is probably due to the small number of ELODIE templates for red stars. It is
likely that the adoption of improved templates from \citet{Bochanski07b} will
yield smaller systematic errors. A similar
analysis for blue stars does not yield a robust detection of the velocity
offset. A detailed comparison of SDSS radial velocities with radial-velocity
standards from the literature arrived at the same null result for blue stars.

As with the vertical component of the solar peculiar motion, if the adopted
value of $v_X^\odot=-10$~\kms\ were incorrect, the median $v_R$ would deviate
from zero. The root-mean-square scatter of the median $v_R$ for subsamples of 
nearby M stars selected by distance and color is $0.5$~\kms, which is an upper limit on
the error in the adopted value of $v_X^\odot$. This result, which is based on
the full three-dimensional velocity distribution, agrees well with results from
indirect statistical deprojection methods using only proper motions
\citep{Dehnen98,Fuchs09}.

For both blue and red disk stars, the extrapolation of the median $v_\phi$ to
$Z=0$ yields $-205$~\kms. Since we corrected stellar velocities for an assumed
solar motion of $-225$~\kms, this implies that the $Y$ component of the solar
velocity relative to the bulk motion of stars in the solar neighborhood is
$20$~\kms, in agreement with recent results obtained by \citet{Fuchs09} for
the same data set. A similar value was obtained by \citet{DB98} for
their\footnote{They extrapolated the mean azimuthal motion of color-selected
samples, which is correlated with the radial-velocity dispersion, to zero
dispersion and obtained $v_Y^{\odot,{\rm pec}}=-5.3$~\kms, used here.} subsample of
red stars within $100$~pc.

%% file: model.tex
\section{ A MODEL FOR THE KINEMATICS OF DISK AND HALO STARS}
\label{sec:models}

Informed by the results from the preceding two sections, we introduce a
model that aims to describe the global behavior of the observed stellar
kinematics. In our model, we do not attempt to account for kinematic
substructure (e.g., the Monoceros stream), or the Galactic bulge region, nor do
we incorporate any complex kinematic behavior close to the Galactic plane.
Nevertheless, we attempt to capture the gross properties of the data in the
volume probed by SDSS, including the bulk kinematic trends and the kinematic
differences between high-metallicity disk stars and low-metallicity halo stars.
We describe the model in
\S~\ref{subsec:modelkin}, then test it in \S~\ref{subsec:modeltests} 
using both the proper-motion and radial-velocity samples.

\subsection{The Kinematic Model }
\label{subsec:modelkin}

For halo stars, a single velocity ellipsoid (expressed in a Galactocentric
spherical coordinate system, see \S~\ref{subsec:Blue6D}) is a good description
of the gross halo kinematics within the $10$~kpc distance limit of our sample.
Our model assumes that the halo has no net rotation (see below for a test of
this assumption), and that the principal axes of the velocity ellipsoid are
aligned with a spherical coordinate system. The velocity dispersions measured in
\S~\ref{sec:aPM} and \S~\ref{sec:a6D}, $\sigma_{R}^H=135$~\kms,
$\sigma_{\phi}^H=85$~\kms, and $\sigma_{Z}^H=85$~\kms, are expressed in a
cylindrical coordinate system. The interplay between proper-motion, 
radial-velocity, and distance measurement errors is complex, so we use Monte Carlo
simulations to translate them to spherical coordinates; the results of this
exercise are 
$\sigma_{r}^H=141$~\kms\ and $\sigma_{\theta}^H=75$~\kms\ ($\sigma_\phi^H$ is
unchanged), with uncertainties of $\sim5$~\kms.

For disk stars within $\sim 1-2$~kpc from the Sun, the velocity-measurement
errors are sufficiently small, and the samples are sufficiently large, to
resolve rich kinematic substructure (e.g., Figure~\ref{fig:vPhivRred}). This
behavior is quantified in detail in Kowalski et al. (in preparation), but here
we simply use the two-component model given by equations~(\ref{pDvPhi}) and (\ref{wZ}) to
describe the non-Gaussian $v_\phi$ distribution and velocity shear seen for disk
stars. Furthermore, we assume that $v_R$ and $v_Z$ have uncorrelated Gaussian
distributions, with zero mean and the $Z$-dependent intrinsic dispersion
parameters listed in Table~\ref{Tab:kinD}. As discussed in \S~\ref{sec:a6D},
there is no compelling evidence for a tilt in the velocity ellipsoid of blue disk
stars in the $v_R-v_Z$ plane, so we model the disk velocity ellipsoid in
cylindrical coordinates.

\subsection{Global Model Tests}
\label{subsec:modeltests}

Our model predicts distributions of the three measured kinematic
quantities, $v_{{\rm rad}}$, $\mu_l$, and $\mu_b$, for an arbitrary 
control volume defined by color, magnitude, and sky coordinates. 
We can test the consistency of this model, $m$, with our
data, $d$, by computing the residuals of each kinematic
quantity, 
\begin{equation} 
        \chi = { d - m \over (\sigma_d^2 + \sigma_m^2)^{1/2}},
\end{equation}
where $\sigma_d$ is the measurement error of the data and $\sigma_m$
is the dispersion predicted by the model. For all three kinematic
quantities, we find $\bar{\chi}\simeq0$ to within $0.05$ for all three
quantities, with dispersions of $\sim 1.05-1.1$.  While this result is
a necessary condition for the model to be acceptable, it is not
sufficient.  A stronger test of the model was shown in
Figure~\ref{fig:RZpanelsGPdiskModel} for disk stars, but in what
follows, we perform additional tests that cover all sky regions
with available data.

\subsubsection{Tests With the Radial Velocity Sample}

Figure~\ref{fig:LambPanelsHvrad} compares the medians and dispersions
for the measured and modeled radial velocities of halo stars. The
alignment of the velocity ellipsoid with spherical Galactocentric
radius is clearly seen in the bottom left panel, where the data show a
dispersion gradient moving away from the Galactic Center.  The
increased dispersion towards $l=180^\circ$ was misinterpreted by
\citet{Ivezic06} to be a sign of substructure.  There are no large
discrepancies between the measured and predicted behavior -- the
median value of the difference between observed and modeled values is
$4.6$~\kms, with a scatter of $19$~\kms\ (see top right panel). A
similar scatter is obtained between two model realizations with the
same number of stars and measurement errors.  The dispersion ratio,
$\sigma_{\rm rad}^d/\sigma_{\rm rad}^m$, is centered on $1.13$, with a
scatter of $0.2$ (see bottom right panel). For pairs of model
realizations, the ratio is always centered on one to within $0.02$,
with a scatter of $0.2$, suggesting that the observed velocity
dispersion is about $10$\% larger than predicted by our smooth
model. If the residuals were due to halo rotation, we would expect a
spatial coherence.  Similarly, for disk stars at $1<d<2.5$~kpc, the
median radial velocity residual is $2.8$~\kms, with a scatter of
$6.6$~\kms\ (not shown).

\subsubsection{Tests Based on the Proper Motion Sample}

The large size of the proper-motion sample enables a much higher
spatial resolution when searching for structure in the model
residuals.  We have compared the observed and modeled proper-motion
distributions in narrow bins of distance, across the sky, and
separately for disk and halo subsamples. As an illustration,
Figures~\ref{fig:LambPanelsDpmL} and \ref{fig:LambPanelsDpmB} show the
median longitudinal and latitudinal proper motion observed for disk
stars. There is very little change in the proper-motion distribution
among different distance bins, due to the nearly linear vertical
rotational-velocity gradient.  The residuals for longitudinal proper
motion are shown in Figure~\ref{fig:LambPanelsDpmLresid}.  They
provide weak evidence for either substructure or a radial gradient
that is not modeled, but it is difficult to distinguish between these two
possibilities with these data alone.

A comparison of disk and halo subsamples in a distance bin centered on
$d=4$~kpc is shown in Figure~\ref{fig:LambPanelsDpmLDds5Hds5}.  The
largest data vs. model discrepancy for halo stars, seen in the bottom
left panel towards $l\sim0^\circ$, is also seen from a different
viewing angle in the top left panel of Figure~\ref{fig:RZpanelsGPblue}
($R\sim6$~kpc and $Z\sim2$~kpc).  It is likely that this discrepancy
is due to contamination of the halo sample by metal-poor disk
stars. Figure~\ref{fig:LambPanelsHds4} shows the residuals for halo
stars selected from the $8-10$~kpc distance bin. The residuals for
both proper-motion components exhibit similar morphology to the
systematic proper-motion errors plotted in the two left panels of
Figure~\ref{qsoPMerrors}. In this distance bin, they correspond to
velocity errors of $\sim30$~\kms; as such, kinematic substructure in
the halo will be difficult to discern with this sample at
$\gtrsim10$~kpc. We note that it is tempting to associate the coherent
$\mu_l$ residuals towards $l\sim300^\circ$ and $l\sim60^\circ$ with
the Virgo overdensity \citep[see J08;][]{An09}.  However, the top left panel in
Figure~\ref{qsoPMerrors} clearly shows systematic proper motion errors
of the required amplitude ($\sim1.5$~mas~yr$^{-1}$) in the same sky
region.

\citet{Schlaufman09} use SDSS radial velocity measurements for
metal-poor turnoff stars to search for pieces of cold debris streams (which they
term Elements of Cold Halo substructure, or ECHOs). In $137$ lines of sight
they detect ten ECHOs. The six northern ECHOs from their Table 2 (class I peak
detections) are shown as white circles in the bottom two panels of
Figure~\ref{fig:LambPanelsHds4}. It seems plausible that four of these ECHOs
might be associated with the Monoceros stream ($b\sim30^\circ$), while one of the
two remaining detections (at $l=162.4^\circ$ and $l=59.2^\circ$) is associated
with the \citet{GD06} stream. The ECHO at $l=100.7^\circ$ and $b=56.8^\circ$
remains unassociated with any known substructure.

We conclude that our model reproduces the first and second moments of the
velocity distributions reasonably well for both disk and halo stars. Except in
the region close to the Monoceros stream, the non-Gaussian $v_\phi$ distribution for
disk stars is also well-described. On average, the model agrees with the data to
within $\sim1$~mas~yr$^{-1}$ for proper motions, and $\sim10$~\kms\ for radial
velocities.

\subsection{Constraints on $v_{{\rm LSR}}$ from Large-Scale Halo Kinematics} 
\label{sec:vLSR}

The proper-motion distribution for halo stars towards the Galactic poles depends
only on the difference between the velocity of the local standard of rest (given
that the solar peculiar motion is known to $\sim1$~\kms), $v_{\rm LSR}$, and
$v_\phi$ for halo stars. At least in principle, samples that extend over a large
sky area can be used to provide constraints on both the halo rotation and
$v_{{\rm LSR}}$. Figure~\ref{fig:LambPanelsModelComparison} compares the radial
velocity and longitudinal proper-motion residuals between two models with
($v_\phi^{\rm halo}$,$v_{{\rm LSR}}$)= ($-20$,$180$)~\kms, and ($20$,$220$)~\kms\ (for
both we fixed $v_{\rm LSR}-v_\phi^{\rm halo}=200$~\kms, to make the models agree
with the data towards the North Galactic Pole; see Figure~\ref{fig:VvsZphi}). In
order to distinguish these two models observationally, systematic errors in the
radial velocities must be below $10$~\kms, and systematic errors in the observed
proper motions must be below $1$~mas~yr$^{-1}$. These requirements are
comparable to the systematic errors in our data set, so we can only state that
the data are consistent with $v_{{\rm LSR}}\sim200$~\kms\ (or similarly, no halo
rotation) to an accuracy of $\sim 20$~\kms. The proper motion measurements from
the Gaia survey
\citep{Perryman01} will be sufficiently accurate to exploit the full
potential of this method.

\subsection{ The Kinematic Parallax Relation }

Any constraint on our best-fit model parameters that uses proper
motions is sensitive to the systematic errors in the distance scale
obtained from our photometric-parallax relationship.  However, since
our model is a good fit to the radial velocity data alone (see
Figure~\ref{fig:LambPanelsHvrad}), it is possible to estimate distance
errors by minimizing the differences between the observed and modeled
proper-motion distributions. Such a kinematic-parallax relation,
derived from a combination of radial velocity and proper-motion data
sets, was proposed for the solar neighborhood by \citet{BT87}. Unlike
in the solar neighborhood, our volume is sufficiently large that
kinematics vary with position. Nevertheless, it is conceptually the
same method; a dipole (in our case a more complex angular function) is
fit to the radial velocity and proper-motion distributions, and the
ratio of the best-fit dipole magnitudes constrains the distance scale.

Using only low-metallicity halo stars, we obtain a distance-scale
error of $\sim 5$\% -- our adopted absolute magnitudes should be
$\sim0.1$~mag brighter. This offset is consistent with the expected
systematic errors in the calibration of the photometric-parallax relation
(see I08).  In other words, the adopted distance scale properly
connects the radial and proper-motion distributions. This method
provides a much weaker distance-scale constraints for disk stars than
for halo stars, because the vertical velocity gradient in the disk leads to
a degeneracy between errors in the distance scale and errors in the
adopted disk velocity.  The radial-velocity measurements do not
constrain the velocity scale because the reference point depends on
distance.

%% file: conclusions.tex
\section{ SUMMARY AND DISCUSSION}

This is the first analysis based on SDSS data that simultaneously studies the
kinematics of the halo and disk populations. Past studies of halo stars alone were
performed by \citet{Sirko04}, \citet{Allende06}, \citet{Carollo07} and \citet{Smith09}, while disk 
samples ranging from nearby M stars to distant F/G stars have been studied by
\citet{Bochanski07a}, \citet{West08}, and \citet{Fuchs09}. Throughout this paper, 
we have quantified the
probability distribution function $p_3(\mu_l, \mu_b, v_{{\rm rad}}|u-g,g,g-r,l,b)$,
introduced in I08, that describes proper-motion and radial-velocity measurements
in the $g$ vs. $g-r$ color-magnitude diagram as a function of position on the
sky and $u-g$ color. We have developed a simple empirical model with disk and
halo components that map well to populations detected in the stellar density
distribution (J08) and the metallicity distribution (I08). At distances
accessible to the HIPPARCOS survey ($<100$ pc), we obtain encouraging agreement
with results from \citet{DB98}, \citet{Dehnen98}, and \citet{Nordstrom04}. The
extension of kinematic mapping to distances up to $\sim10$~kpc with millions of
stars represents a significant observational advance, and delivers powerful
new constraints on the dynamical structure of the Galaxy. In less than two decades,
the observational material for such {\it in situ} mapping has progressed from
first pioneering studies based on only a few hundred objects \citep{Majewski92},
to over a thousand objects \citep{CB00}, to the massive data set
discussed here.

\subsection{ Disk Kinematics }

The disk kinematics of the Milky Way are dominated by rotation with a smooth vertical
gradient. Our analysis extends the early measurements of this gradient (e.g.,
\citealt{Murray86, Majewski92, CB00}) to vastly larger sky area and to a much larger
distance range. The mean rotational velocity and the three velocity dispersions
for disk stars can be modeled as simple functions of the form $a+b|Z|^c$ (see
equations~(\ref{pDvPhi}) and (\ref{wZ}), and Table~\ref{Tab:kinD}). The rotational
velocity distribution for the disk component is non-Gaussian, and can be formally
modeled as a sum of two Gaussian components with fixed normalization ratio for
$0.1 \la |Z|/{\rm kpc} \la 4$. The fact that the normalization ratio of these two components
does not vary with $Z$ is at odds with the standard disk decomposition into
thin- and thick-disk components (see also sections 3.4.4 and 4.2.1 in I08). Based on
N-body simulations performed by \citet{Roskar08},
\citet{Loebman08} argued that the absence of a velocity-metallicity
correlation at the thin/thick disk boundary, pointed out by I08, may
be due to a combination of a strong vertical age gradient and the radial
migration of stars \citep[see also][]{SB09}. A more detailed 
study will be presented in Loebman
et al. (in preparation).  A significant vertical age gradient for disk
stars is also supported by an analysis of active M dwarfs presented in
\citet{Bochanski07a}. Such an age gradient, together with the measured
velocity dispersion--age correlations for local disk stars 
(e.g., \citealt{Nordstrom04,Helio04,West08}), 
may be responsible for the measured increase of velocity dispersions 
with distance from the Galactic plane. 

Close to the plane, the proper-motion data imply a complex multi-modal
velocity distribution that is inconsistent with a description based
on a simple Schwarzschild ellipsoid.  It is reassuring that we obtained 
a velocity-distribution morphology very similar to that obtained by 
\citet{Dehnen98} using statistical deprojection of the HIPPARCOS data,
and directly by \citet{Nordstrom04} 
using a subsample of HIPPARCOS stars with full three-dimensional velocity 
information. In addition, our results for the
first and second moments of the velocity distribution for nearby M
stars agree with analogous results obtained recently by
\citet{Fuchs09}. The orientation of the velocity ellipsoid is strongly
affected by multi-modal structure, so one should take care in its
interpretation. We discuss these issues in more detail in Kowalski et
al. (in preparation).

\subsection{ Halo Kinematics }

Our results for the velocity distribution of halo stars are in excellent
agreement with \citet{Smith09}. They used proper-motion measurements 
based only on SDSS data, and thus have significantly different, and most likely
much smaller, systematic errors than the SDSS-POSS proper-motion measurements
analyzed here. The much larger size of the \citet{Munn04} catalog analyzed here
allows us to rule out the possibility that the Smith et al. result was biased by local
substructure. The close agreement of our results for the orientation and size of
the halo velocity ellipsoid (we obtained $\sigma_{r}^H=141$~\kms,
$\sigma_{\theta}^H=75$~\kms, and $\sigma_{\phi}^H=85$~\kms, while their values are
$142$~\kms, $77$~\kms, and $81$~\kms, respectively) are encouraging 
\citep[see also][]{Carollo09}. Their estimated errors of $2$~\kms\ apparently do 
not include systematic effects (such as errors in photometric parallax; both 
studies used the same calibration from I08) -- based on our Monte Carlo simulations, 
we believe that the true errors cannot be smaller than $\sim5$~\kms. Additional 
independent evidence for the tilt of the velocity ellipsoid comes from the RAVE 
survey \citep{Siebert08}. These measurements of the velocity ellipsoid for halo 
stars represent a strong constraint for the shape of gravitational potential, 
as discussed by, e.g., \citet{AC91}, \citet{KdZ91} and \citet{Smith09}.

We note that \citet{Majewski92} measured a retrograde halo rotation
using stars observed towards the North Galactic Pole (in our 
nomenclature, he obtained a mean rotational velocity $v_\Phi=50\pm16$ \kms). 
A star-by-star comparison of his data and the data analyzed here showed 
that photometric and proper-motion measurements agree within the stated
errors. The main reason for different conclusions about halo rotation
are the different distance scales: his distances are on average 30\% larger than our
distances, resulting in larger tangential velocities. 
The \citet{Carollo07} claim for a large outer-halo retrograde rotation, further refined by
\citet{Carollo09} (in our nomenclature, a mean rotational velocity
$v_\Phi=80 \pm13$ \kms), remains intact, as their distance scale is similar to
ours. 

The kinematic measurements for halo stars presented here should not be
extrapolated beyond the sample distance limit of $10$~kpc. For
example, using $241$ halo objects, including stars, globular clusters,
and satellite galaxies, \citet{Battaglia05} detected a continuous
decline of the radial velocity dispersion beyond a Galactocentric
radius of $\sim30$~kpc, from about $120$~\kms\ to $50$~\kms\ at
$\sim120$~kpc. In addition, the distance limit of our sample, together
with the decreasing sensitivity of the photometric metallicity indicator 
for $[{\rm Fe/H}]<-2.5$, prevent us from robustly testing the possible halo
dichotomy discussed by \citet{Carollo07, Carollo09}. 

\subsection{ Kinematic Substructure }

The model developed here can be used to search for kinematic
substructure with a low contrast level. For example, \citet{Schlaufman09}
had to generate a background model when
searching for cold streams using radial velocity data; a similar
study for the solar neighborhood was performed by \citet{Klement09}. A
global model-based description is especially important when using
large numbers of proper-motion measurements to search for substructure. While 
radial-velocity data are superior at large distances, searches based on
proper motions should be better within a few kpc, due to the high
completeness and much larger sample size.  A user-friendly interface
to our model code {\it Galfast}\footnote{Please see www.mwscience.net/galfast}, 
which allows generation of mock catalogs in an
arbitrary direction (or across the entire sky) and to an arbitrary
depth, will be described elsewhere (Juri\'{c} et al., in preparation).

The Monoceros stream is clearly detected as a major outlier from the
smooth model presented here. We have not found any other large kinematic
substructure within $10$~kpc that deviates at a detectable level. Other deviations
from the smooth model predictions are likely due to systematic errors
in the proper motion and radial velocity measurements. However, since
the main goal of this paper was to quantify the overall kinematic
behavior, the emphasis of our analysis was on the first and second
moments of the kinematic quantities. It is likely that more
sophisticated statistical methods, such as those discussed by
\citet{Schlaufman09} and \citet{Klement09}, will
be more efficient in searching for substructure. As an example, 
we have verified that a moving group discovered by \citet{Majewski92}
is reproducible with our data. 

\subsection{ Future Surveys }

The results presented here will be greatly extended by several upcoming large-scale, deep optical surveys, including the Dark Energy Survey \citep{Flaugher08}, Pan-STARRS \citep{Kaiser02}, and the Large Synoptic Survey Telescope \citep{Ivezic08b}.  These surveys will extend the faint limit of this sample and that of the upcoming Gaia mission \citep{Perryman01,Wilkinson05} by $4-6$~mag.  For example, LSST will obtain proper motion measurements of comparable accuracy to those of Gaia at their faint limit, and smoothly extend the error vs. magnitude curve deeper by $5$~mag (for details see Eyer et al., in preparation).  With its $u$-band data, LSST will enable studies of metallicity and kinematics using the {\it same} stars out to a distance of $\sim100$~kpc ($\sim 200$~million F/G main sequence stars brighter than $g=23.5$, for a discussion see I08).  By comparison, the best measurements of the outer-halo radial velocity dispersion to date are based on from several hundred \citep{Battaglia05} to several thousand \citep{Xue08} objects. These upcoming studies are thus certain to provide valuable new information about the formation and evolution of our Galaxy.

\acknowledgements

We thank Vladimir Korchagin for making his code for calculating 
equilibrium velocity dispersion profile available to us. 
\v{Z}. Ivezi\'{c} and B. Sesar acknowledge support by NSF grants AST-615991 
and AST-0707901, and by NSF grant AST-0551161 to LSST for design and
development activity. J. Dalcanton acknowledges NSF CAREER grant AST-02-38683. 
D. Schneider acknowledges support by NSF grant AST-06-07634. Allende Prieto 
acknowledges support by NASA grants NAG5-13057 and NAG5-13147.
T.C. Beers, Y.S. Lee, and T. Sivarani acknowledge partial support from
PHY 08-22648: Physics Frontier Center/Joint Institute for Nuclear
Astrophysics (JINA), awarded by the U.S. National Science Foundation.
P. Re Fiorentin acknowledges support through the Marie Curie
Research Training Network ELSA (European Leadership in Space Astrometry) 
under contract MRTN-CT-2006-033481.
We acknowledge the hospitality of the KITP at the University of California, 
Santa Barbara, where part of this work was completed (supported by NSF grant 
PHY05-51164). Fermilab is
operated by Fermi Research Alliance, LLC under Contract No. DE-AC02-07CH11359 
with the United States Department of Energy.

Funding for the SDSS and SDSS-II has been provided by the Alfred
P. Sloan Foundation, the Participating Institutions, the National
Science Foundation, the U.S. Department of Energy, the National
Aeronautics and Space Administration, the Japanese Monbukagakusho, the
Max Planck Society, and the Higher Education Funding Council for
England. The SDSS Web Site is http://www.sdss.org/.
The SDSS is managed by the Astrophysical Research Consortium for the
Participating Institutions. The Participating Institutions are the
American Museum of Natural History, Astrophysical Institute Potsdam,
University of Basel, University of Cambridge, Case Western Reserve
University, University of Chicago, Drexel University, Fermilab, the
Institute for Advanced Study, the Japan Participation Group, Johns
Hopkins University, the Joint Institute for Nuclear Astrophysics, the
Kavli Institute for Particle Astrophysics and Cosmology, the Korean
Scientist Group, the Chinese Academy of Sciences (LAMOST), Los Alamos
National Laboratory, the Max-Planck-Institute for Astronomy (MPIA),
the Max-Planck-Institute for Astrophysics (MPA), New Mexico State
University, Ohio State University, University of Pittsburgh,
University of Portsmouth, Princeton University, the United States
Naval Observatory, and the University of Washington.

%% file: appendix.tex
\appendix

\section{\bf The Revised SDSS Metallicity Scale}

Analysis of the metallicity and kinematics of halo and disk 
stars by I08 utilized photometric metallicity estimates for F/G 
stars with $0.2 < g-r < 0.6$. Their mapping function from the 
$g-r$ vs. $u-g$ color-color diagram to metallicity was calibrated 
using stars with spectroscopic metallicities distributed in SDSS 
Data Release 6. At that time, high-metallicity stars required for 
the calibration of methods implemented in the automated spectroscopic 
pipeline \citep[SEGUE Stellar Parameters Pipeline,][]{Beers06} were 
not available. Between Data Releases 6 and 7, the required data 
were collected and the new calibration resulted in the revised 
spectroscopic metallicity values distributed with Data Release 7 
\citep{Lee08a,Lee08b,Allende08}.

Here, we recalibrate the photometric-metallicity estimator using
updated spectroscopic metallicities from Data Release 7.  In addition,
we re-derive the parts of the I08 analysis that are most affected by
this change in the metallicity scale.

\subsection{ The Updated Photometric-Metallicity Estimator} 

As shown in Figure~\ref{Fig:App1}, the largest difference between the
SDSS spectroscopic metallicity values distributed with Data Releases 6
and 7 is, as expected, at the high-metallicity end. In particular, the
abrupt cutoff in the metallicity distribution at $[{\rm Fe/H}]\sim-0.5$ (see
figure~9 in I08) is no longer present, and the distribution extends to
values as high as $[{\rm Fe/H}]\sim-0.2$ (the distances for the stars shown
range from $\sim1$~kpc to $\sim7$~kpc).

We proceed to re-derive the photometric-metallicity calibration using
the same selection criteria and the same methodology as in I08.  The
new data set admits a slightly simpler function -- the double
definition of the $x$ axis is no longer required, and the new
expression is
\begin{equation}
\label{Zphotom}
  [{\rm Fe/H}]_{{\rm ph}}= A + Bx + Cy + Dxy  + Ex^2 + Fy^2 + Gx^2 y + Hxy^2 + Ix^3 + Jy^3,
\end{equation}
with $x=(u-g)$ and $y=(g-r)$. The best-fit coefficients are ($A-J$) =
($-13.13$, $14.09$, $28.04$, $-5.51$, $-5.90$, $-58.68$, $9.14$,
$-20.61$, $0.0$, $58.20$).  Note that the coefficient for $x^3$, $I$, is
to zero. We removed this term because it was producing too much curvature
at the right end (red $u-g$) of the best-fit map.

We estimate that an upper limit on the intrinsic metallicity scatter
for fixed noiseless $u-g$ and $g-r$ colors (presumably due to limited
sensitivity of broadband colors to metallicity variations) is about
$0.1$~dex. This value is estimated from the scatter in the difference
between spectroscopic and photometric metallicities, discussed below.
Unlike I08, who simply adopted the median metallicity value given by
the above expression for each star, we draw photometric metallicity
estimates from a Gaussian distribution centered on the best-fit median
value, and with a width of $0.1$~dex. The main benefit to this is that
we avoid hard edges in the photometric metallicity distribution for
stars close to the edges of the calibration region in the $g-r$
vs. $u-g$ diagram.

The performance of the new map is qualitatively similar to that of the
old map. The mean and rms values for the difference between spectroscopic 
and photometric metallicities as functions of the $g-r$ and $u-g$ colors are shown in
the top two panels of Figure~\ref{Fig:App2}. Typical systematic errors
in the map (i.e. the median difference per pixel) are $\sim0.1$~dex or
smaller, and the scatter varies from $\sim0.2$~dex at the
high-metallicity end to $\sim0.3$~dex at the low-metallicity end (note
that this scatter includes a contribution from errors in both
spectroscopic and photometric metallicities).

The above photometric metallicity estimator is applicable for stars
with $0.2 < g-r < 0.6$ and $-0.25 + 0.5(u-g) < g-r < 0.05 +
0.5(u-g)$; that is, for main-sequence F and G stars in the calibration
region of the $g-r$ vs. $u-g$ color-color diagram (top two panels of
Figure~\ref{Fig:App2}). For stars with spectroscopic metallicity
$[{\rm Fe/H}]>-2.2$, the distribution of the difference between
spectroscopic and photometric metallicities is well-described by a
Gaussian with a width of $0.26$~dex (see the bottom right panel in
Figure~\ref{Fig:App2}).

It should be noted that the performance of the photometric-metallicity
estimator deteriorates at the low-metallicity end because the $u-g$
color becomes insensitive to decreases in metallicity. As shown in the
bottom left panel of Figure~\ref{Fig:App2}, the photometric
metallicity saturates at $[{\rm Fe/H}] \sim -2$ for smaller values of
spectroscopic metallicity.  Even at $[{\rm Fe/H}]=-2$, the true metallicity
is overestimated by $0.2-0.3$~dex, and by $[{\rm Fe/H}]=-3$ this bias is as
large as $1$~dex (the photometric-metallicity values never become
significantly lower than $[{\rm Fe/H}]=-2$). This shortcoming could be
partially alleviated by employing more accurate $u$-band photometry
(say, with errors of $0.01$~mag instead of $0.03$~mag, as used here),
but probably not for metallicities lower than $[{\rm Fe/H}] = -2.5$.
Fortunately, the low-metallicity inner-halo stars within SDSS reach have a
median metallicity of $[{\rm Fe/H}]\sim-1.5$ (I08), so for the majority of
stars the photometric metallicities are robust.  Another important
note is that, despite the improvement at the high-metallicity end, the
calibration range only extends to $[{\rm Fe/H}]\sim-0.2$. Our calibration
sample did not include young, metal-rich main-sequence stars ($u-g>1.1$
and $g-r<0.3$). For this reason, our polynomial model underestimates
true metallicities by $0.2-0.3$~dex at $[{\rm Fe/H}]=0$, and probably more
for $[{\rm Fe/H}]>0$.  Any result relying on higher metallicities
should be interpreted with caution (especially at low Galactic
latitudes where the uncertain ISM extinction may strongly affect the
estimated metallicities). We plan to extend our calibration further
into the metal-rich domain by employing data from the SDSS-III
SEGUE-2 survey \citep{Yanny09}, and further refinements of the SEGUE Stellar Parameters
Pipeline, which are now underway. 

\subsection{    Tomography II Reloaded    } 

I08 pointed out several aspects of their analysis that may have been affected by
the metallicity ``compression'' at the high-metallicity end in DR6. We repeated
their full analysis and report here on those aspects where differences warrant
discussion.

The ``hard'' upper limit on photometric metallicity estimates at the
high-metallicity end ($[{\rm Fe/H}]\sim-0.5$) with the DR6 calibration is best seen in
the bottom left panel in figure~9 from I08. We reproduce that map of the
conditional metallicity distribution in the top left panel of
Figure~\ref{Fig:App3}. As expected, the metallicity distribution of disk stars
within $2$~kpc of the Galactic plane now extends to $[{\rm Fe/H}]\sim 0$.

In the new calibration, the parameters describing the variation of the median
metallicity for disk stars as a function of the distance from the Galactic
plane,
\begin{equation}
  \mu_D(Z) = \mu_\infty + \Delta_\mu \,\exp(-|Z|/H_\mu) \,\, {\rm dex},
\label{muZApp}
\end{equation}
are also changed.  The updated values are $H_\mu=0.5$~kpc,
$\mu_\infty=-0.82$ and $\Delta_\mu=0.55$ (the old values were
$H_\mu=1.0$~kpc, $\mu_\infty=-0.78$ and $\Delta_\mu=0.35$).  The
best-fit values of $\mu_\infty$ and $\Delta_\mu$ are accurate to
$\sim0.05$~dex. Values of $H_\mu$ in the range $350-700$~pc are
consistent with the data -- the decrease in $H_\mu$ is required by the
local constraint, $\mu_D(Z) = -0.2$ \citep{Nordstrom04,Allende04}.

An interesting result from I08 was the detection of disk stars at a distance
from the Galactic plane as large as $\sim6$~kpc (see their Figure~10). A peak at
$[{\rm Fe/H}]=-0.5$ in the metallicity distribution of stars at those distances was
another manifestation of the metallicity ``compression.'' As demonstrated in the
top right panel of Figure~\ref{Fig:App3}, this peak is not present when using
the revised calibration. However, there is still statistical evidence that disk
stars exist at these large distances from the plane: about 5\% of stars in the
$5<Z<7$~kpc bin are consistent with disk stars, in agreement with extrapolation
of the exponential profile derived from stellar counts. \citet{LB09} and
\citet{Carollo09} have also commented on signatures of thin-disk-like
chemistry and kinematics for a small fraction of stars several kpc above the
Galactic plane.

Perhaps the most intriguing result of I08 was the non-detection of a correlation
between rotational velocity and metallicity for disk stars at $Z\sim1$~kpc. At
such distances from the Galactic plane, the counts of thin- and thick-disk stars
are expected to be similar. Since traditionally the thick disk component is
associated with a larger velocity lag and lower metallicities, a fairly strong
and detectable correlation was expected (see I08 for details). The two bottom
panels in Figure~\ref{Fig:App3} demonstrate that such a correlation is still
undetected, although the photometric metallicity range now extends to higher
values (up to $[{\rm Fe/H}]\sim-0.2$).

The higher metallicity values obtained with the re-calibrated relation
have a quantitative effect on the best-fit metallicity distributions
shown in Figure~7 of I08.  Using the same methodology, we reproduce
the metallicity distributions with the new calibration in
Figure~\ref{Fig:App4}. I08 modeled the non-Gaussian disk metallicity
distribution using a sum of two Gaussians with a fixed amplitude ratio
($1.7$:$1$), fixed difference of the mean values ($0.14$~dex), and
fixed widths ($0.21$~dex and $0.11$~dex), which ``slides'' as a
function of $Z$, according to equation~(\ref{muZApp}).  We find that the
only significant change is an increase to the width of the second
Gaussian to $0.21$~dex, which accounts for the extension of the
metallicity distribution to higher values. Only minor changes are
required for the best-fit halo metallicity distribution (see Table~3
in I08): the median halo metallicity is now $-1.46$ in the first three
bins, and $-1.56$ in the most distant $Z$ bin, and its width changed
from $0.32$~dex to $0.36$~dex in the last bin. We note somewhat less
scatter of the data points around the best-fit functions with the
re-calibrated data set. To summarize, the revised best-fit parameters
that describe halo and disk metallicity distributions are:
\begin{itemize}
\item The halo metallicity distribution is spatially invariant and
  well-described by a Gaussian distribution centered on
  $[{\rm Fe/H}]=-1.46$, with a width $\sigma_H=0.30$~dex (not including measurement
  errors).  For $|Z|\la10$~kpc, an upper
  limit on the halo radial-metallicity gradient is
  $0.005 $~dex~kpc$^{-1}$.
\item The disk metallicity distribution varies with $Z$ such that its
  shape remains fixed, while its median, $\mu_D$, varies as given by
  equation~(\ref{muZApp}) (with best-fit parameter values
  $H_\mu=0.5$~kpc, $\mu_\infty=-0.82$, and $\Delta_\mu=0.55$). The
  shape of the disk metallicity distribution can be modeled as
\begin{equation} 
\label{pDmetalApp}
  p_D(x=[{\rm Fe/H}]|Z)= 0.63\,G[x|\mu=a(Z), \sigma=0.2] + 0.37\,G[x|\mu=a(Z)+0.14, \sigma=0.2], 
\end{equation} 
where the position $a$ and the median $\mu_D$ are related via
$a(Z)=\mu_D(Z)-0.067$ (unless measurement errors are very large).
\end{itemize}

We point out that the asymmetry of the metallicity distribution for
disk stars is now less pronounced (as implied by the same widths of
the two best-fit Gaussian components). Nevertheless, due to our large
sample size, the non-Gaussianity is detected at high significance. A
remaining uncertainty is the error distribution of the photometric
metallicities, which itself could account for such a deviation from
Gaussianity. However, to the extent possible using a highly incomplete
spectroscopic sample (c.f. the bottom right panel in
Figure~\ref{Fig:App2} and discussion in I08), we are unable to
quantitatively explain the observed deviation from Gaussianity as an
artifact of the photometric-metallicity method.

\section{\bf {\tt SQL} Query Example}

The following {\tt SQL} query was used to select and download data
for all SDSS stars with spectroscopic and proper-motion measurements
(see http://casjobs.sdss.org/CasJobs). 

\begin{verbatim}
SELECT
  round(p.ra,6) as ra, round(p.dec,6) as dec, 
  p.run, p.rerun, round(p.extinction_r,3) as rExt, 
  round(p.psfMag_u,2) as upsf,  --- comments are preceded by ---
  round(p.psfMag_g,2) as gpsf,  --- rounding up 
  round(p.psfMag_r,2) as rpsf, 
  round(p.psfMag_i,2) as ipsf, 
  round(p.psfMag_z,2) as zpsf, 
  round(p.psfMagErr_u,2) as uErr, 
  round(p.psfMagErr_g,2) as gErr, 
  round(p.psfMagErr_r,2) as rErr, 
  round(p.psfMagErr_i,2) as iErr,
  round(p.psfMagErr_z,2) as zErr,
  round(s.pmL,2) as pmL, round(s.pmB,2) as pmB, pmRaErr, 
  t.specObjID, t.plate, t.mjd, t.fiberid, t.feha, t.fehaerr, 
  t.logga, t.loggaerr, t.elodierv, t.elodierverr
INTO mydb.pmSpec
FROM star p, propermotions s, sppParams t, specobjall q
WHERE 
  p.objID = s.objID and s.match = 1 --- must have proper motion   
  and t.specobjid = q.specobjid and q.bestobjid = p.objid 
  and s.sigra < 350 and s.sigdec < 350  --- quality cut on pm
  and (p.flags & '4295229440') = 0  --- see text for flag list
  and p.psfMag_r > 14.5  --- avoid saturation                 
  and p.psfMag_r < 20    --- practical faint limit for pm
--- the end of query
\end{verbatim}

The following {\tt SQL} query was used to select and download data
for all spectroscopically confirmed quasars with proper-motion 
measurements and redshifts in the range $0.5-2.5$. 

\begin{verbatim}
SELECT
  round(p.ra,6) as ra, round(p.dec,6) as dec, 
  p.run, p.rerun, round(p.extinction_r,3) as rExt, 
  round(p.psfMag_u,2) as upsf, 
  round(p.psfMag_g,2) as gpsf, 
  round(p.psfMag_r,2) as rpsf, 
  round(p.psfMag_i,2) as ipsf, 
  round(p.psfMagErr_u,2) as uErr, 
  round(p.psfMagErr_g,2) as gErr, 
  round(p.psfMagErr_r,2) as rErr, 
  round(p.psfMagErr_i,2) as iErr,
  round(s.pmL,2) as pmL, round(s.pmB,2) as pmB, pmRaErr, 
  q.specObjID, q.plate, q.mjd, q.fiberID, 
  q.z, q.zErr, q.zConf, q.zWarning, q.specClass
INTO mydb.pmQSO
FROM star p, propermotions s, specobjall q
WHERE 
  p.objID = s.objID and s.match = 1      
  and q.bestobjid = p.objid 
  and s.sigra < 350 and s.sigdec < 350 -- per \citet{Munn04}
  and (p.flags & '4295229440') = 0
  and p.psfMag_r > 14.5                   
  and p.psfMag_r < 20
  and q.z > 0.5   --- redshift limits
  and q.z < 2.5
--- the end of query
\end{verbatim}

%% file: tables.tex
\begin{deluxetable}{rrrr}
\tablenum{1} \tablecolumns{4} \tablewidth{340pt}
\tablecaption{Best-Fit Parameters$^a$ for the Disk Velocity Distribution$^b$.}
\tablehead{Quantity  & $a$ & $b$ & $c$  }
\startdata
      $\bar{v_\phi}^1$ & $-$194  & 19.2 & 1.25 \\
   $\sigma_{\phi}^1$   & 12  & 1.8  & 2.0  \\ 
   $\sigma_{\phi}^2$   & 34  & 1.2  & 2.0  \\
   $\sigma_{\phi}^D$   & 30  & 3.0  & 2.0  \\ 
       $\sigma_{R}$    & 40  & 5.0  & 1.5  \\ 
       $\sigma_{Z}$    & 25  & 4.0  & 1.5  \\ 
\enddata
\tablenotetext{a}{
All listed quantities are modeled as $a+b|Z|^c$, with $Z$ in kpc, 
and velocities in \kms.} The uncertainties are typically $\sim 10$ \kms\ 
for $a$, $\sim$20\% for $b$ and 0.1-0.2 for $c$. 
\tablenotetext{b}{
The $v_\phi$ distribution is non-Gaussian, and can be formally 
described by a sum of two Gaussians with a fixed normalization ratio 
$f_k$:1, with $f_k=3.0$. The mean value for the second Gaussian has a 
fixed offset from the first Gaussian, 
$\bar{v_\phi}^2 = (\bar{v_\phi}^1 - \Delta \bar{v_\phi})$,
with $\Delta \bar{v_\phi}=34$ km/s. Extrapolation beyond 
$Z>5$ kpc is not reliable. The velocity 
dispersion for the second Gaussian is given by $\sigma_{\phi}^2$.
If this non-Gaussianity is ignored, the $v_\phi$ dispersion
is given by $\sigma_{\phi}^D$.}
\label{Tab:kinD}
\end{deluxetable}

\begin{deluxetable}{rrrr}
\tablenum{2} \tablecolumns{4} \tablewidth{340pt}
\tablecaption{A Model Prediction for Velocity Ellipsoid Tilt$^a$}
\tablehead{ $R$ (kpc) & $Z$ (kpc) & $\theta_{RZ}$ (deg) & arctan($Z/R$) (deg) }
\startdata
 8.0 & 0.7 &  3 &  5.0 \\
 8.0 & 2.0 & 11 & 14.0 \\
 8.0 & 3.5 & 26 & 23.6 \\
 7.0 & 4.0 & 32 & 29.7 \\
 9.0 & 4.0 & 25 & 24.0 \\
11.0 & 4.0 & 21 & 20.0 \\ 
\enddata
\tablenotetext{a}{
The first two columns determine the position in the Galaxy.
The third column lists the predicted orientation of the 
velocity ellipsoid computed as described in \S~\ref{sec:tilt}. 
The fourth column lists the orientation of a velocity ellipsoid 
pointed towards the Galactic center.}
\label{Tab:tilt}
\end{deluxetable}

%% file: figuresDR7.tex
\clearpage

\begin{figure}                      
\figurenum{1}
\vskip -3.1in
\plotone{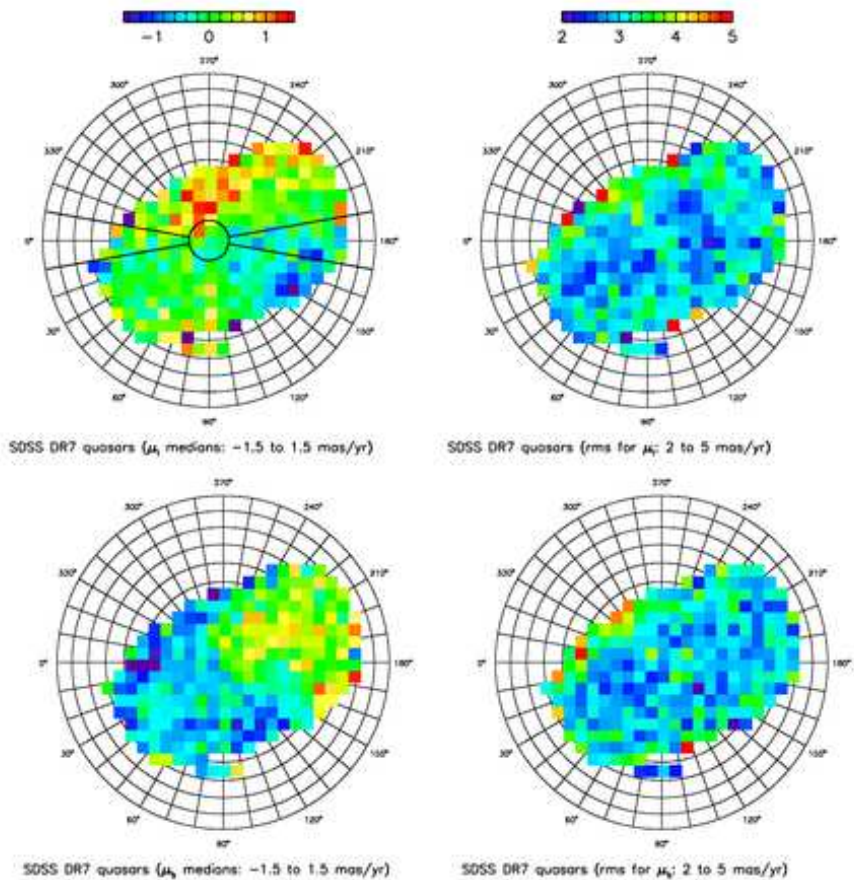}
\vskip -1.4in
\caption{
Behavior of proper-motion measurements for 
$60,000$ spectroscopically confirmed SDSS quasars with
$b>0^\circ$. The color-coded maps (see the 
legend on top, units are mas~yr$^{-1}$) show the 
distribution of the median (left) and rms (right) for 
the longitudinal (top) and latitudinal (bottom) proper motion 
components in a Lambert projection of the northern Galactic cap. 
The median number of quasars per pixel is $\sim250$. 
For both components, the scatter across the sky is $0.60$~mas~yr$^{-1}$. 
The median proper motion for 
the full quasar sample is $0.15$~mas~yr$^{-1}$ in the longitudinal 
direction, and $-0.20$~mas~yr$^{-1}$ in the latitudinal direction.
The thick line in the top left panel shows the 
selection boundary for the ``meridional plane'' sample. 
\label{qsoPMerrors}}
\end{figure}

\begin{figure}                      
\figurenum{2}
\plotone{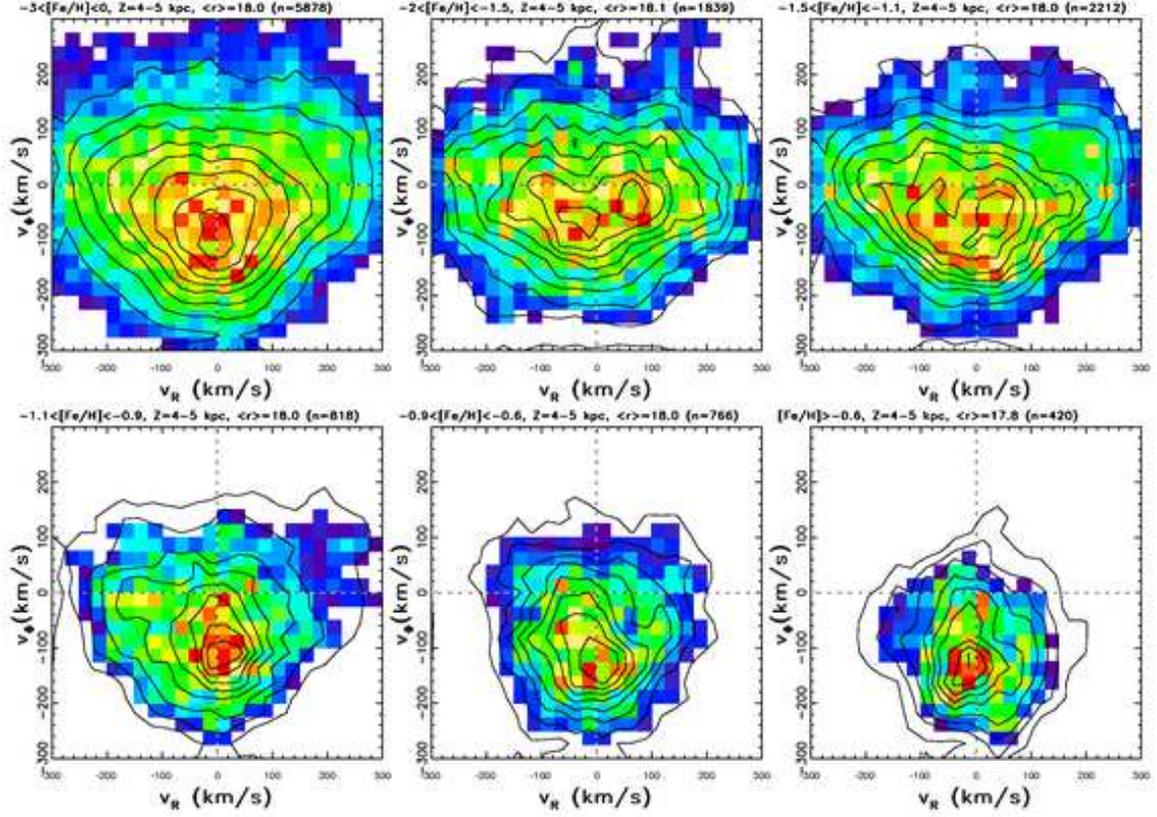}
\vskip -0.5in
\caption{Change of the $v_\Phi$ vs. $v_R$ velocity distribution
  with metallicity, at an approximately constant $R$ and
  $Z$. Velocities are determined from proper-motion measurements.  The
  top left panel shows the $v_\Phi$ vs. $v_R$ diagram for $\sim6,000$
  blue ($0.2<g-r<0.4$) stars at $Z=4-5$~kpc and detected towards the
  North Galactic Pole ($b>80^\circ$). The distribution is shown using
  linearly-spaced contours, and with color-coded maps showing smoothed
  counts in pixels (low-to-high from blue-to-red). The other five panels
  are analogous, and show subsamples selected by metallicity, with the
  $[Fe/H]$ range listed above each panel (also listed are the median
  $r$-band magnitude and subsample size). The measurement errors are
  typically $70$~\kms~(per star). Note the strong variation of median $v_\Phi$
  with metallicity.
\label{fig:vPhivRblue}}
\end{figure}

\begin{figure}                      
\figurenum{3}
\vskip -1in
\plotone{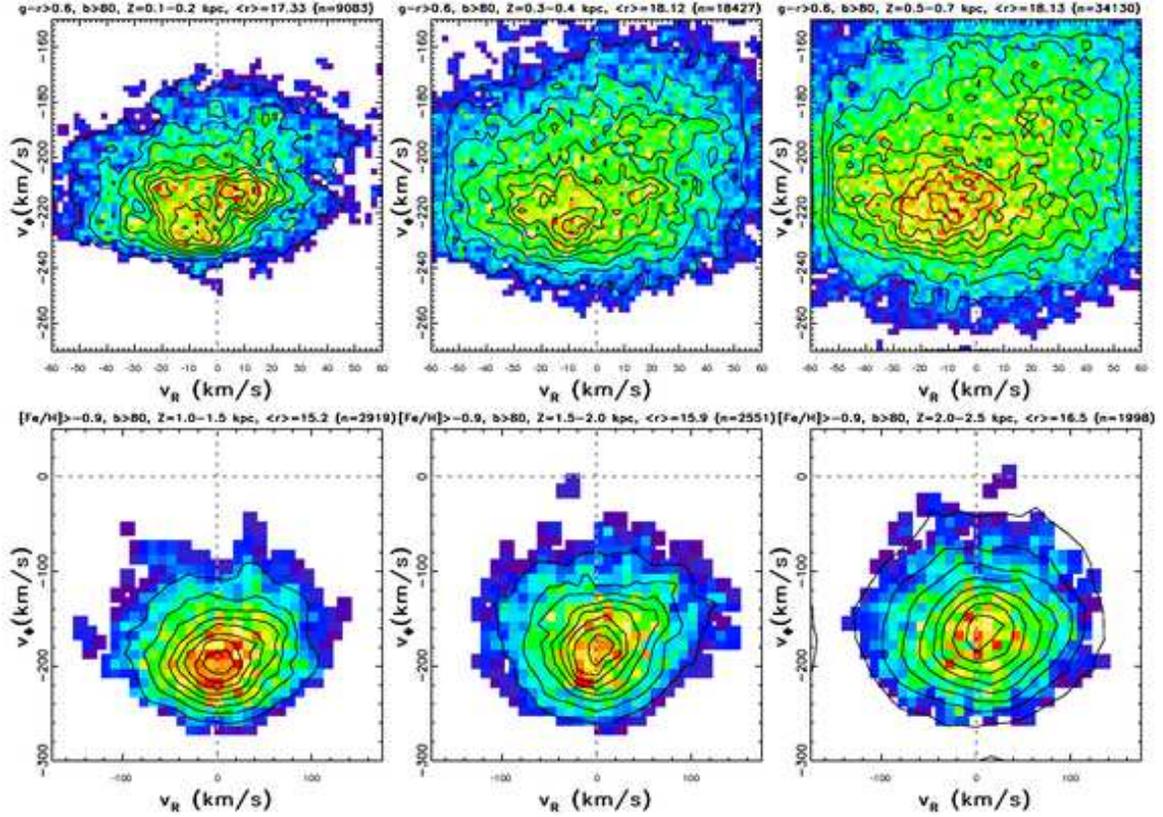}
\vskip -0.5in
\caption{Similar to Figure~\ref{fig:vPhivRblue}, 
except that the $v_\Phi$ vs. $v_R$ velocity distribution 
is plotted for a range of $Z$. The top row shows the $v_\Phi$ vs. 
$v_R$ diagrams for $\sim60,000$ red ($g-r>0.6$) stars at $Z=100-700$~pc and observed towards the North Galactic Pole. Each 
panel corresponds to a narrow range in 
$Z$, given above each panel. The measurement 
errors vary from $\sim3$~\kms\ in the closest
bin to $\sim12$~\kms\ in the most distant bin. Note the 
complex multi-modal substructure in the top left panel. 
The bottom three panels are analogous, and show the $v_\Phi$ vs. 
$v_R$ diagrams for $\sim7,000$ blue ($0.2<g-r<0.4$) stars
with high metallicity ($[Fe/H]>-0.9$). The measurement 
errors vary from $\sim20$~\kms in the closest
bin to $\sim35$~\kms in the most distant bin. Note that
the median $v_\Phi$ approaches zero as $Z$ increases.
\label{fig:vPhivRred}}
\end{figure}

\begin{figure}                      
\figurenum{4}
\vskip -1in
\plotone{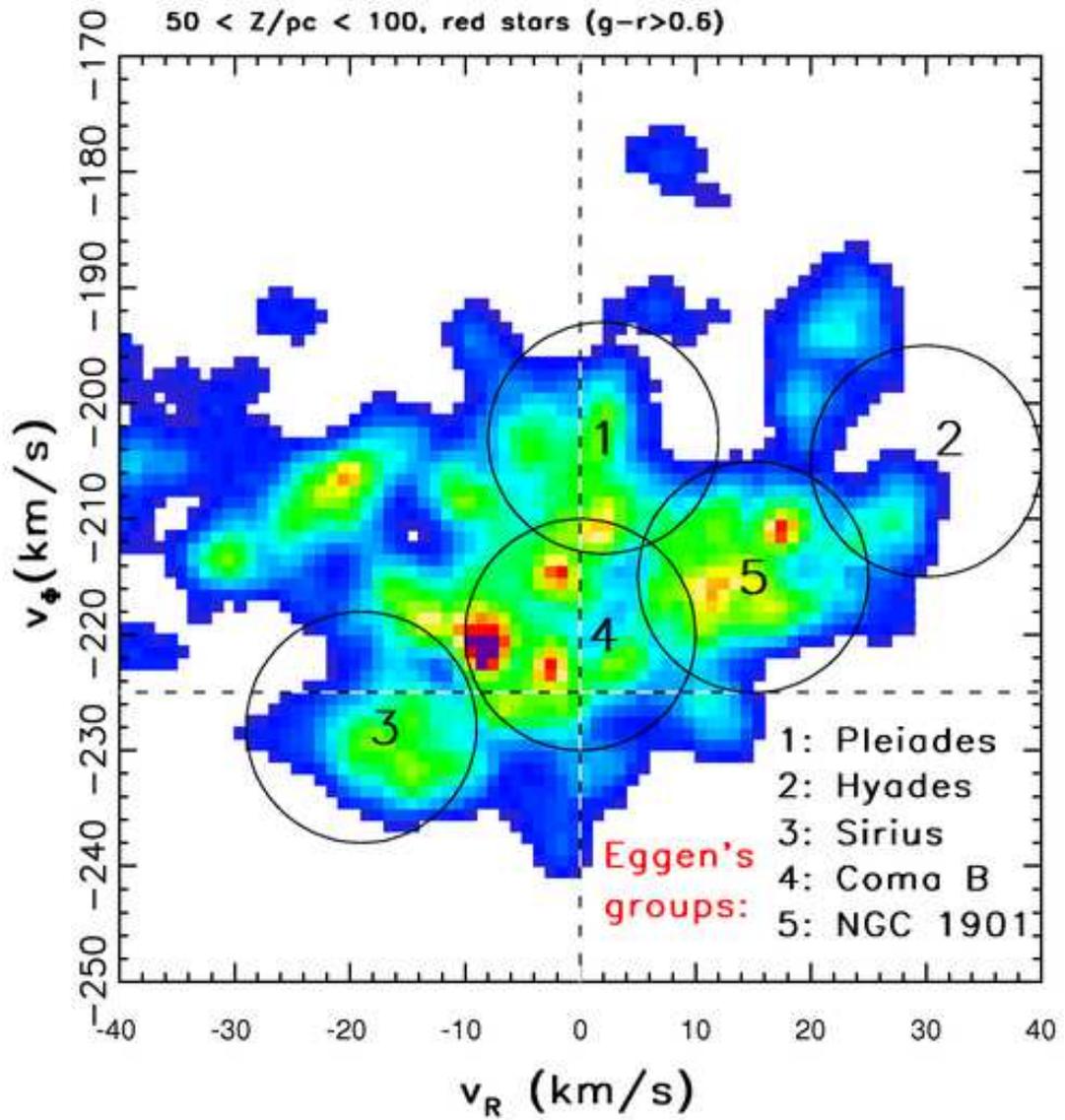}
\vskip -1.5in
\caption{Similar to the top left panel in Figure~\ref{fig:vPhivRred},
except that stars are selected from a distance bin that corresponds
to HIPPARCOS sample ($Z=50-100$~pc). The positions of Eggen's moving
groups \citep{Egg96} are marked by circles, according to the legend
in the bottom right corner. The horizontal line at $v_\phi=-225$~\kms\
corresponds to vanishing heliocentric motion in the rotational
direction. 
\label{fig:Eggen}}
\end{figure}

\begin{figure}        
\figurenum{5}
\vskip -2.2in        
\plotone{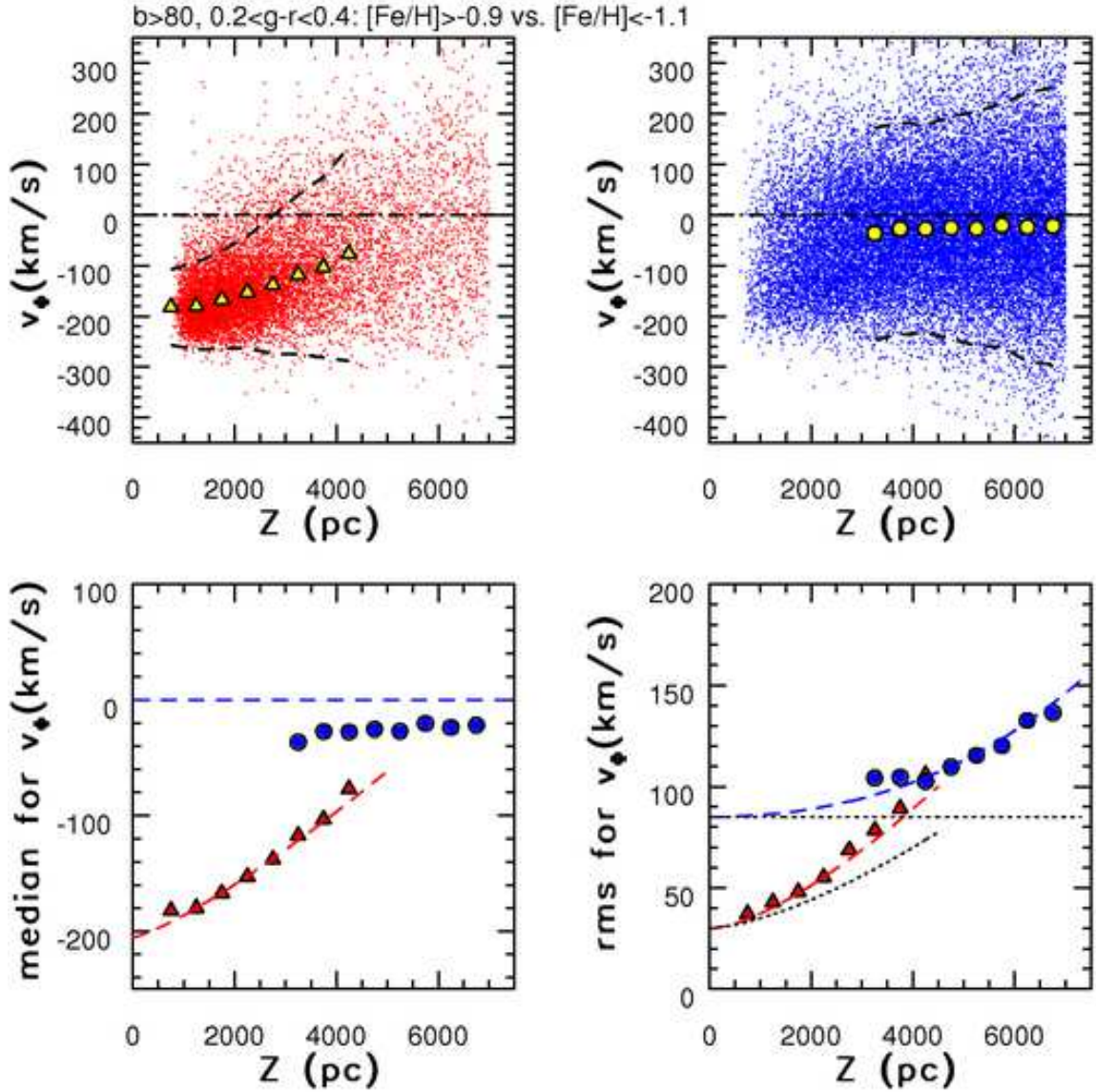}
\vskip -0.5in        
\caption{Dependence of the rotational velocity, $v_\Phi$, 
on distance from the Galactic plane for $14,000$ high-metallicity ($[Fe/H]>-0.9$;
top left panel) and $23,000$ low-metallicity ($[Fe/H]<-1.1$, top right) 
stars with $b>80^\circ$. In the top two panels, individual stars are plotted 
as small dots, and the medians in bins of $Z$ are plotted as large 
symbols. The $2\sigma$ envelope around the medians is shown by dashed 
lines. The bottom two panels compare the medians (left) and dispersions 
(right) for the two subsamples shown in the top panels and the dashed
lines in the bottom two panels show predictions of a kinematic model
described in text. The dotted lines in the bottom right panel show
model dispersions {\bf without} a correction for measurement errors
(see Table~\ref{Tab:kinD}).  
\label{fig:VvsZphi}}
\end{figure}

\begin{figure}
\figurenum{6}
\plotone{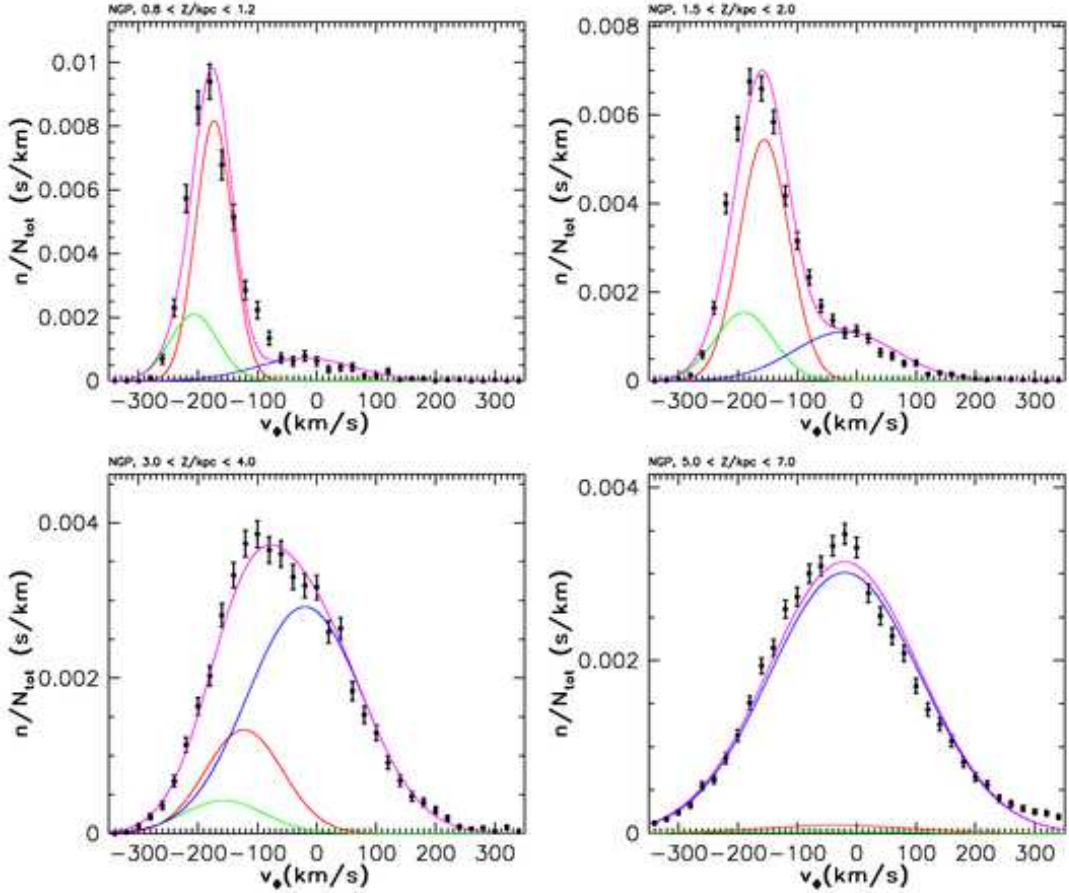}
\vskip -1.5in
\caption{
Symbols with error bars are the measured rotational-velocity
distribution, $v_\Phi$, for stars with $0.2 < g-r < 0.4$, $b>80^\circ$,
and $Z=0.8-1.2$~kpc 
(top left, $\sim1,500$ stars), 
$1.5-2.0$~kpc (top right, $\sim4,100$ stars), $3.0-4.0$~kpc (bottom left,
$\sim6,400$ stars) and $5.0-7.0$~kpc (bottom right, $\sim12,500$ stars).
The red and green curves show the contribution of a two-component disk 
model (see Equations~\ref{pDvPhi} and \ref{wZ}), the blue curves show the Gaussian
halo contribution, and the magenta curves are their sum.  Note the difference in the scale of the $y$ axis between the top two panels.}
\label{fig:VphiHist}
\end{figure}

\begin{figure}                
\figurenum{7}
\vskip -3in        
\plotone{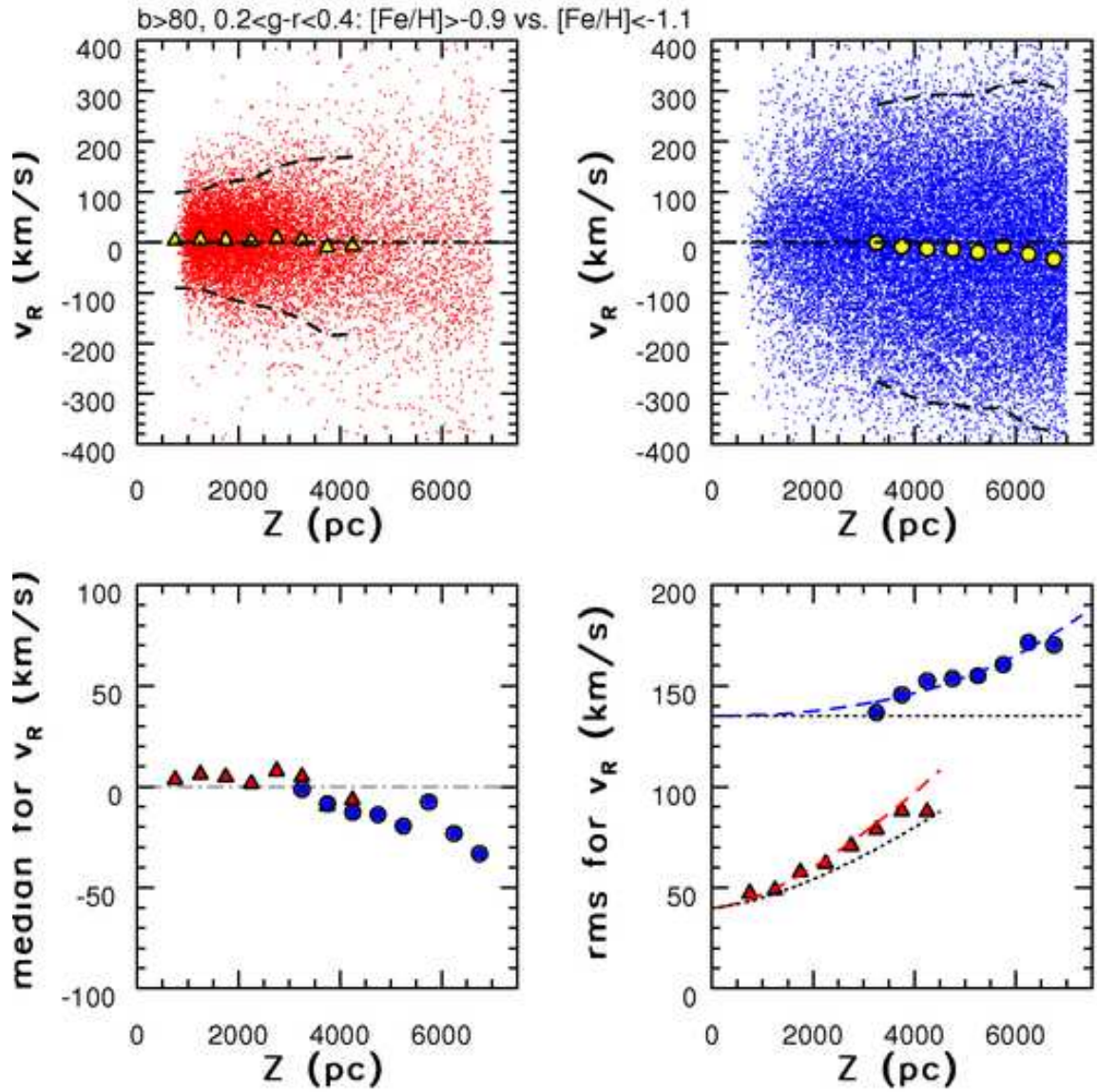}
\vskip -0.5in        
\caption{Analogous to Figure~\ref{fig:VvsZphi}, but for
the radial-velocity component, $v_R$. 
\label{fig:VvsZR}}
\end{figure}

\begin{figure}                      
\figurenum{8}
\vskip -2.5in
\epsscale{0.85}        
\plotone{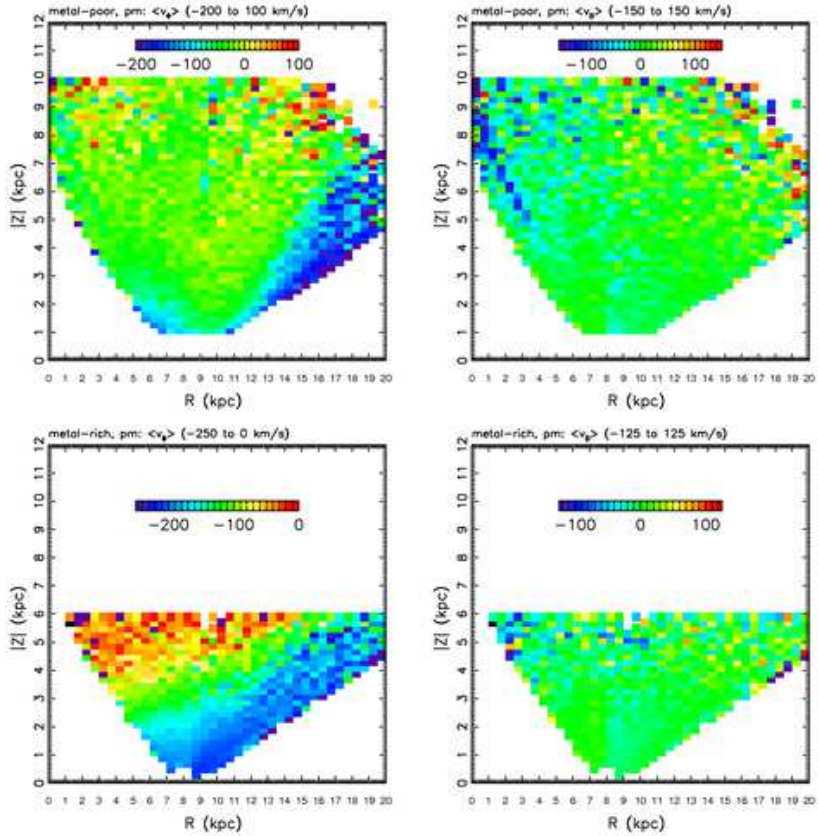}
\epsscale{1.0}        
\vskip -1.1in        
\caption{Dependence of velocity, measured using proper
motions, on cylindrical
Galactocentric coordinates for $172,000$ metal-poor halo stars ($[Fe/H]<-1.1$; top panels) 
and $205,000$ metal-rich disk-like stars ($[Fe/H]>-0.9$; bottom 
panels).  Stars are selected from three regions: $b>80^\circ$
(the North Galactic pole), $170^\circ < l < 190^\circ$ (anticenter), 
and $350^\circ < l < 10^\circ$ (Galactic center).  The left column plots rotational velocity, $v_\Phi$, while the
right column plots $v_B=\sin(b)v_R + \cos(b)v_Z$.  To aid visualization
of these boundaries, see the thick line in the top left
panel in Figure~\ref{qsoPMerrors}. The median values of 
velocity in each bin are color-coded according to the legend 
shown in each panel. The measurements are reliable 
for distances up to about $7$~kpc, but regions beyond this limit
are shown for halo stars for completeness. The fraction
of disk stars is negligible at such distances; their
velocity distribution is shown for $Z<6$~kpc. The region
with negative velocity on the right side of top left panel is
due to contamination of the halo sample by stars from the Monoceros stream. 
The thin region with negative velocity on the left side of top 
right panel is a data artifact (see text).
\label{fig:RZpanelsGPblue}}
\end{figure}

\begin{figure}                      
\figurenum{9}
\vskip -2.5in        
\plotone{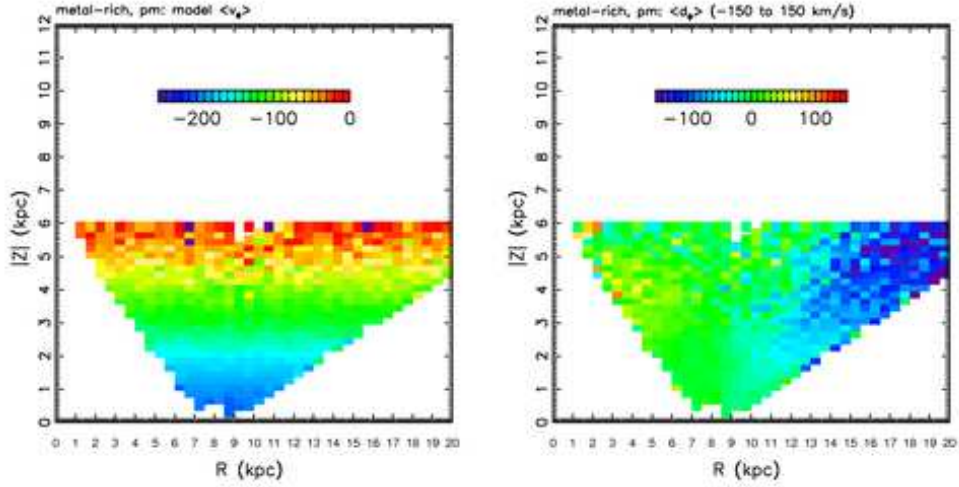}
\vskip -0.7in        
\caption{Left panel is analogous to the bottom 
left panel in Figure~\ref{fig:RZpanelsGPblue}, but for the
model described in the text. The right panel shows
the median difference between the data and model.
Large discrepancies at $R>12$~kpc are due to the Monoceros 
stream (at $R=18$~kpc and $Z=4$~kpc; disk stars rotate
with a median $v_\phi\sim -100$~\kms, while for the Monoceros 
stream, $v_\phi\sim -200$~\kms). 
\label{fig:RZpanelsGPdiskModel}}
\end{figure}

\begin{figure}
\figurenum{10}
\vskip -0.6in        
\plotone{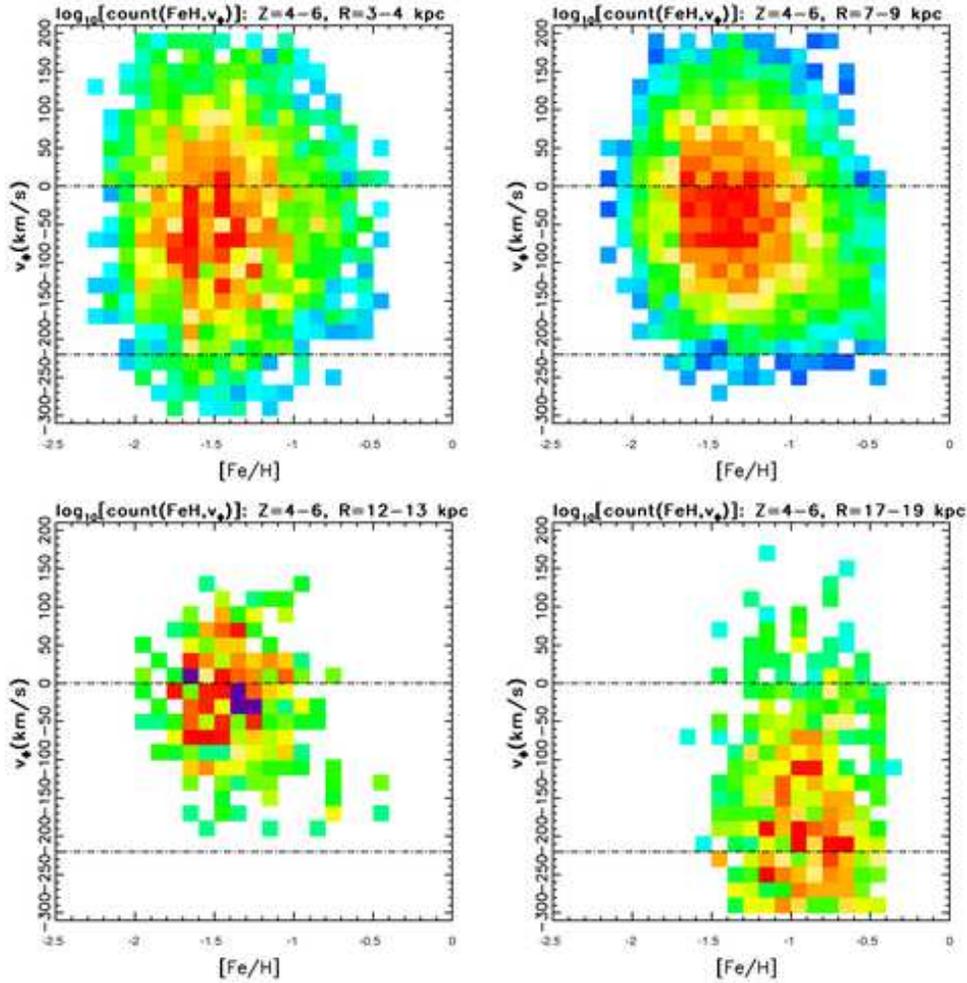}
\vskip -1in        
\caption{
Distribution of stars with $0.2 < g-r < 0.4$ and $Z=4-6$~kpc in the rotational velocity 
vs. metallicity plane, for four ranges of Galactocentric cylindrical
radius, $R$ (top left: $3-4$~kpc; top right: $7-9$~kpc; bottom left:
$12-13$~kpc; bottom right: $17-19$~kpc). In each panel, the color-coded
map shows the logarithm of counts in each pixel, scaled by the total 
number of stars. The horizontal lines at  $v_\Phi=0$~\kms\ and  $v_\Phi=-220$~\kms\ are added to guide the eye. High-metallicity ($[Fe/H]\sim -1$)
stars with fast rotation ($v_\Phi \sim -220$~\kms) visible in the 
bottom right panel belong to the Monoceros stream, and are responsible
for the features seen at $R>15$~kpc in the two left panels in 
Figure~\ref{fig:RZpanelsGPblue}.
\label{fig:vPhiFeHpanelsGPblue}}
\end{figure}

\begin{figure}                
\figurenum{11}
\vskip -3in        
\plotone{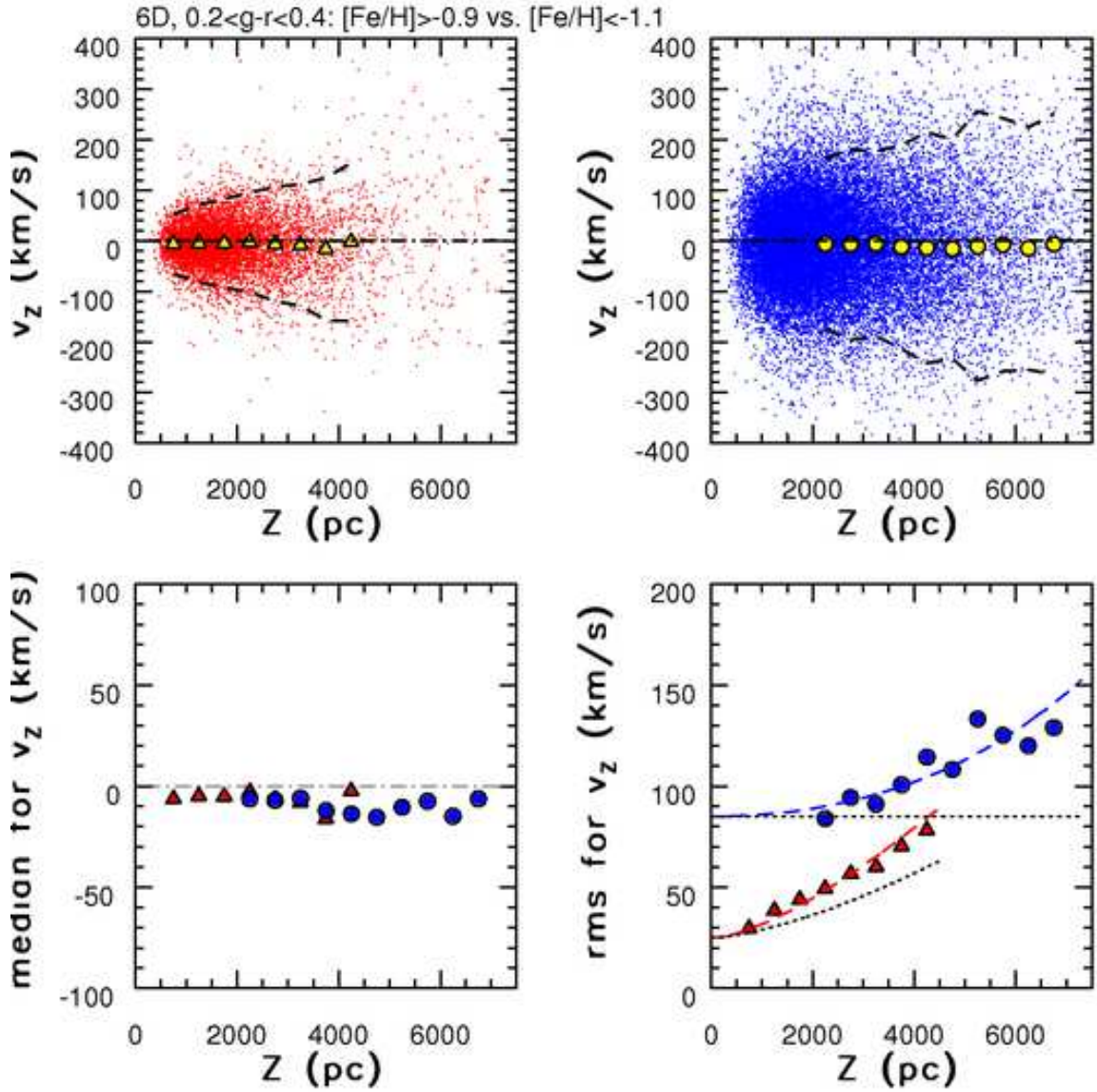}
\vskip -0.5in        
\caption{Similar to Figure~\ref{fig:VvsZphi}, but for the vertical-velocity component, $v_Z$, and using 
a sample of stars with SDSS radial-velocity measurements,
$0.2<g-r<0.4$ and $b>20^\circ$ ($12,000$ stars in the high-metallicity 
subsample, and $38,000$ stars in the low-metallicity subsample). 
An analogous figure for extended samples of $53,000$ disk stars and 
$47,000$ halo stars with $0.2<g-r<0.6$ has a similar appearance. 
The behavior of the rotational- and radial-velocity components
in this sample is consistent with that shown in 
Figures~\ref{fig:VvsZphi} and \ref{fig:VvsZR}. 
\label{fig:VvsZZ}}
\end{figure}

\begin{figure}                      
\figurenum{12}
\vskip -1in
\plotone{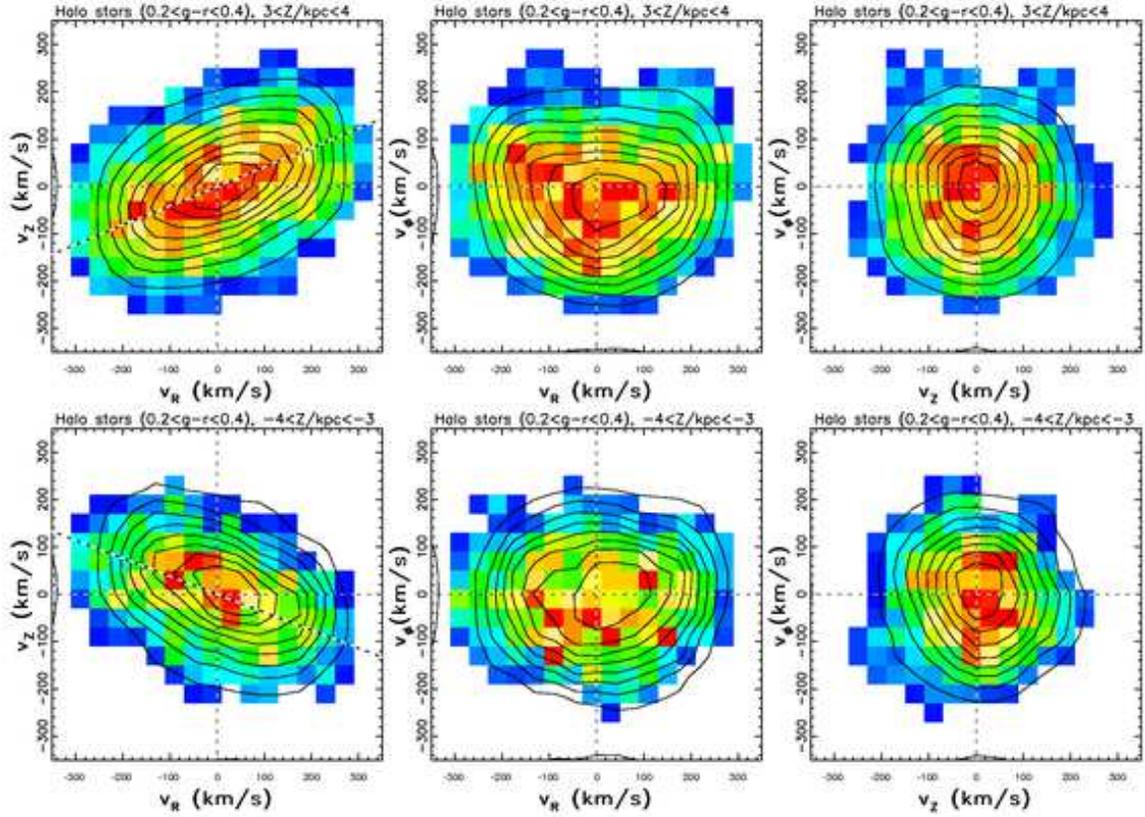}
\vskip -0.5in
\caption{Three two-dimensional projections
of the velocity distribution for two subsamples 
of candidate halo stars selected using spectroscopic 
metallicity ($-3<[Fe/H]<-1.1$) and with $6<R/{\rm kpc}<11$. 
The top row corresponds to $2,700$ stars with distances,
$3<Z/{\rm kpc}<4$, and the bottom row 
to $1,300$ stars with $-4<Z/{\rm kpc}<-3$. The distributions 
are shown using linearly-spaced contours, and with a color-coded
map showing smoothed counts in pixels (low-to-high 
from blue-to-red). The measurement errors are typically
$60$~\kms. Note the strong evidence for a velocity-ellipsoid
tilt in the top and bottom left panels (see also Fig.~\ref{fig:VVHVT2}). 
The two dashed lines in these panels show the median direction 
towards the Galactic center. 
\label{fig:VVHVT}}
\end{figure}

\begin{figure}                      
\figurenum{13}
\vskip -1in
\plotone{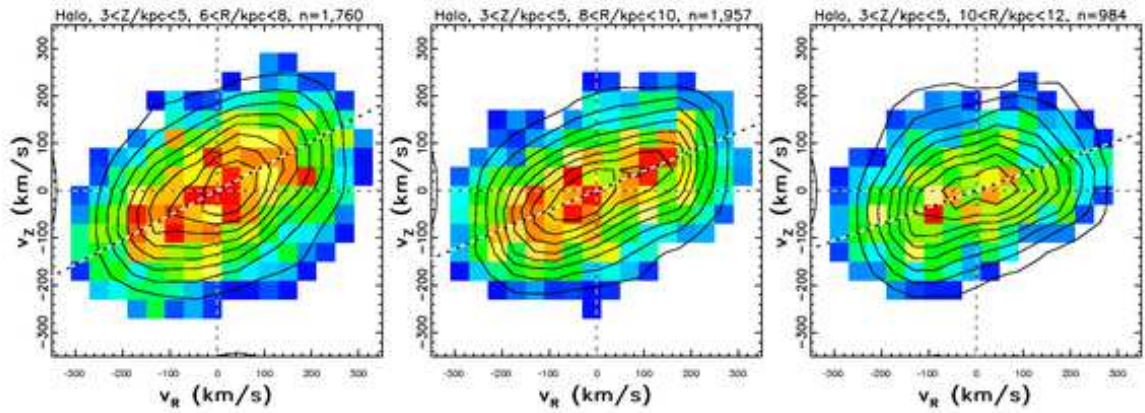}
\vskip -0.5in
\caption{Illustration of the velocity-ellipsoid tilt- 
angle variation. Analogous to Fig.~\ref{fig:VVHVT}, 
except that only the $v_Z$ vs. $v_R$ projection is shown
for a constant $Z$, for three ranges of $R$, as marked 
on the top of each panel. 
\label{fig:VVHVT2}}
\end{figure}

\begin{figure}                      
\figurenum{14}
\vskip -1in
\plotone{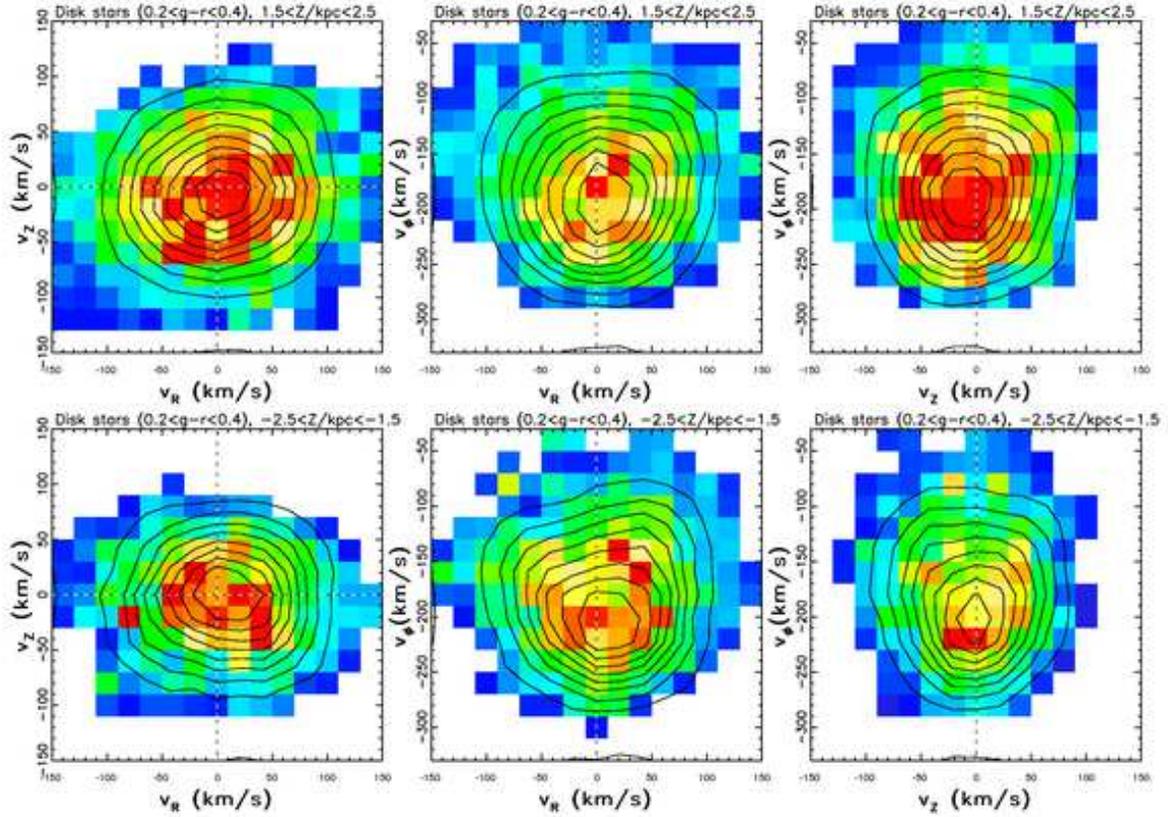}
\vskip -0.5in
\caption{Analogous to Figure~\ref{fig:VVHVT}, except
that the velocity distribution is shown for two subsamples 
of candidate disk stars selected using spectroscopic 
metallicity ($[Fe/H] > -0.9$). The top row corresponds
to $2,200$ stars with distances from the Galactic plane 
$1.5<Z/{\rm kpc}<2.5$, and the bottom row to $1,500$ stars 
with $-2.5<Z/{\rm kpc}<-1.5$. The measurement errors are 
typically $35$~\kms. Note the absence of velocity-ellipsoid
tilt in the top and bottom left panels. 
\label{fig:VVDVT}}
\end{figure}

\begin{figure}                      
\figurenum{15}
\vskip -1in
\plotone{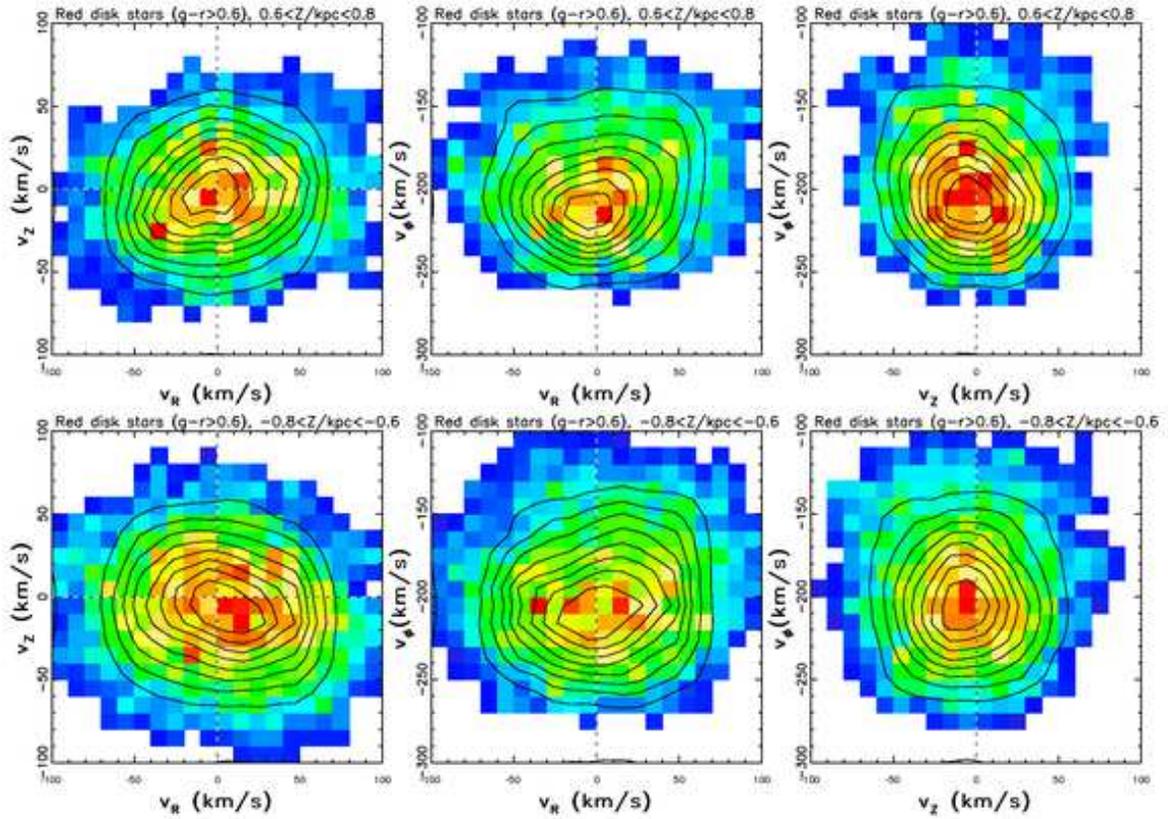}
\vskip -0.5in
\caption{Analogous to Figure~\ref{fig:VVDVT}, except
that the velocity distribution is shown for two subsamples 
of red stars ($g-r>0.6$): the top row corresponds
to $3,000$ stars with distances from the Galactic plane 
$0.6<Z/{\rm kpc}<0.8$, and the bottom row to $4,600$ stars 
with $-0.8<Z/{\rm kpc}<-0.6$. The measurement errors are 
typically $\sim 15$~\kms.  There is no strong evidence in these panels for 
a tilt in the velocity ellipsoid.  
\label{fig:VVDredVT}}
\end{figure}

\clearpage



\begin{figure}                      
\figurenum{16}
\vskip -3.5in
\epsscale{0.9}        
\plotone{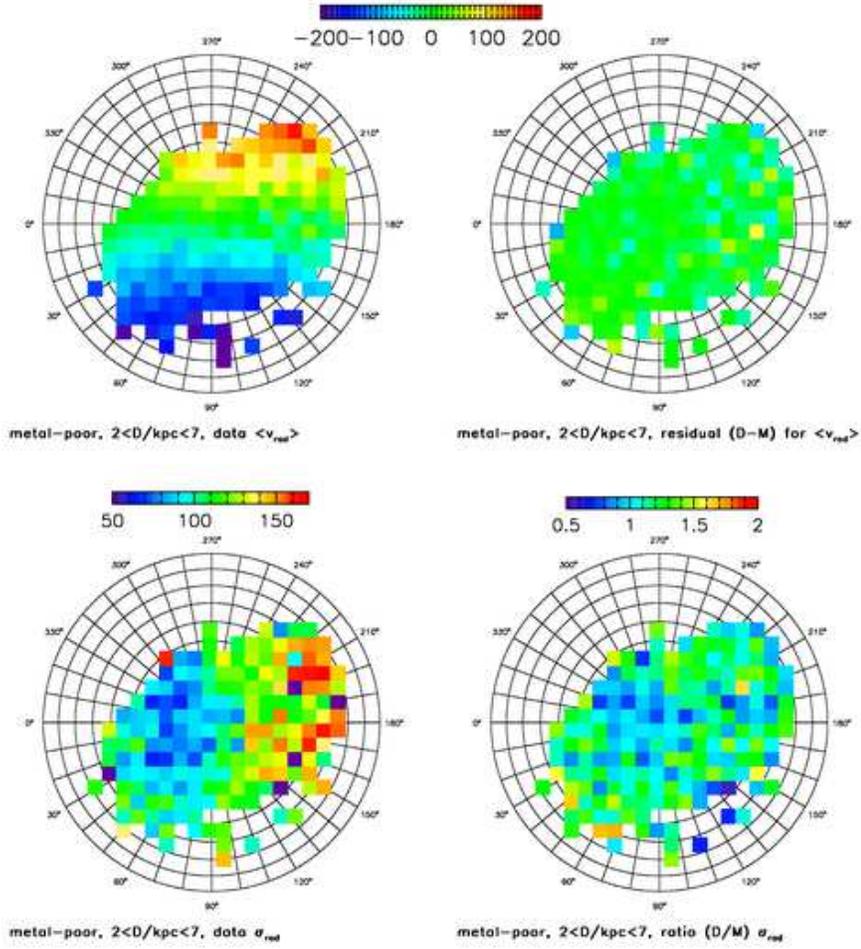}
\epsscale{1.0}        
\vskip -1.4in
\caption{Comparison of medians and dispersions for the measured and modeled 
radial velocities of $20,000$ blue ($0.2<g-r<0.4$) halo stars (spectroscopic
$[Fe/H]<-1.1$) at distances, $D=2-7$~kpc, and $b>20^\circ$. 
The top left panel shows the median measured radial velocity in each pixel, 
color-coded according to the legend shown at the top (units are \kms).
The top right panel shows the difference between this map and an
analogous map based on model-generated values of radial velocity,
using the same scale as in the top left panel. The bottom left panel 
shows the dispersion of measured radial velocities, color-coded according 
to the legend above it. The bottom right panel
shows the ratio of this map and an analogous map based on 
model-generated values of radial velocity, color-coded 
according to the legend above it. When the 
sample is divided into $1$~kpc distance shells, the behavior
is similar. 
\label{fig:LambPanelsHvrad}}
\end{figure}

\begin{figure}                      
\figurenum{17}
\vskip -1in
\plotone{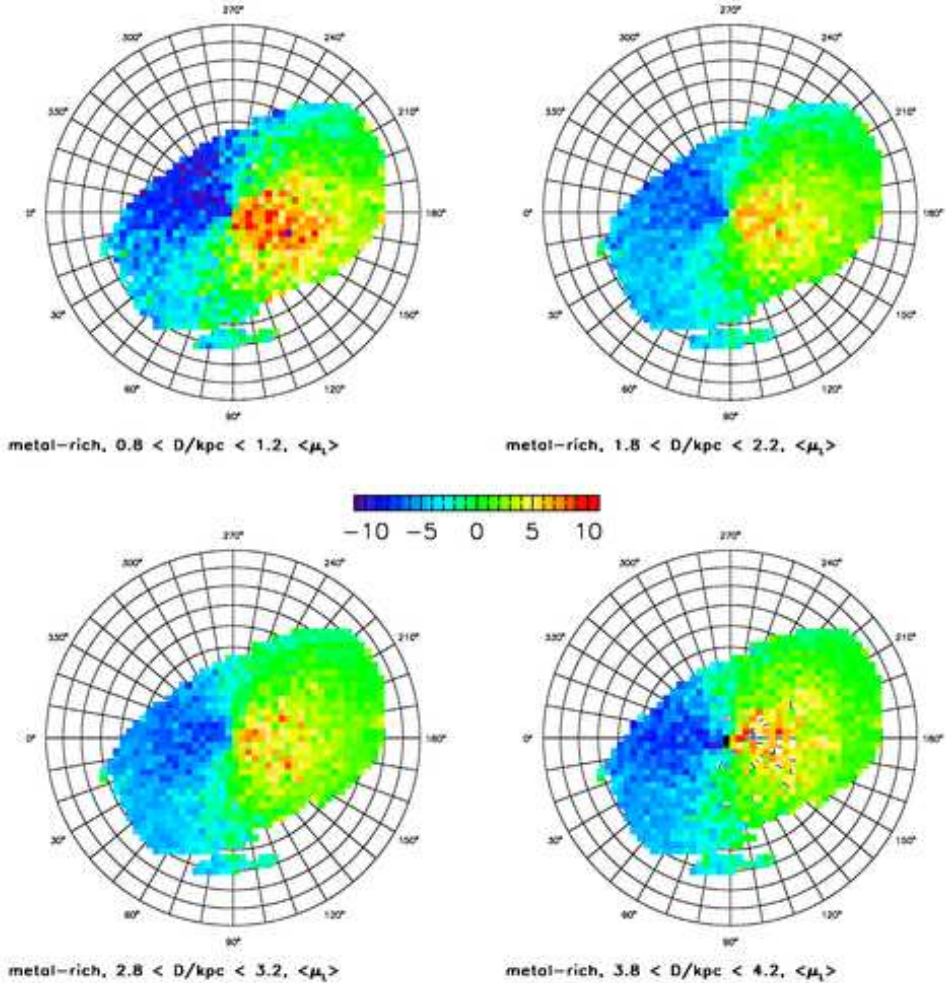}
\vskip -1.4in
\caption{Distribution of the median longitudinal proper motion
in a Lambert projection of the North Galactic cap for 
high-metallicity ($[Fe/H]>-0.9$), blue ($0.2<g-r<0.4$) stars,
in several distance bins (top left: $58,000$ stars with $D=0.8-1.2$~kpc;
top right: $119,000$ stars with $D=1.8-2.2$~kpc; bottom left: $72,000$ 
stars with $D=2.8-3.2$~kpc; bottom right: $43,000$ stars with 
$D=3.8-4.2$~kpc). All maps are color-coded using the same 
scale, shown in the middle (units are mas~yr$^{-1}$). Note that the magnitude of the
proper motion does not change appreciably as the distance 
varies from $\sim2$~kpc to $\sim4$~kpc; this is due to a vertical 
gradient in the rotational velocity for disk stars (see 
Figure~\ref{fig:vPhivRred}). 
\label{fig:LambPanelsDpmL}}
\end{figure}

\begin{figure}                      
\figurenum{18}
\vskip -1in
\plotone{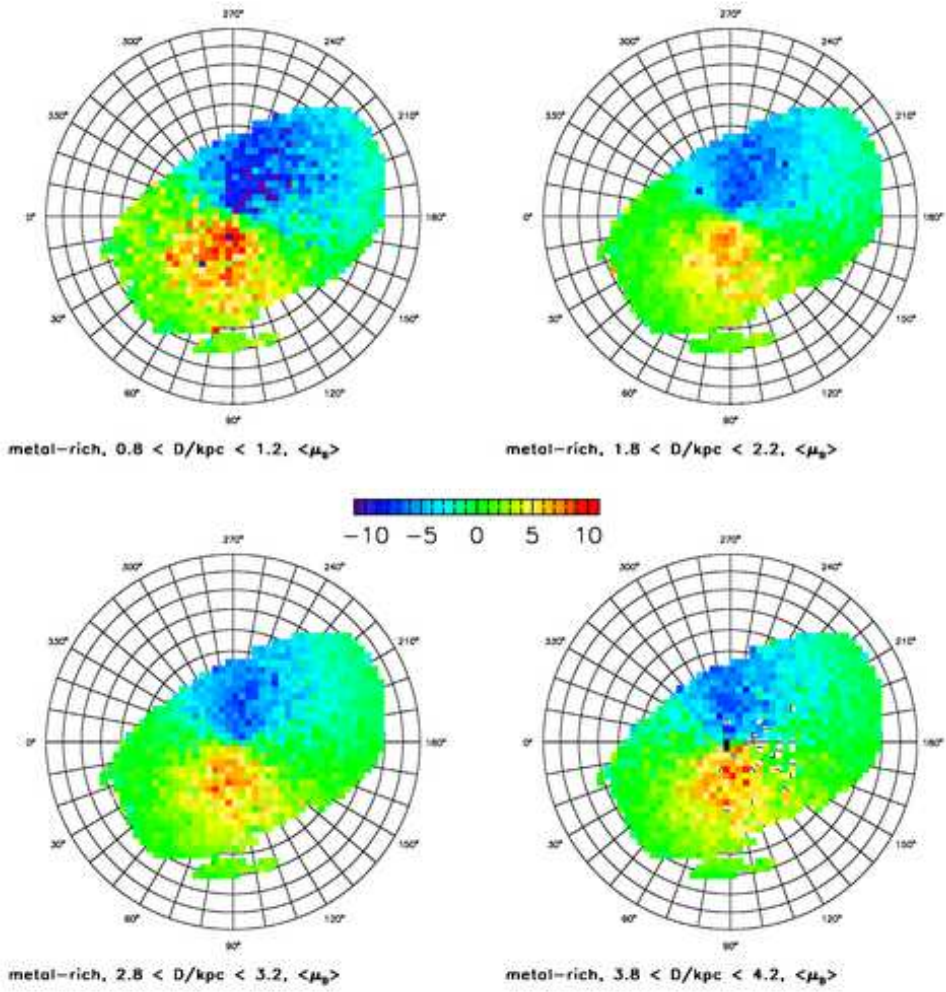}
\vskip -1in
\caption{Analogous to Figure~\ref{fig:LambPanelsDpmL}, except
that the latitudinal proper motion is shown.
\label{fig:LambPanelsDpmB}}
\end{figure}

\begin{figure}                      
\figurenum{19}
\vskip -1in
\plotone{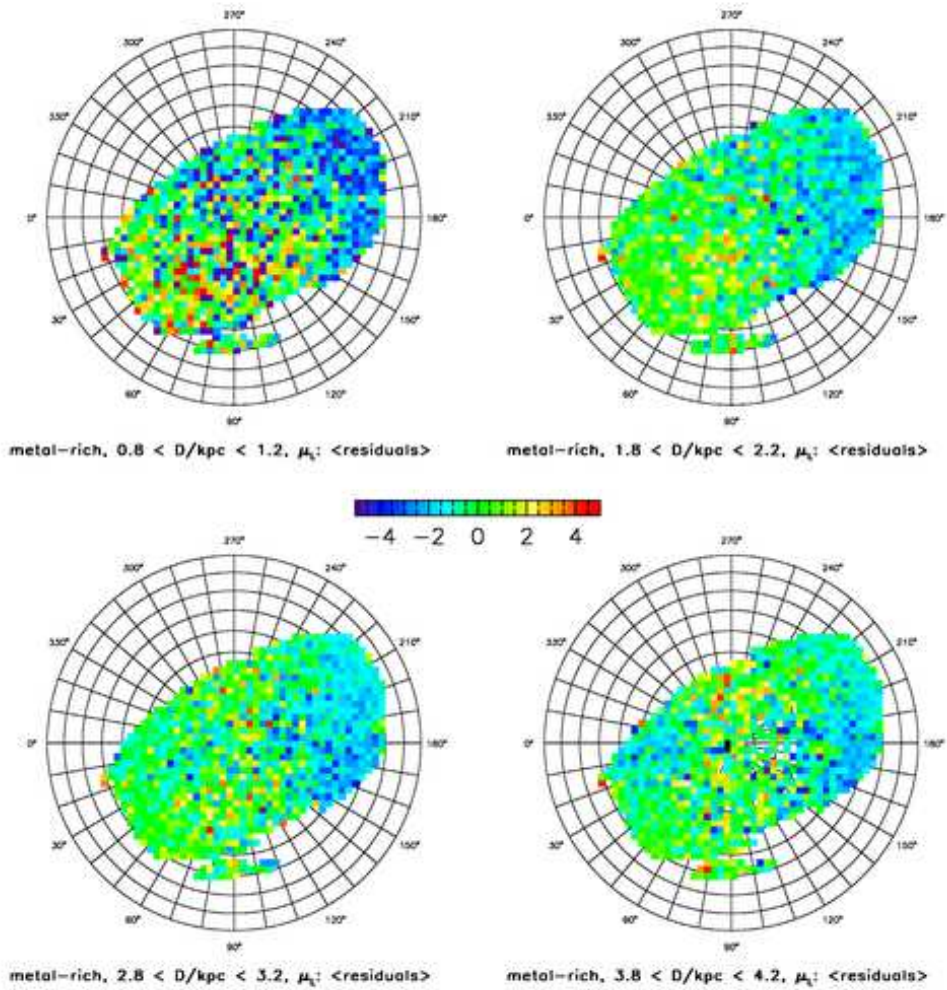}
\vskip -1in
\caption{Similar to Figure~\ref{fig:LambPanelsDpmL}, except
that the median difference between the observed value of longitudinal 
proper motion and a value predicted by the model described
in the text is shown. All maps are color-coded using the same 
scale, shown in the middle. Note that the displayed scale is
stretched by a factor of two compared to the scale from 
Figure~\ref{fig:LambPanelsDpmL}, in order to emphasize
discrepancies.
\label{fig:LambPanelsDpmLresid}}
\end{figure}

\begin{figure}                      
\figurenum{20}
\vskip -1in
\plotone{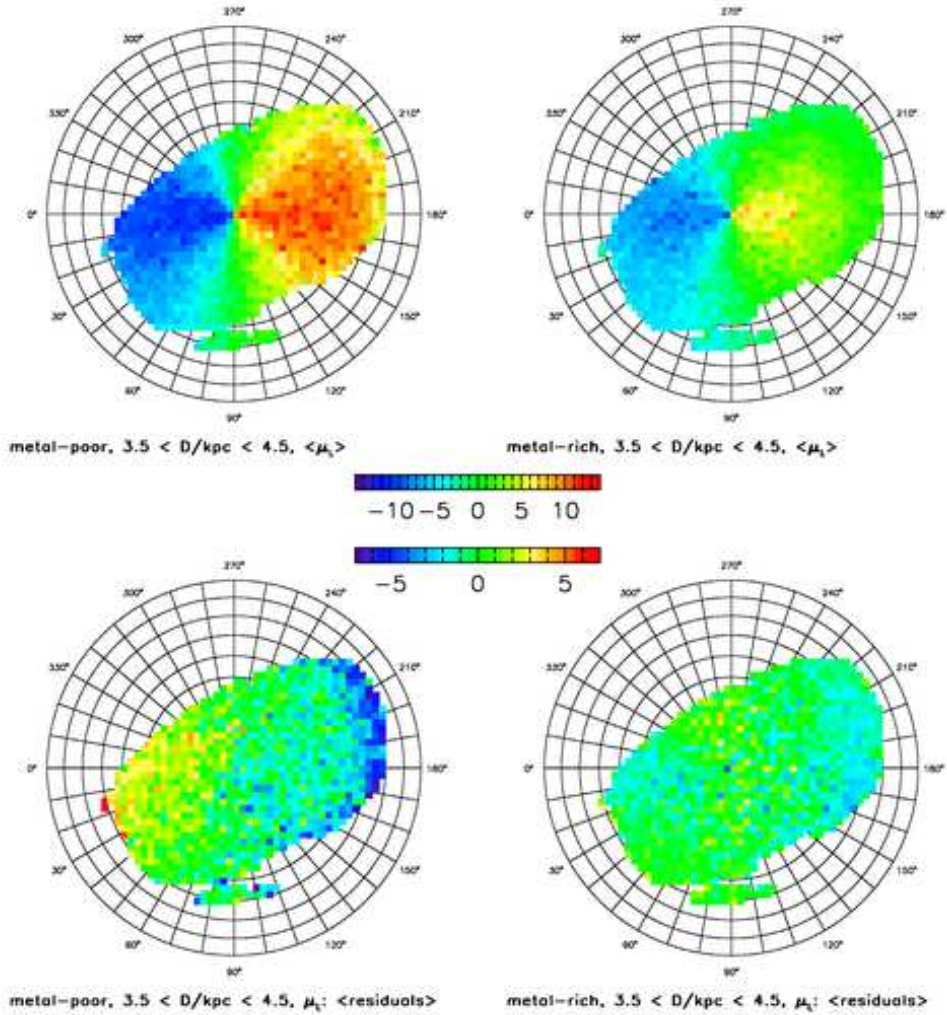}
\vskip -1in
\caption{Similar to Figures~\ref{fig:LambPanelsDpmL} and
\ref{fig:LambPanelsDpmLresid}, except that the behavior of 
high-metallicity (left) and low-metallicity (right) stars is 
compared in a single distance bin ($3.5-4.5$~kpc). The top 
two panels show the median longitudinal proper motion,
and the two bottom panels show the median difference between the 
observed and model-predicted values. An analogous figure for the
latitudinal proper motion has similar characteristics.
\label{fig:LambPanelsDpmLDds5Hds5}}
\end{figure}

\begin{figure}                      
\figurenum{21}
\vskip -1.4in
\plotone{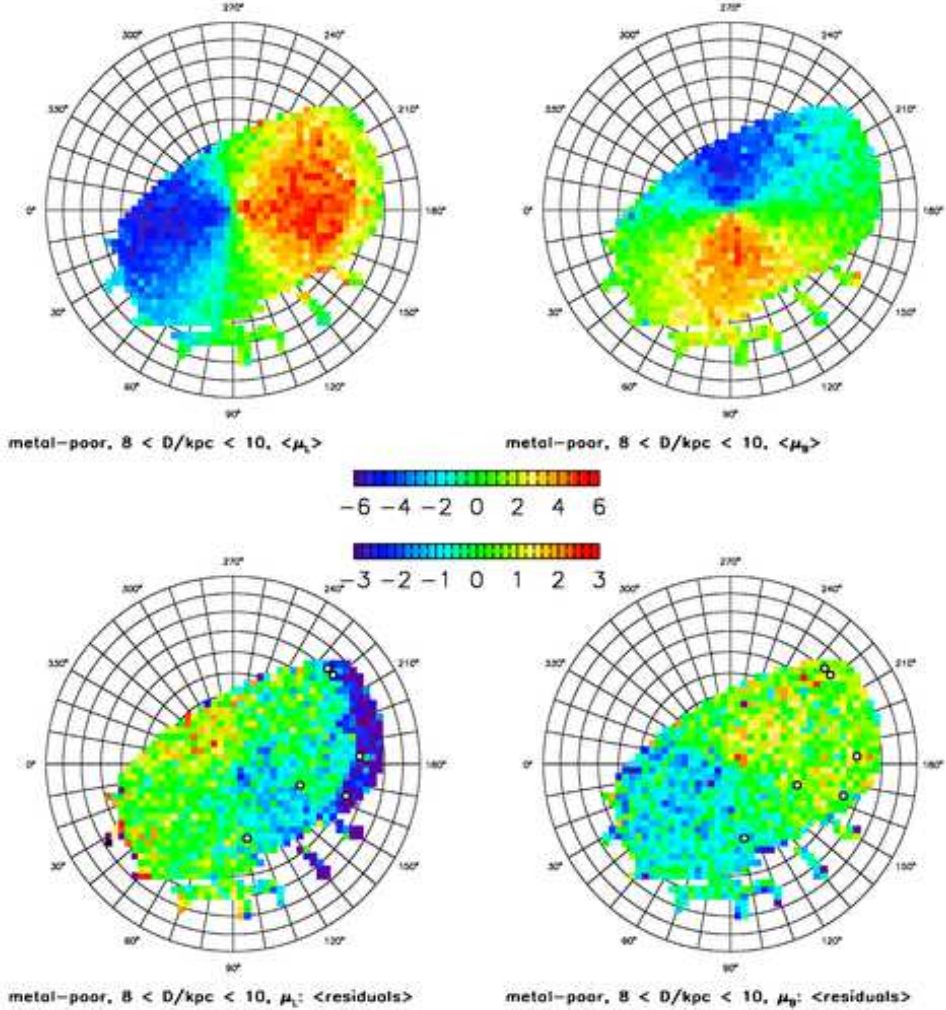}
\vskip -1.7in
\caption{Similar to Figure~\ref{fig:LambPanelsDpmLDds5Hds5}, 
except that the behavior of low-metallicity stars from
the $8-10$~kpc distance bin is analyzed. The top 
two panels show the median longitudinal (left) and latitudinal
(right) proper motions, and the two bottom panels show 
the median difference between the observed and model-predicted 
values. The maps are color-coded according to the legends in
the middle (mas~yr$^{-1}$; note that the bottom scale has a harder
stretch to emphasize structure in the residual maps). 
The two bottom panels display very similar morphology to 
systematic proper-motion errors shown in the two left panels
in Figure~\ref{qsoPMerrors}. 
In the bottom panels, the white symbols show the
positions of the six northern cold substructures identified by Schlaufman et al. (2009). 
\label{fig:LambPanelsHds4}}
\end{figure}

\begin{figure}                      
\figurenum{22}
\vskip -1in
\plotone{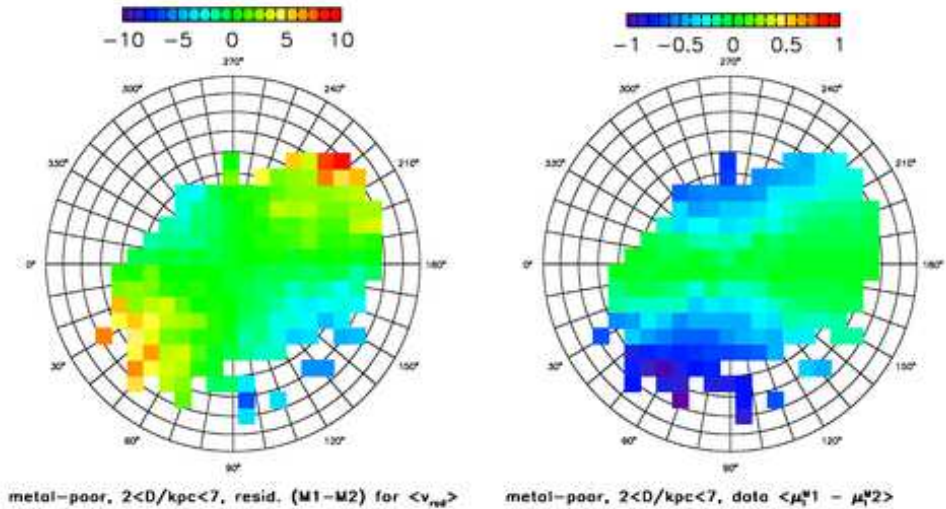}
\vskip -1in
\caption{Comparison of radial-velocity residuals (left panel;
analogous to the top right panel in Fig.~\ref{fig:LambPanelsHvrad})
and longitudinal proper-motion residuals (right panel; 
analogous to the bottom left panel in
Fig.~\ref{fig:LambPanelsDpmLDds5Hds5}, except for the larger
distance range) for two halo models with ($v_\phi^{halo}$,$v_{LSR}$)=
($-20$,$180$)~\kms, and ($20$,$220$)~\kms.  Note that we set 
$v_{LSR}-v_\phi^{halo}=200$~\kms. The residuals are color-coded
according to the legend above each panel (units are \kms\ for
the left panel and mas~yr$^{-1}$ for the right panel). In order to distinguish 
these models, systematic errors in radial velocity must be below
$10$~\kms, and systematic errors in proper motion must be below 
$1$~mas~yr$^{-1}$.
\label{fig:LambPanelsModelComparison}}
\end{figure}

%% file: figuresApp.tex
\begin{figure}
\figurenum{A1}
\vskip -0.2in
\plotone{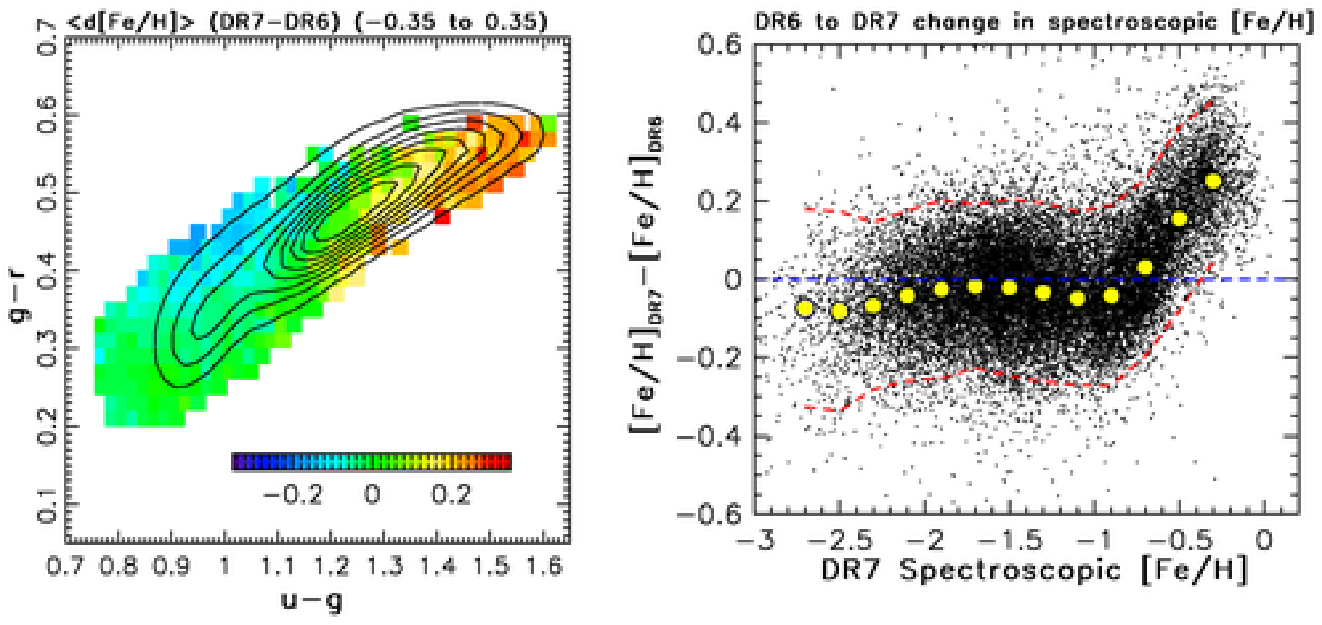}
\vskip -0.6in
\caption{Summary of the differences between SDSS spectroscopic-metallicity
values distributed with Data Releases 6 and 7. The left panel shows
the median difference between the DR7 and DR6 values for $0.02\times0.02$~mag$^2$ bins in the $g-r$ vs. $u-g$ color-color diagram, color-coded 
according to the legend shown in the panel. The largest differences 
of $0.2-0.3$~dex are seen in the top right corner, which corresponds
to high metallicities. The right panel shows the difference in 
metallicities as a function of the new DR7 values. Individual stars
are shown as small dots, and the median values of the difference are 
shown as large circles. The two dashed lines mark the $\pm 2\sigma$ envelope
around the medians, where $\sigma$ is the root-mean-square scatter 
($\sim0.1$~dex, due to software updates) estimated from the inter-quartile 
range. The median differences are larger than $0.1$~dex only at the 
high-metallicity end ($[Fe/H]>-0.6$).}
\label{Fig:App1}
\end{figure}

\begin{figure}
\figurenum{A2}
\vskip -1.4in
\epsscale{0.9}        
\plotone{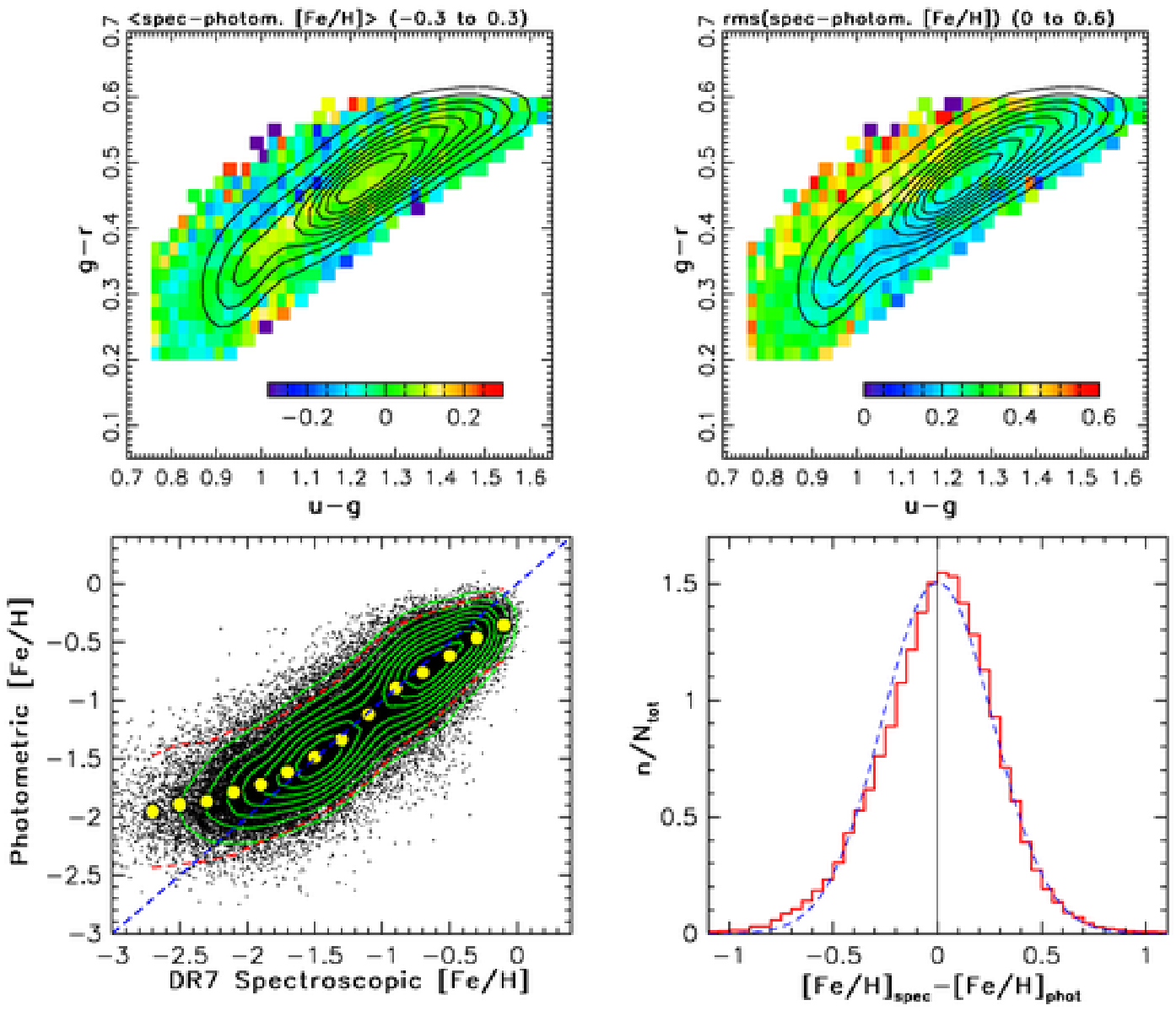}
\epsscale{1.0}        
\vskip -1.6in
\caption{Color-coded map in the top left panel shows the 
median difference between spectroscopic and revised photometric 
metallicities for $\sim50,000$ stars from SDSS Data Release 7. 
The median value is zero, and the root-mean-square scatter is $0.07$~dex. 
The contours show the distribution of stars with $r<20$ and at
high Galactic latitudes. The top right panel shows the 
root-mean-square scatter of the difference between spectroscopic 
and photometric metallicities in each pixel. The top two panels
are analogous to the bottom two panels in figure~2 from I08. The bottom 
left panel shows the photometric metallicity as a function of the
spectroscopic metallicity. Individual stars are shown by small dots, 
and the median values of the difference are shown by large circles. 
The distribution of stars is shown as linearly-spaced contours. 
Note that the photometric metallicity saturates at $[Fe/H]\sim -2$
at the low-metallicity end. The histogram in the bottom right panel 
shows the distribution of the difference between spectroscopic 
and photometric metallicities for stars with spectroscopic metallicity
$[Fe/H]>-2.2$. A best-fit Gaussian centered on zero and with 
a width of $0.26$~dex is shown by the dashed line.}
\label{Fig:App2}
\end{figure}

\begin{figure}
\figurenum{A3}
\vskip -2.9in
\plotone{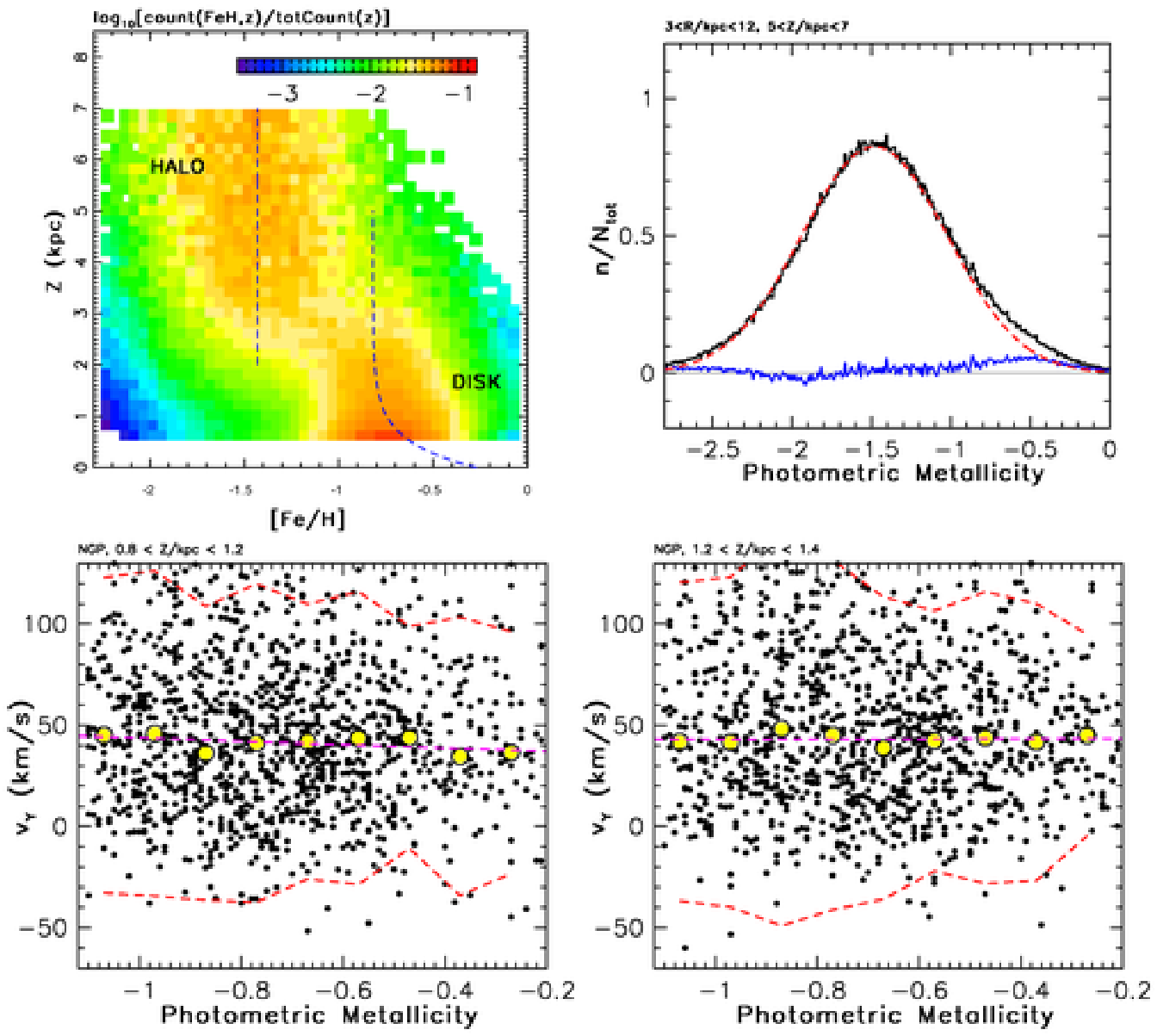}
\vskip -1.9in
\caption{Conditional metallicity
  distribution as a function of distance from the Galactic plane (top left), analogous
  to the bottom left panel of figure~9 from I08.  Note
  that the new photometric metallicities include $[Fe/H]>-0.4$.  The
  dashed line, which shows the median disk metallicity, is also
  revised (see text). The top right panel is analogous to figure~10
  from I08, and shows the metallicity distribution for stars with
  $5<Z<7$~kpc, where $Z$ is the distance from the Galactic plane. Note
  that the photometric-metallicity artifact at $[Fe/H] = -0.5$
  discussed by I08 is no longer present.  However, there is still
  evidence that disk stars exist at such large distances from the
  plane. The bottom two panels show the heliocentric rotational
  velocity for disk stars in two thin $Z$ slices, and are analogous to
  the bottom right panel in figure~16 from I08. Note that the
  correlation between velocity and metallicity is still absent.}
\label{Fig:App3}
\end{figure}

\begin{figure}
\figurenum{A4}
\vskip -0.2in
\plotone{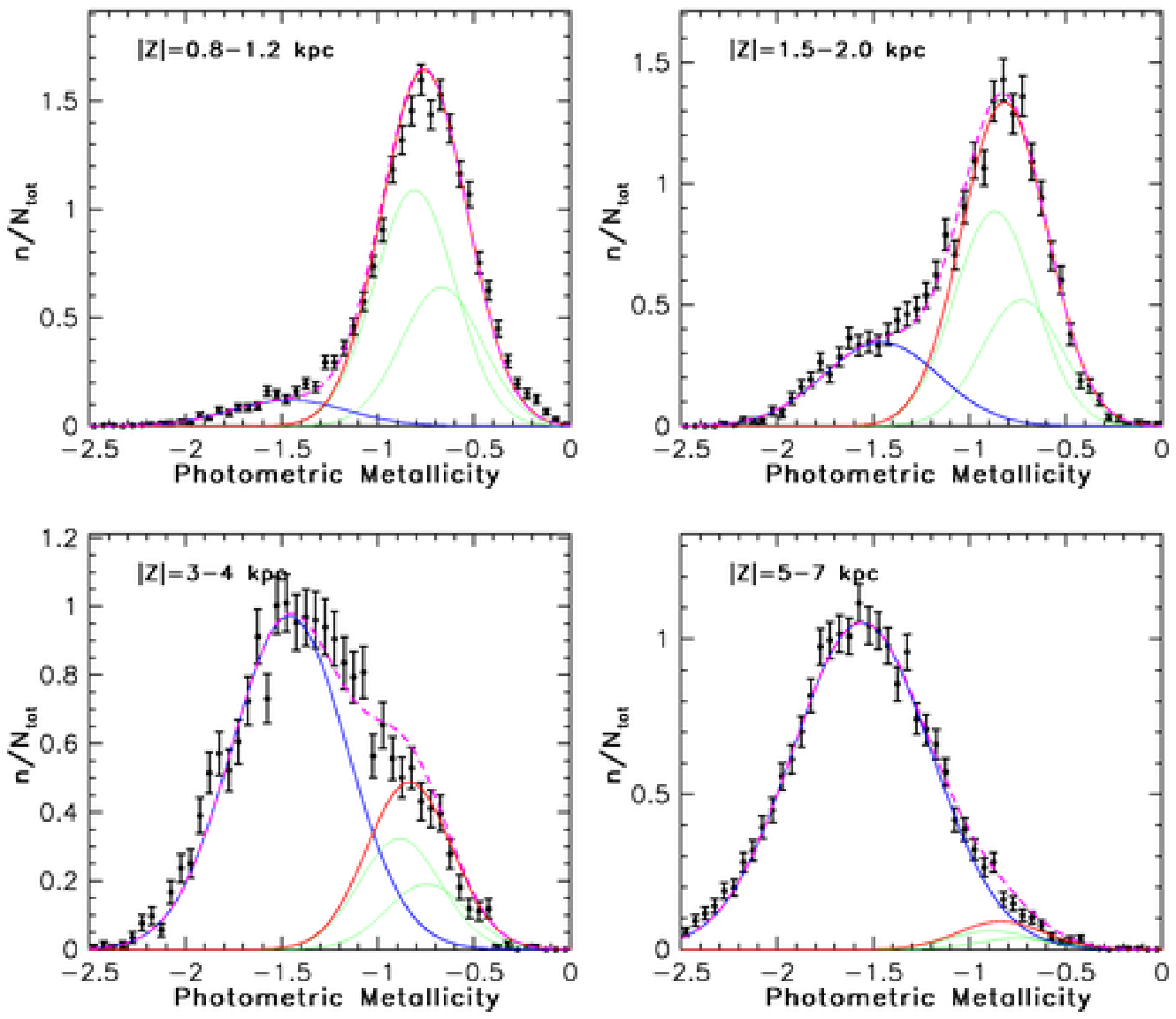}
\vskip -1.2in
\caption{Analogous to figure~7 from I08. The symbols with error
bars show the metallicity distribution for stars with $0.2 < g-r < 0.4$, 
$7<R<9$~kpc, and distances from the Galactic plane 
as marked, where $R$ is the Galactocentric cylindrical radius. 
The behavior is qualitatively similar to that seen in 
I08. The only significant quantitative difference is in the 
model for the metallicity distribution of disk stars (see 
text).}
\label{Fig:App4}
\end{figure}